\documentclass{aa} 
\usepackage{comment}
\usepackage{mathtools}
\usepackage{hyperref}
\hypersetup{
    colorlinks=true,
    linkcolor=blue,
    citecolor=blue,
    urlcolor=green,
    }
\usepackage{natbib}
\usepackage{graphicx}
\usepackage{txfonts}
\usepackage{ulem}
\usepackage{threeparttable}
\usepackage{mathrsfs}
\usepackage{amsmath}
\usepackage{placeins}
\usepackage{afterpage}
\usepackage{caption}
\usepackage[rightcaption]{sidecap}
\begin{document}

\title{Resurging from the ashes: A spectral study of seven candidate revived radio fossils in nearby low-mass galaxy clusters}

   \author{L. Bruno
          \inst{1},
          A. Botteon
          \inst{1},
          D. Dallacasa
          \inst{2,1},
          T. Venturi
          \inst{1,3},
          M. Balboni
          \inst{2,4},
          N. Biava
          \inst{5},
          M. Brienza
          \inst{1},
          M. Br\"uggen
          \inst{6},
          G. Brunetti
          \inst{1},
          F. de Gasperin
          \inst{1},
          E. De Rubeis
          \inst{2,1},
          G. Di Gennaro
          \inst{1},
          F. Gastaldello
          \inst{4},
          A. Ignesti
          \inst{7},
          T. Pasini
          \inst{1},
          K. Rajpurohit
          \inst{8},
        A. Shulevski
          \inst{9,10,11,12},
          K. S. L. Srikanth
          \inst{2,1},
          R. J. van Weeren
          \inst{13},
          X. Zhang
          \inst{14}
          }

   \institute{
       Istituto Nazionale di Astrofisica (INAF) - Istituto di Radioastronomia (IRA), via Gobetti 101, 40129 Bologna, Italy 
       \and
   Dipartimento di Fisica e Astronomia (DIFA), Universit\`a di Bologna, via Gobetti 93/2, 40129 Bologna, Italy
        \and
    Center for Radio Astronomy Techniques and Technologies, Rhodes University, Grahamstown 6140, South Africa
        \and
     Istituto Nazionale di Astroﬁsica – Istituto di Astroﬁsica Spaziale e Fisica cosmica (IASF), Via A. Corti 12, 20133 Milano, Italy
    \and 
      Th\"uringer Landessternwarte, Sternwarte 5, 07778 Tautenburg, Germany
      \and 
    Hamburger Sternwarte, Universit\"at Hamburg, Gojenbergsweg 112, 21029 Hamburg, Germany
         \and
     Istituto Nazionale di Astroﬁsica (INAF) – Osservatorio Astronomico di Padova (OAPD), Vicolo dell’Osservatorio 5, I-35122 Padova, Italy  
    \and      
    Center for Astrophysics | Harvard \& Smithsonian, 60 Garden Street, Cambridge, MA 02138, USA
    \and
    ASTRON, the Netherlands Institute for Radio Astronomy, Postbus 2, 7990 AA, Dwingeloo, The Netherlands
    \and
    Kapteyn Astronomical Institute, University of Groningen, PO Box 800, 9700 AV Groningen, The Netherlands
    \and 
    Anton Pannekoek Institute for Astronomy, University of Amsterdam, Postbus 94249, 1090 GE Amsterdam, The Netherlands
    \and 
    Center for Advanced Interdisciplinary Research, Ss. Cyril and Methodius University in Skopje, Macedonia
    \and 
    Leiden Observatory, Leiden University, PO Box 9513, 2300 RA Leiden, The Netherlands
    \and 
    Max-Planck-Institut fu\"ur Extraterrestrische Physik (MPE), Gießenbachstraße 1, D-85748 Garching bei M\"unchen, Germany
    \\
    \email{luca.bruno@inaf.it}
 }

 
  \abstract
   {Complex energy transfer processes in the intracluster medium (ICM) can revive fossil (with spectral ages $\gg100$ Myr) plasma initially generated by radio galaxies. This leads to the re-ignition of faint radio sources with irregular and filamentary morphologies, and ultra-steep ($\alpha \gtrsim 1.5$) synchrotron spectra, which can be more easily detected at low frequencies ($\sim 100$ MHz). These sources offer the opportunity to investigate the microphysics of the ICM and its interplay with radio galaxies, the origin of seed relativistic electrons, the merging history of the host cluster, and the phenomenology of radio filaments.}
   {The study of revived sources has so far been hampered by the requirement of sensitive and high-resolution multi-frequency radio data at low frequencies to characterise their spatial properties and provide a proper classification. We aim to perform the analysis of a sample of candidate revived sources identified among nearby ($z\leq0.35$) and low-mass ($M_{500}\leq5\times 10^{14} M_\odot$) \textit{Planck} clusters in the footprint of the Second Data Release of the LOw Frequency ARray (LOFAR) Two Metre Sky Survey (LoTSS-DR2).}
   {By inspecting LoTSS-DR2 images at 144 MHz, we identified 7 targets with patchy and filamentary morphologies, which have been followed-up with the upgraded Giant Metrewave Radio Telescope (uGMRT) at 400 MHz. By combining LOFAR and uGMRT data, we obtained high-resolution images and spectral index maps, which we used to interpret the nature of the sources.}
   {All targets show regions with very steep spectra, confirming the effectiveness of our morphology-based selection in identifying fossil plasma. Based on their morphology, spectral properties, and optical associations, we investigated the origin of the targets. We found a variety of promising revived fossil sources, while also showing that apparently intricate structures can be easily misclassified in the absence of high-resolution and multi-band data.}
   {}


\titlerunning{Revived fossil candidates in galaxy clusters}
\authorrunning{Bruno et al.}
   \maketitle
%

\section{Introduction}

Galaxy clusters form and evolve hierarchically via mergers. These events generate weak shocks and turbulence in the thermal intracluster medium (ICM), which dissipate a fraction of the merger energy into the non-thermal components of the ICM. Populations of pre-existing relativistic electrons  (with Lorentz factor of $\gamma\sim 100$) can be re-accelerated by these shocks and turbulence, thus originating steep-spectrum ($\alpha \sim 1-1.5$) synchrotron diffuse emission on megaparsec-scales in the form of radio relics (RRs) and radio halos (RHs), respectively  \citep[e.g.][for a review]{vanweeren19}. The seed electrons are reasonably injected by AGN, but the details on their origin and the energy transfer mechanisms operating in the ICM are not completely understood \citep[e.g.][for reviews]{brunetti&jones14,vazza&botteon24}. 

In addition to RRs and RHs, more elusive, irregular, and filamentary diffuse sources, extending for a few hundreds of kiloparsecs, and characterised by ultra-steep spectral indices ($\alpha \gtrsim 1.5$), have been revealed in some galaxy clusters \citep[e.g.][]{slee01,cohen&clarke11,shimwell16,degasperin17,mandal20,raja24,shulevski24,rotella25}. These sources are thought to trace fossil lobes and tails of radio galaxies that have been revived by magneto-hydrodynamical processes in the ICM. Indeed, due to synchrotron and inverse Compton losses, the synchrotron spectrum of relativistic electrons in radio galaxies steepens at high frequencies ($\gtrsim 1$ GHz), making them only visible for a few hundreds of megayears after the jets switch off. If their fossil components gain energy, radio emission can be revived and detected at low frequencies (a few hundreds of megahertz), where they are much brighter due to their ultra-steep spectrum.

Examples of revived fossil sources are radio phoenices (RPs) and Gently Re-Energised Tails (GReETs), classified based on their morphological and spectral properties. Phoenices are patchy, irregular, and filamentary sources, extending for $\sim 100-300$ kpc, and exhibiting an integrated ultra-steep spectral index, which is spatially distributed across the source without a specific trend \citep[e.g.][]{kempner04,kale18,mandal20,rahaman22}. They may trace re-ignition of old radio lobes from adiabatic compression after the passage of a shock wave \citep[e.g.][]{ensslin&gopalkrishna01,ensslin&bruggen02,nolting19}. GReETs likely consist of revived plasma from tailed radio galaxies, which include head-tail (HT), narrow-angle tail (NAT), and wide-angle tail (WAT) radio galaxies \citep[e.g.][]{miley72,owen&rudnick76}. While a gradual spectral steepening with increasing distance from the core is observed along the tail, GReETs show a sudden constant brightness trend across hundreds of kpc and a spectral flattening in regions of the tail that are thought to be `gently', that is inefficiently, re-energised by ICM turbulence \citep[e.g.][]{degasperin17,edler22,ignesti22,pasini22,gopal-krishna24,lusetti24}.

Understanding the physics behind revived sources has an impact on our view of the microphysics of the ICM and its interplay with radio galaxies, the injection and lifecycle of cosmic rays, and, ultimately, the dynamics and merging history of galaxy clusters. However, work on diffuse radio sources in clusters has been mostly focused on RRs and RHs, whereas the study of RPs and GReETs has been hampered by the requirement of sensitive, high-resolution, and multi-frequency radio data at low frequencies to derive their spatially-resolved spectral properties. In this respect, the Second Data Release of the LOFAR Two Meter Sky Survey \citep[LoTSS-DR2;][]{shimwell22LOTSS} has enabled the study of a large variety of radio sources with unprecedented capabilities at low frequencies (120-168 MHz). In the present work, we search for promising revived fossil sources in a sample of nearby and low-mass galaxy clusters in LoTSS-DR2, and investigate their spectral properties and nature with follow-up observations conducted with the upgraded Giant Metrewave Radio Telescope (uGMRT) at 300-500 MHz.

Throughout this paper, we adopted a standard $\Lambda$CDM cosmology with $H_0=70\;\mathrm{km\; s^{-1}\; Mpc^{-1}}$, $\Omega_{\rm M}=0.3$, and $\Omega_{\rm \Lambda}=0.7$. We adopted the convention on the spectral index $\alpha$ as defined from the flux density as $S(\nu) \propto \nu^{-\alpha}$. The paper is organised as follows. In Sect. \ref{sect: Sample selection}, we describe the selection criteria of our sample. In Sect. \ref{sect: Observations and data reduction}, we present the radio data and their processing. In Sect. \ref{sect: Morphological and spectral properties}, we report on the results of our analysis and discuss possible scenarios for the origin of the targets. In Sect. \ref{sect: Discussion}, we summarise our findings and future prospects.

\section{Sample selection}   
\label{sect: Sample selection}

\begin{figure*}
        \centering

\includegraphics[width=0.287\textwidth]{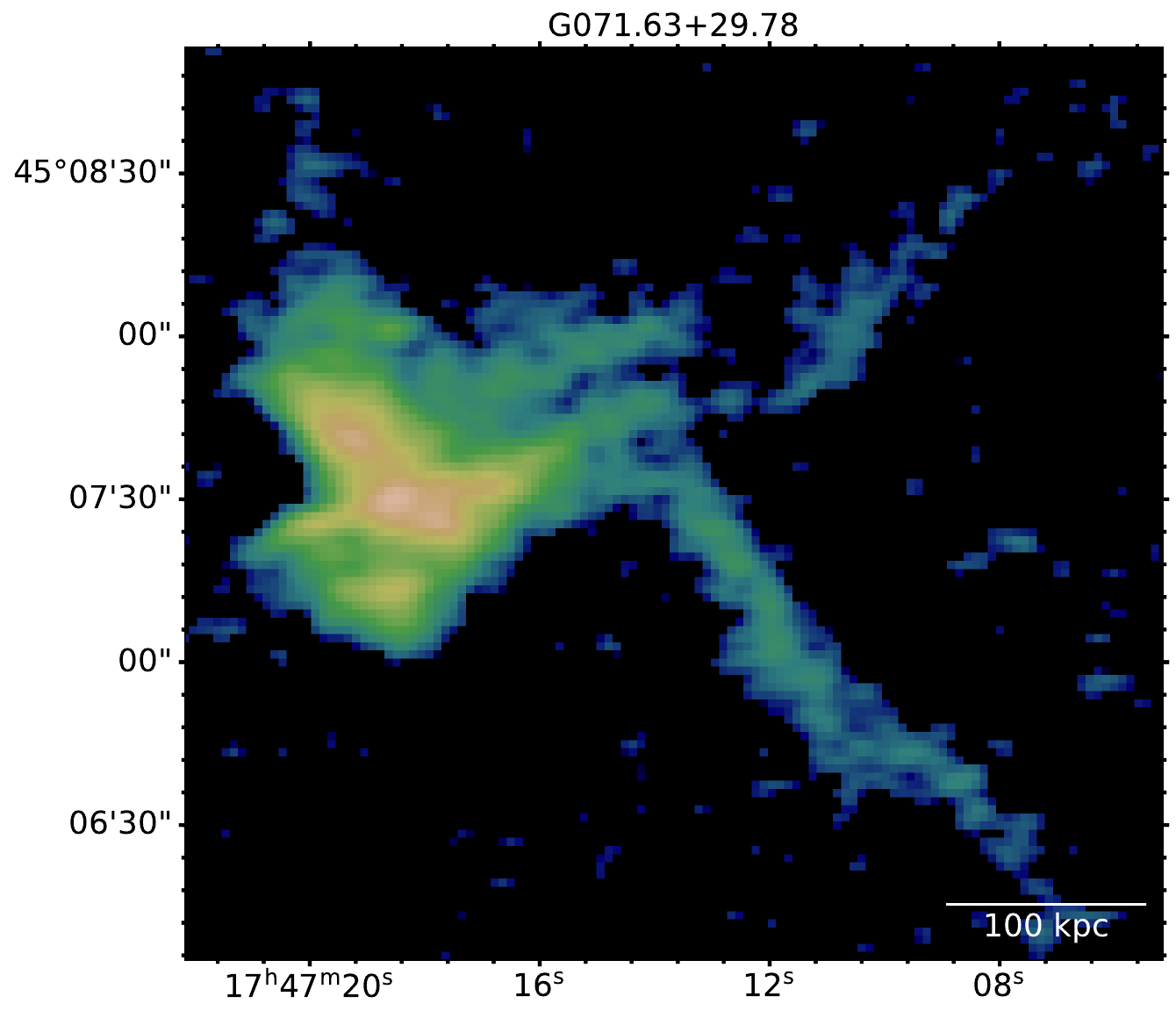}
\includegraphics[width=0.282\textwidth]{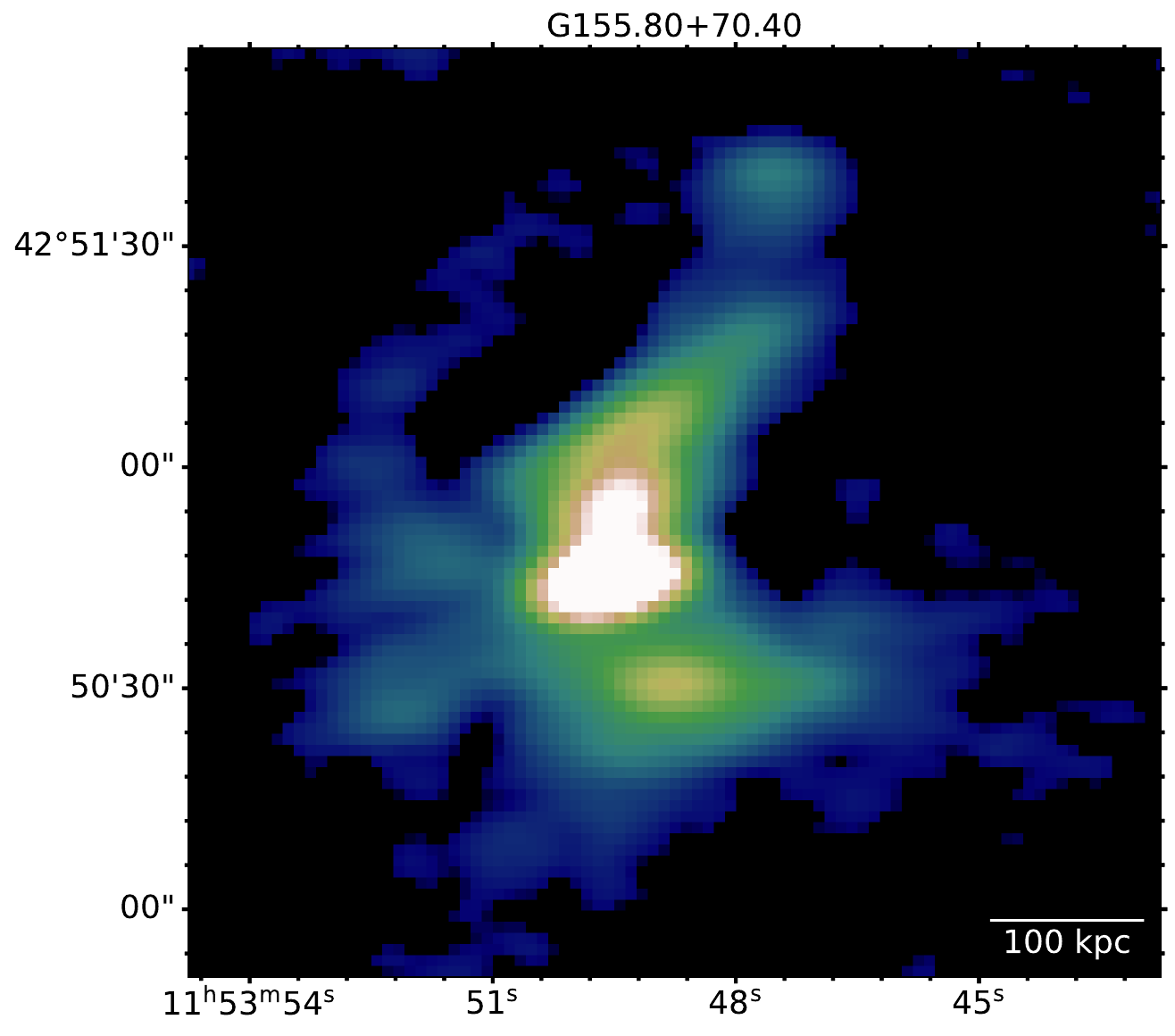}
\includegraphics[width=0.192\textwidth]{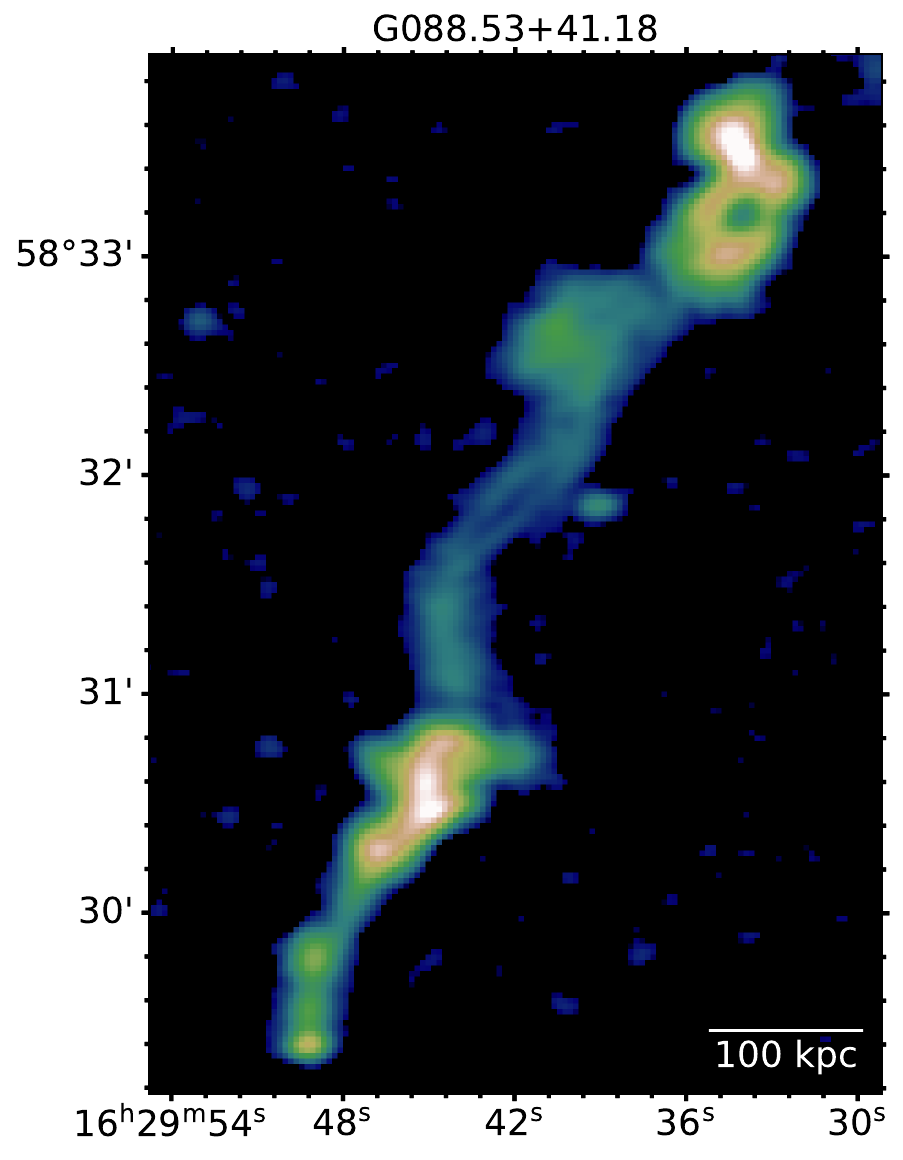}
\includegraphics[width=0.199\textwidth]{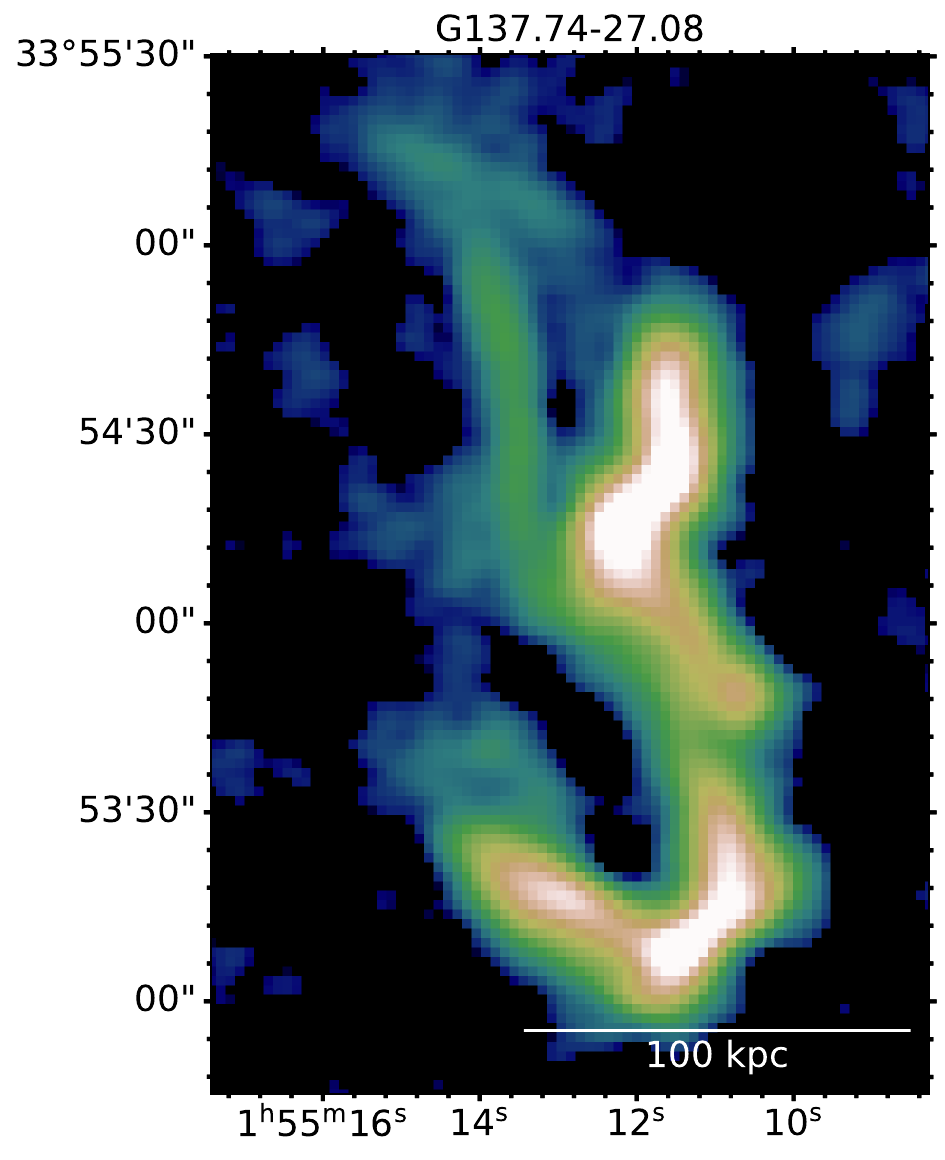}
\includegraphics[width=0.232\textwidth]{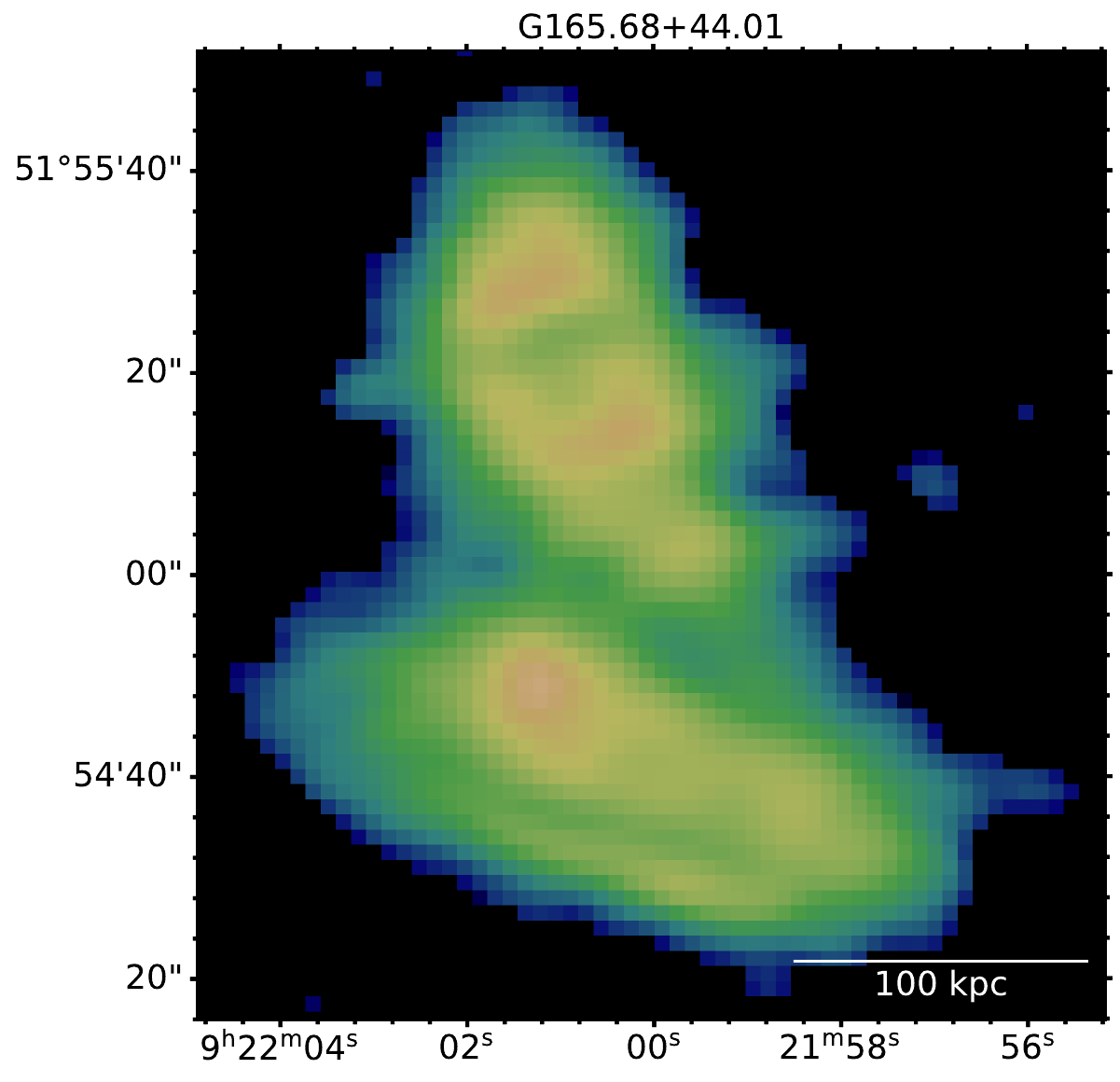}
\includegraphics[width=0.282\textwidth]{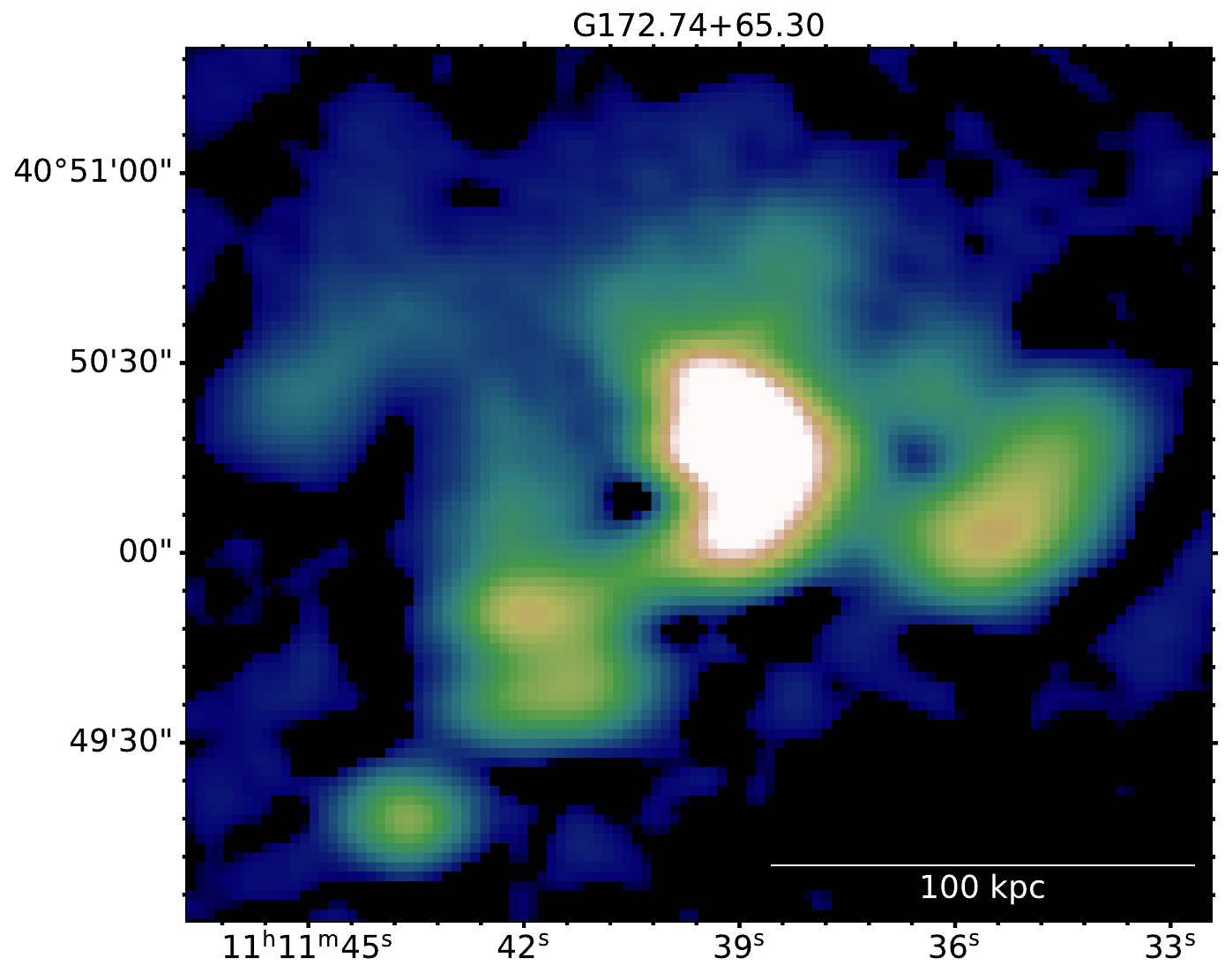}
\includegraphics[width=0.446\textwidth]{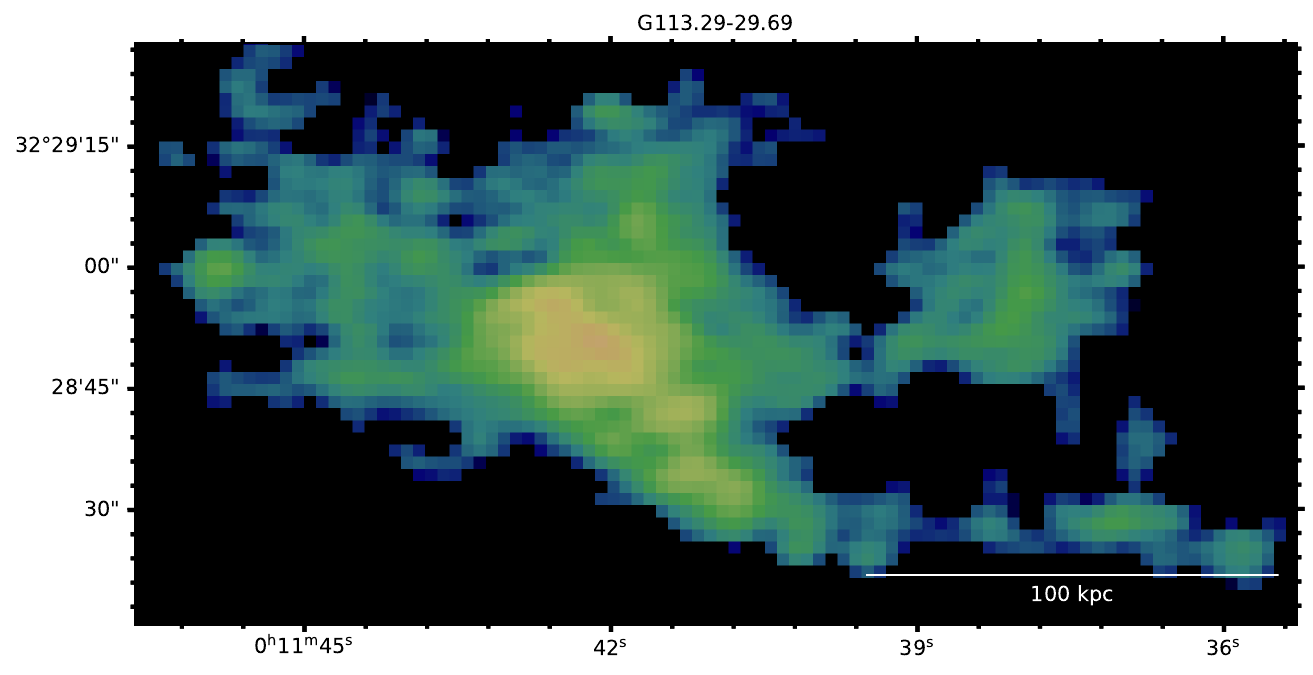}

        \caption{Overview of the candidate revived fossil sources identified by visual inspection among low-$z$ and low-$M_{\rm 500}$ PSZ2 clusters in LoTSS-DR2 based on their irregular and filamentary morphology.}
        \label{fig: collection}
\end{figure*}

\begin{table*}
 \fontsize{7.5}{7.5}\selectfont
\centering
	\caption[]{Properties of the clusters hosting the candidate revived sources analysed in this work.}
	\label{table: data}   
	\begin{tabular}{cccccccccccccc}
	\hline
	\noalign{\smallskip}
	PSZ2 Name & Abell Name & ${\rm RA}_{\rm J2000}$ & ${\rm DEC}_{\rm J2000}$ & $z$ & $M_{500}$ & $R_{500}$ & Scale & Class. & $c$ & $w$  \\   
    & & (deg) & (deg) & &   ($10^{14} \; M_{\odot}$) & (kpc) & (${\rm kpc \; arcsec^{-1}}$) &   & ($10^{-2}$)  & ($10^{-2}$)  \\   
    \hline
	\noalign{\smallskip}
   G071.63+29.78 & --- & 266.8257 & 45.1899  & 0.157 & $4.13 \pm 0.29$ &  $1080 \pm 25$ & 2.715 & NDE & $8.24 \pm 0.33$ & $2.45 \pm 1.38$  \\
   G088.53+41.18 & A2208 & 247.3887 & 58.5338 & 0.133 & $2.56 \pm 0.34$ &  $929 \pm 42$ & 2.363 & NDE & $16.36 \pm 1.09$ & $1.59 \pm 0.28$ \\
   G113.29-29.69 & A7  & 2.9363 & 32.4325 &  0.107 & $3.71 \pm 0.27$ & $1060 \pm 25$ &  1.958 & U & $16.90 \pm 1.02$ & $1.33 \pm 0.50$  \\
   G137.74-27.08 & A272 & 28.7835 &	33.9443 &  0.087 & $2.83 \pm 0.28$ & $975 \pm 32$ & 1.629 & NDE & $14.60 \pm  0.24$ &  $4.31 \pm 0.07$  \\
   G155.80+70.40 & --- & 178.4833	& 42.8600 & 0.333 & $4.42 \pm 0.56$ & $1036 \pm 44$ & 4.781 & NDE & --- & ---  \\
   G165.68+44.01 & --- & 140.5859 & 51.8876 & 0.21 & $3.76 \pm 0.50$ & $1027 \pm 46$ & 3.427 & NDE & --- & ---  \\
   G172.74+65.30 & A1190 & 167.9029 &	40.8574 & 0.079 & $2.45 \pm 0.21$ & $932 \pm 27$ & 1.493 & U & $21.80 \pm  1.10$ & $2.44 \pm 1.69$  \\
\noalign{\smallskip}
	\hline
	\end{tabular}  
	\begin{tablenotes}
\item    {\small \textbf{Notes}. Cols. 1-4: host cluster name and coordinates (\textit{Planck} centre). Cols. 5-8: cluster redshift, mass, radius, and angular to linear scale conversion. Col. 9: cluster classification as No Diffuse Emission (NDE) or Uncertain (U). Cols. 10, 11: concentration ($c$) and centroid shift ($w$) morphological X-ray parameters (see \citealt{botteon22}, \citealt{zhang23}, and Appendix \ref{sect: Xray data for G088}). } 
 \end{tablenotes}	
\label{tab: targets}
\end{table*}

\cite{botteon22} presented a sample of 309 galaxy clusters selected from the second \textit{Planck} catalogue of Sunyaev-Zel'dovich detections (PSZ2; \citealt{planckcollaboration16}) that are in the footprint of LoTSS-DR2\footnote{Data products are publicly available on the project website: \url{https://lofar-surveys.org/planck_dr2.html}}. This represents the largest homogeneous sample of mass-selected galaxy clusters observed at radio frequencies to date. This sample was analysed in a series of works \citep{bruno23,cassano23,cuciti23,jones23,zhang23} to derive the statistical properties of clusters with and without RHs and RRs, and test the predictions of theoretical formation scenarios. The project did not include the analysis of revived fossil sources, owing to the lack of spectral measurements that are essential for a proper classification.

In the present work, we inspected the clusters of the sample to search for targets that potentially host revived fossil sources based on their irregular/filamentary morphology and the non-trivial connection with an optical host counterpart. In general, if spectral measurements are not available, the only a priori criterion to identify such sources is the visual inspection of their morphology. To avoid contamination from RHs and RRs that are also detected in some clusters, we considered only clusters classified as `NDE' (No Diffuse Emission, that is no detection of RRs or RHs) or `U' (Uncertain, that is detection of a diffuse source of unclear nature) as defined in \cite{botteon22}. As high-resolution spectral index maps are crucial to study revived fossil sources, we set a redshift limit of $z\leq 0.35$ that provides a spatial resolution $\lesssim 50$ kpc at $\sim 10''$. Currently, no firm relation between the presence of revived sources and the mass or dynamical state of the host cluster is known. Here we considered only low-mass systems having $M_{500}\leq 5\times 10^{14} \; M_\odot$; in this regime, the occurrence of RHs drops with respect to higher masses, e.g. \citealt{cassano23}, thus reducing the probability of possible contaminating emission.

Among 92 NDE and U clusters in the considered $z$ and $M_{500}$ ranges out of 309 total entries, we identified 9 promising revived sources. In two of these clusters, namely PSZ2 G080.41-33.24 \citep{cohen&clarke11,clarke13} and PSZ2 G100.45-38.42 \citep{gitti04,gitti13,kale&gitti17,ignesti17,ignesti20b}, the presence of a revived source was confirmed in previous studies and we excluded them from our sample. Therefore, here we will focus on the remaining 7 PSZ2 galaxy clusters, being G071.63+29.78 (G071), G088.53+41.18 (G088), G113.29-29.69 (G113), G137.74-27.08 (G137), G155.80+70.40 (G155), G165.68+44.01 (G165), and G172.74+65.30 (G172). Their properties are reported in Table \ref{tab: targets}. Among the radio sources within these clusters, we will focus on the targets displayed in Fig. \ref{fig: collection}.



\section{Observations and data processing}
 \label{sect: Observations and data reduction}

In this section, we summarise the LOFAR data processing. We then present the follow-up uGMRT observations of the targets and their processing.

\subsection{LOFAR data}
\label{sect: LOFAR HBA data}

For details on the LoTSS-DR2 data reduction and post-processing, we refer to \cite{botteon22}, \cite{shimwell22LOTSS}, and references therein. These include steps of direction-independent and direction-dependent calibration, and additional self-calibration towards a smaller extracted portion of the field of view (see \citealt{vanweeren21} for details on this procedure). The typical noise level is $\sigma\sim 100$ ${\rm \mu Jy \; beam^{-1}}$ at the nominal resolution of $\theta\sim 6''$. 

In the present work, we re-imaged the extracted and self-calibrated visibilities with {\tt WSClean} \citep{offringa14,offringa17} v. 3.6, using wide-field, multi-frequency, and multi-scale synthesis algorithms, and testing additional weighting schemes. The flux density scale of all images was aligned to the LoTSS-DR2 scale following \cite{botteon22} and \cite{shimwell22LOTSS}.

\subsection{uGMRT data}
\label{sect: uGMRT data}

Follow-up observations of the targets were carried out with the uGMRT in band 3 (300-500 MHz) between December 2023 and March 2025 (project codes: 45$\_$035, 47$\_$027; PI: L. Bruno) for $\sim 5.5-6.5$ hours on-source each. Data were recorded in 4096 channels of width 48.8 kHz each. All observations include pointings ($\sim 20$ min) on 3C48 and/or 3C286, which were used as absolute flux density scale calibrators.

We processed the data by means of the Source Peeling and Atmospheric Modeling ({\tt SPAM}; \citealt{intema09}) pipeline. {\tt SPAM} computes flux density and bandpass solutions for the calibrator, which are then transferred to the target. Ionospheric effects are corrected by performing rounds of direction-dependent calibration by means of bright sources in the field. The total bandwidth was split into six (33.3 MHz) sub-bands, which were processed independently, and afterwards recombined for imaging with {\tt WSClean}, using the same algorithms considered for LOFAR. The reached noise level is typically $\sigma\sim 25$ ${\rm \mu Jy \; beam^{-1}}$ at a resolution of $\theta \sim 8''$, but it can increase due to calibration artefacts around bright sources. In the next sections, we will consider the noise level close to the target as reference.

\subsection{Spectral index maps}
\label{sect: Spectral index maps}

A proper classification of the targets requires spatially-resolved spectral index maps. We produced spectral index maps from LOFAR and uGMRT images obtained by matching the \textit{uv}-range (60 $\lambda$ - 35 k$\lambda$) to sample the same angular scales, and convolving these images to a common resolution of $ 8''-10''$. Regions having surface brightness below a threshold of $3\sigma$ of each input image were masked, and afterwards the 144-400 MHz spectrum and errors were computed pixel-by-pixel with the {\tt immath} task of the Common Astronomy Software Applications ({\tt CASA}; \citealt{mcmullincasapaper07}) v. 6.4 as
\begin{equation}
 \alpha = - \frac{\ln{(S_1)}-\ln{(S_2)}}{\ln{(\nu_1)}-\ln{(\nu_2)}} \; \pm \; \left\lvert \frac{1}{\ln{ \left( \frac{\nu_{\rm 1}}{\nu_{\rm 2} } \right) }}\right\lvert \sqrt{ \left( \frac{\Delta S_{\rm 1}}{S_{\rm 1}}\right)^2 + \left( \frac{\Delta S_{\rm 2}}{S_{\rm 2}}\right)^2 } \; \; \; .
\label{eq: spectralindexerrorformula}
\end{equation}
In Eq. \ref{eq: spectralindexerrorformula}, uncertainties $\Delta S$ on the flux density are computed as 
\begin{equation}
\Delta S= \sqrt{ \left( \sigma \cdot \sqrt{N_{\rm b}} \right)^2 + \left(  \xi_{\rm cal} \cdot S \right) ^2}, \; \; \;
\label{eq: erroronflux}
\end{equation}
where $\sigma$ is the reference image noise, $N_{\rm b}$ is the number of independent beams within the target area, and $\xi_{\rm cal}$ is the systematic calibration error. We assumed standard calibration errors of $\xi_{\rm cal}=10\%$ for LOFAR \citep{shimwell22LOTSS} and $\xi_{\rm cal}= 6\%$ for uGMRT in band 3 \citep{Lal22}.


\section{Properties and classification}
\label{sect: Morphological and spectral properties}

In this section we present the analysis of our targets. We show the LOFAR and uGMRT radio images in Figs. \ref{fig: mappefullres1}-\ref{fig: mappefullres7}. The same figures include panels with optical images from the Panoramic Survey Telescope \& Rapid Response System (Pan-STARRS; \citealt{flewelling20PANSTARRS}), 144-400 MHz spectral index maps (see the corresponding errors maps in Fig. \ref{fig: errspixmap} in Appendix \ref{sect: errspixmap}), and, if available, X-ray images. We used the optical images to search for host galaxies, whose redshifts are listed in Table \ref{table: galaxies} in Appendix \ref{sect: Properties of selected clusters}. For details on the considered X-ray data, we refer to \citealt{botteon22}, \citealt{zhang23}, and Appendix \ref{sect: Xray data for G088}. In addition, we reported surface brightness and  spectral index profiles in Appendix \ref{sect: Surface brightness and spectral index profiles} to better interpret some targets.

\subsection{G071.63+29.78}
\label{sect: G071}

\begin{figure*}
        \centering

\includegraphics[width=0.33\textwidth]{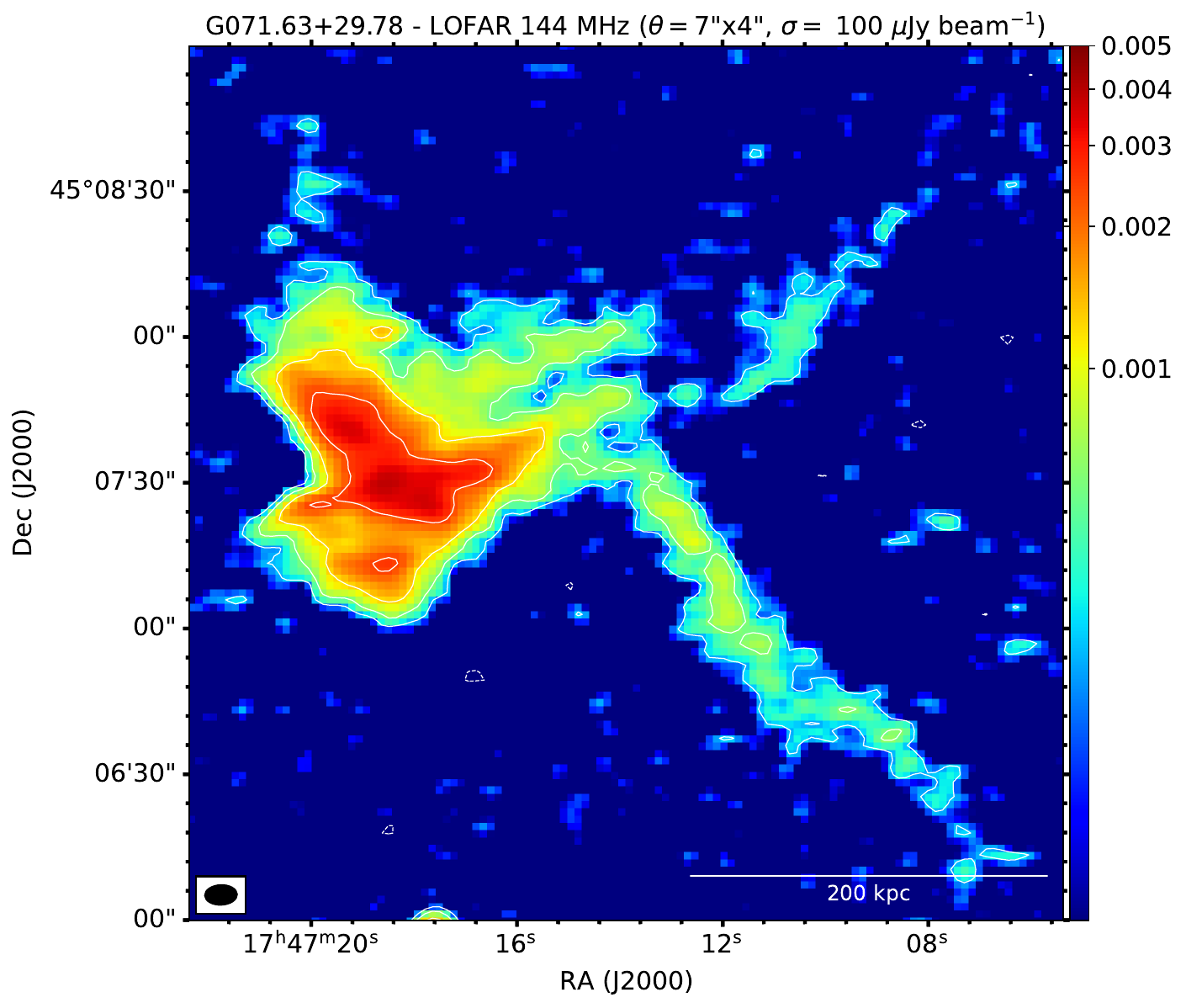}
\includegraphics[width=0.33\textwidth]{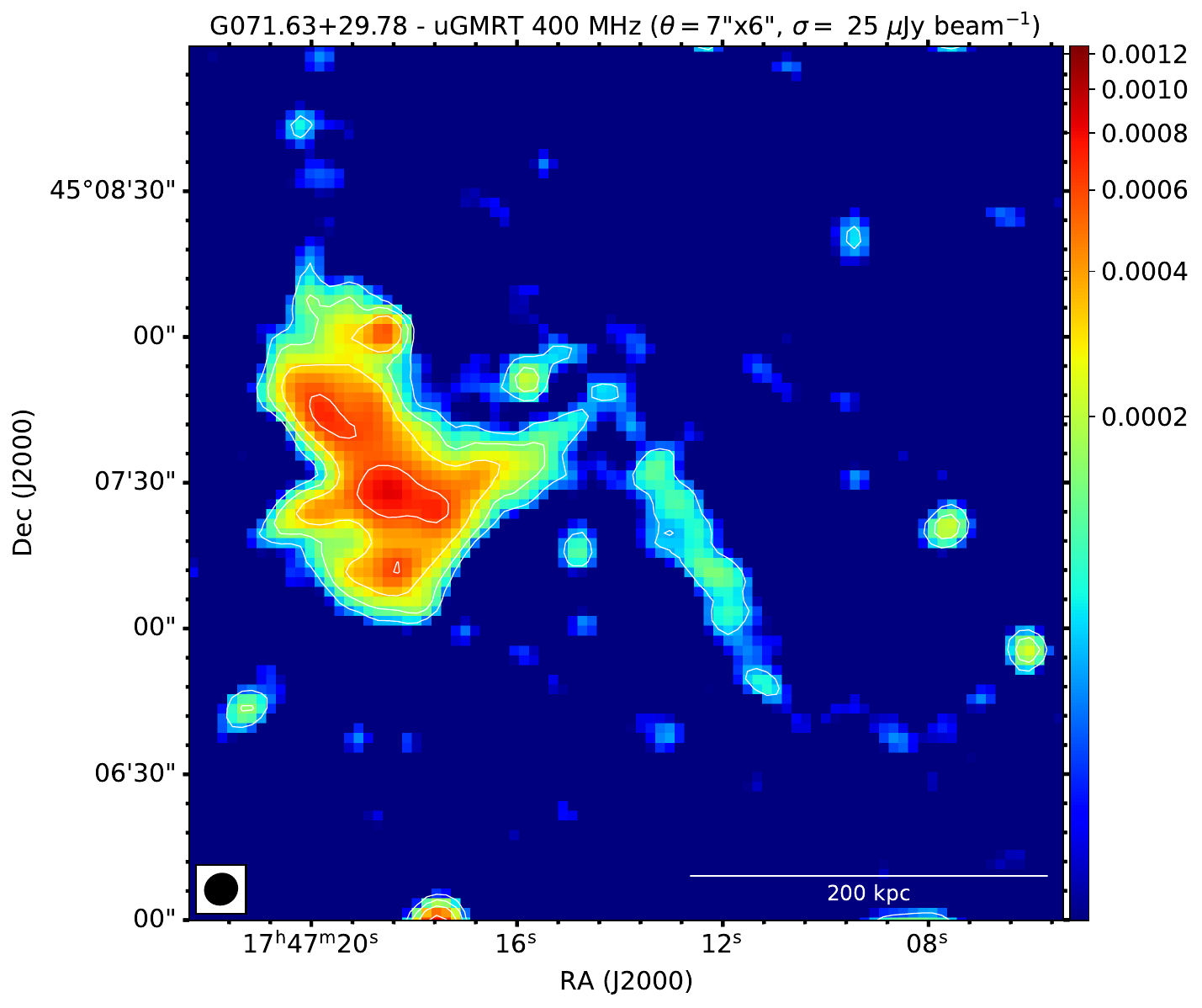} \\
\includegraphics[width=0.31\textwidth]{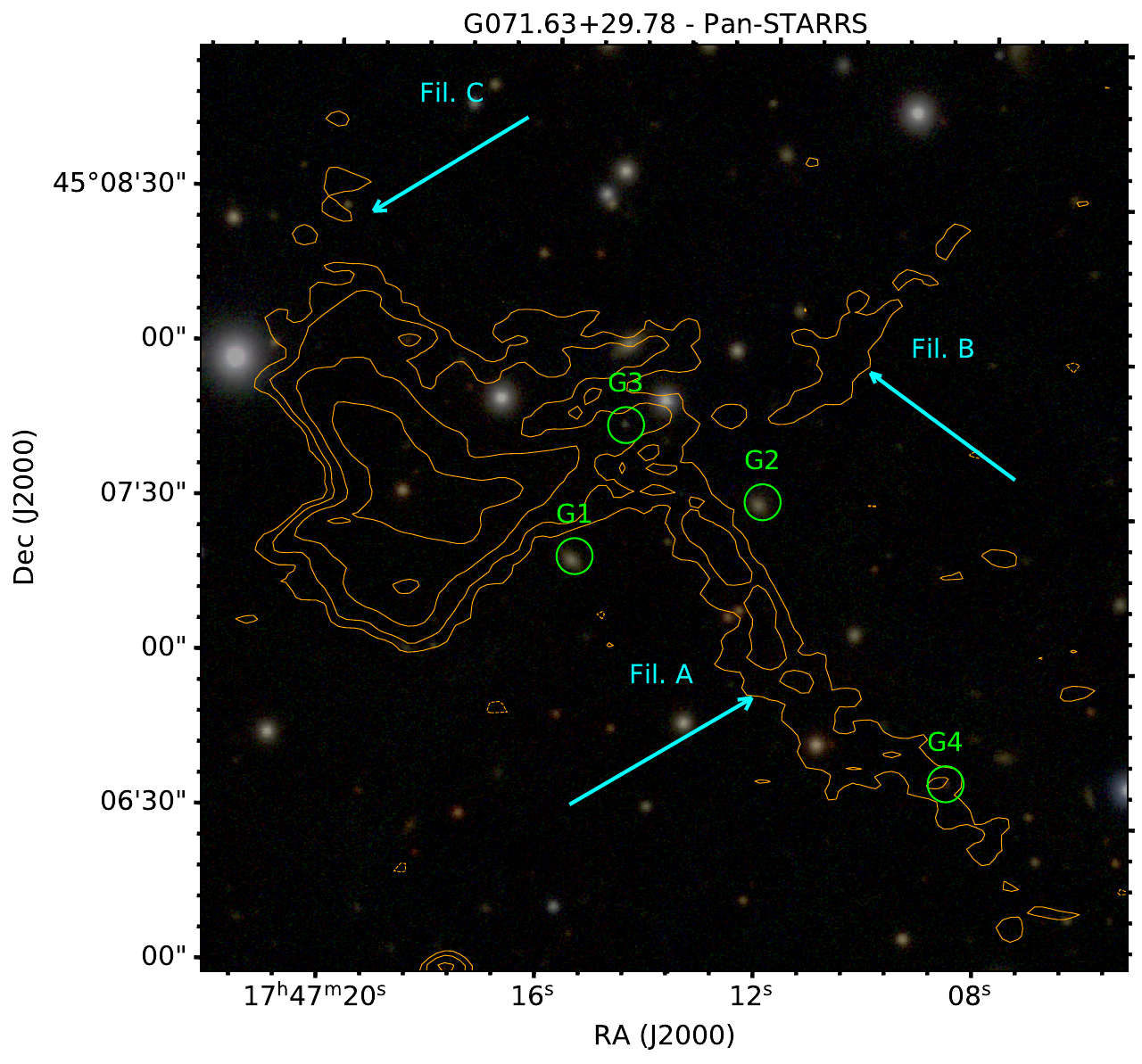}
\includegraphics[width=0.33\textwidth]{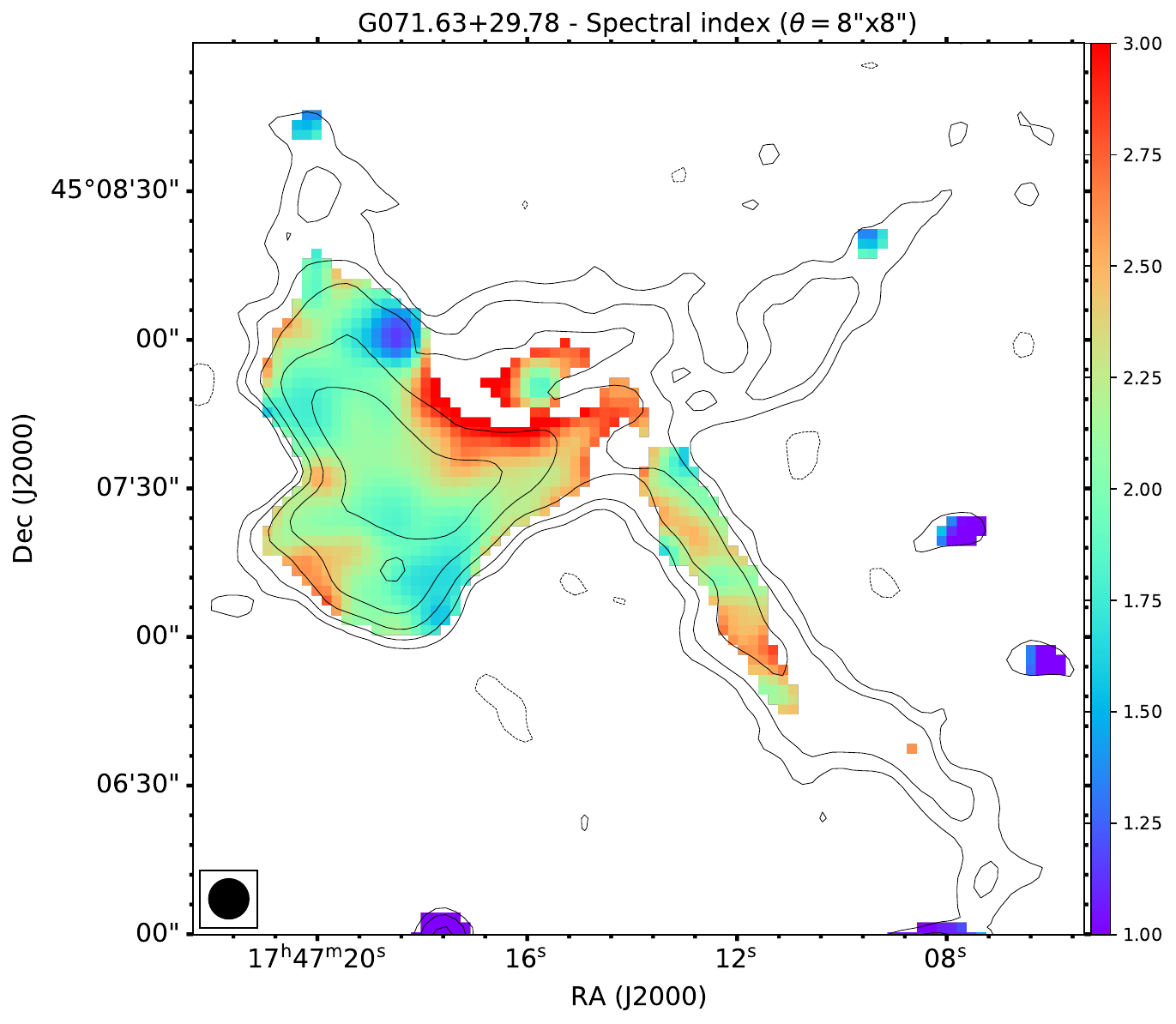}
\includegraphics[width=0.33\textwidth]{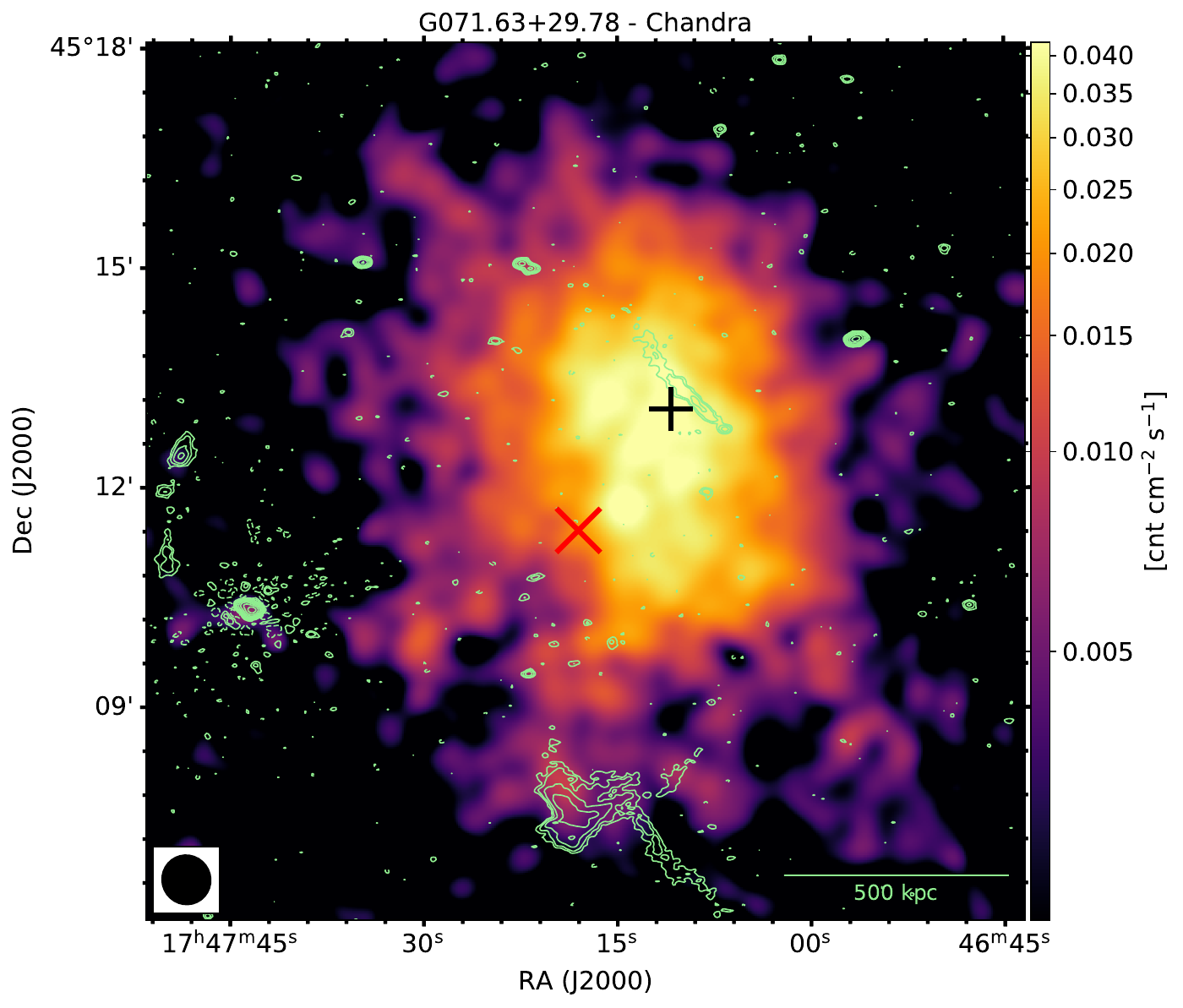}

        \caption{Images of G071.63+29.78. {\it Top}:  LOFAR at 144 MHz and uGMRT at 400 MHz radio images (units are ${\rm Jy \; beam^{-1}}$). Resolution and noise are reported on top of each panel. The contour levels are $[\pm3, \;6, \;12,\; 24,\; ...]\times \sigma$. {\it Bottom left}: Pan-STARRS optical (composite i, r, g filters) image. Orange contours are from radio images on top panels. Green circles and cyan lines indicate optical sources and regions discussed in the text. {\it Bottom centre}: 144-400 MHz spectral index map (the corresponding error map is shown in Fig. \ref{fig: errspixmap} in Appendix \ref{sect: errspixmap}). The contour levels are $[\pm3, \;6, \;12,\; 24,\; ...]\times \sigma$ from the 144 MHz image. {\it Bottom right}: X-ray image of the cluster. Green contours are from the LOFAR image. Red and black crosses indicate the \textit{Planck} centre and X-ray peak, respectively.   }
        \label{fig: mappefullres1}
\end{figure*}

The radio source in G071 (Fig. \ref{fig: mappefullres1}) consists of a main region having a patchy morphology, which extends for  $\sim 200$ kpc and exhibits various brightness peaks. Three thin filaments are connected to this main region. The brightest of them, filament A, extends straight along the NE-SW axis for $\sim 250$ kpc at 144 MHz, whereas it is partially detected at 400 MHz. Interestingly, the surface brightness increases (or remains roughly constant) with the distance from the main region for $\sim 100$ kpc and then rapidly drops for $\sim 150$ kpc (Fig. \ref{fig: profiles G071}). In the northern region of the target, the two faint filaments B and C are only detected in the 144 MHz image (see also Fig. \ref{fig: profiles G071}), where we measured lengths of $\sim 130$ kpc and $\sim 60$ kpc, respectively. The surface brightness of filament B decreases in the initial $\sim 40$ kpc, increases for $\sim 70$ kpc, and finally drops (Fig. \ref{fig: profiles G071}).

The spectral index map shows a uniform distribution of $\alpha\sim 2$ across the main region, confirming the ultra-steep spectrum nature of the source. A similar spectral index is measured along filament A, exhibiting a constant trend (at least for the initial 100 kpc). Filaments B and C are exclusively detected at 144 MHz, therefore we expect $\alpha > 2$.

By using the Pan-STARRS optical image and available redshift from the literature, we identified four cluster members (G1, G2, G3, G4) close to the radio source that might be their host. None of these galaxies are co-located with a compact radio peak, not allowing us to determine a preferential host. 

In the X-ray images, the radio source is located in the cluster outskirts, at a projected distance of $\sim 900$ kpc from the X-ray peak. The ICM does not show a defined X-ray core, consistent with a disturbed dynamical state, as indicated by the morphological concentration and centroid shift parameters (Table \ref{tab: targets}).

The amorphous and filamentary shape, ultra-steep spectrum, and non-trivial host galaxy are textbook features of radio phoenices.  We therefore classify the source in G071 as a likely RP. Our target exhibits a uniform spectral index distribution, consistent with the existing variety of spectral behaviours reported for RPs  \citep[e.g.][]{mandal20,raja24}.

\subsection{G088.53+41.18}
\label{sect: G088}

\begin{figure*}
        \centering

\includegraphics[width=0.33\textwidth]{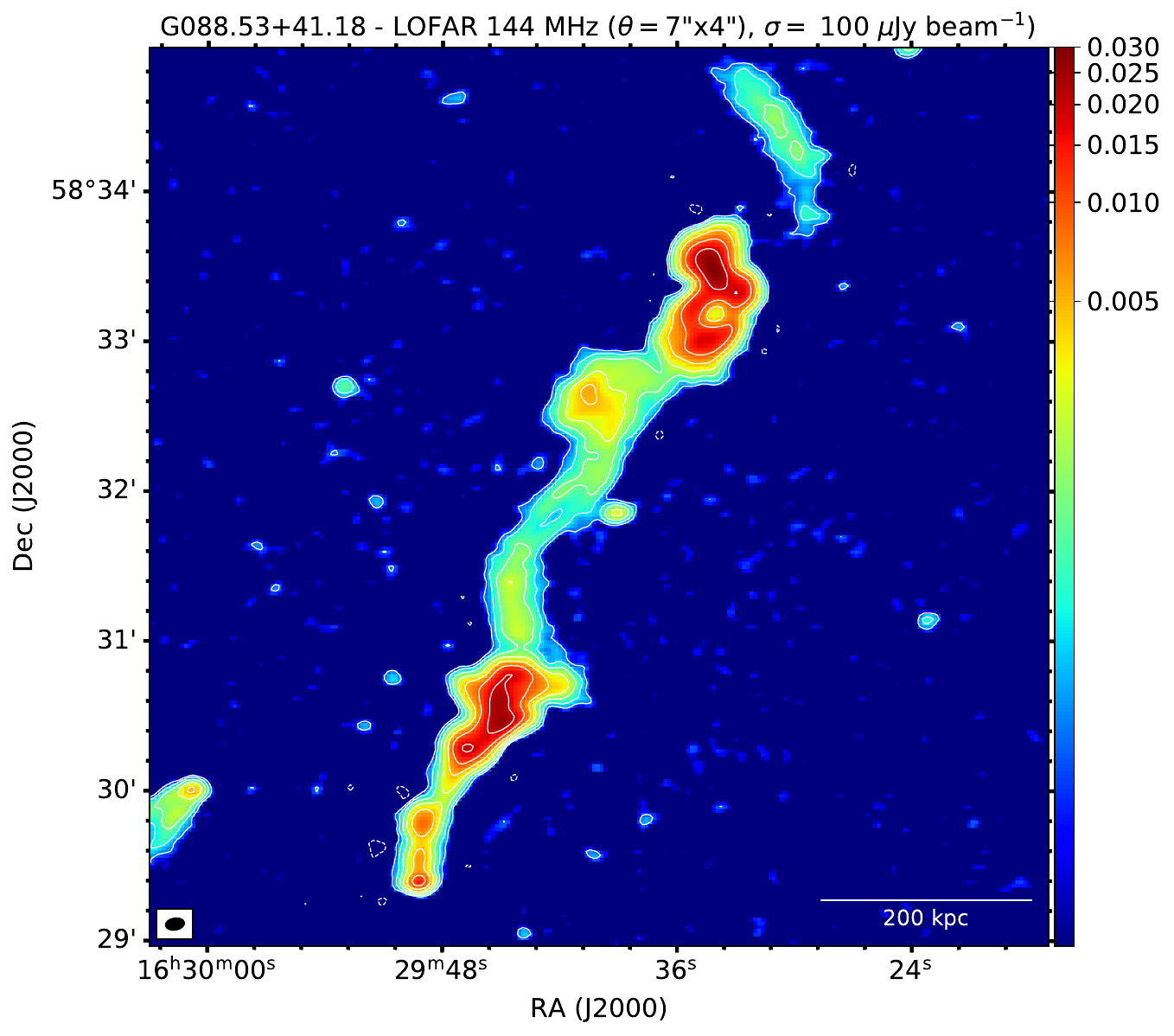}
\includegraphics[width=0.33\textwidth]{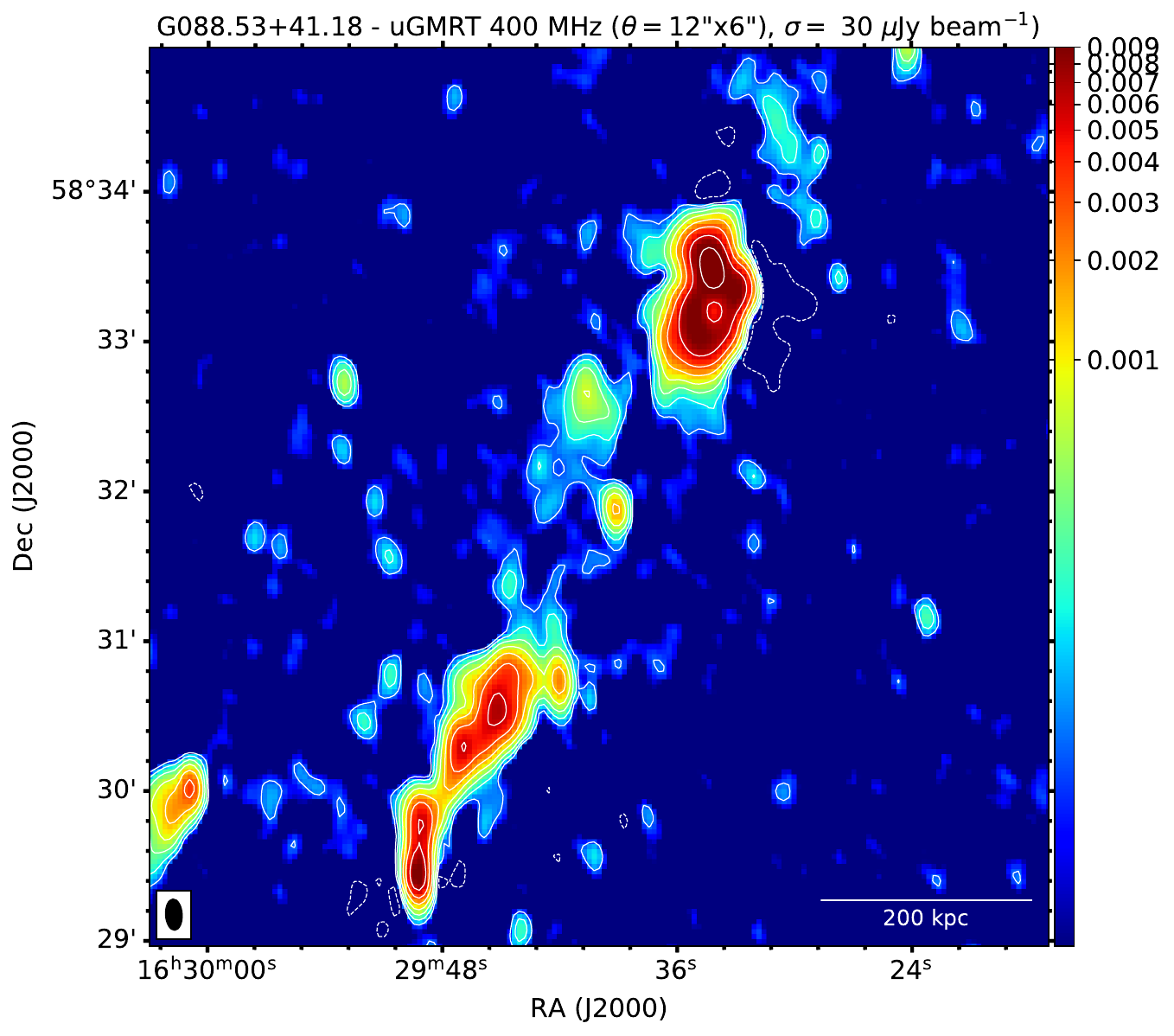} \\
\includegraphics[width=0.31\textwidth]{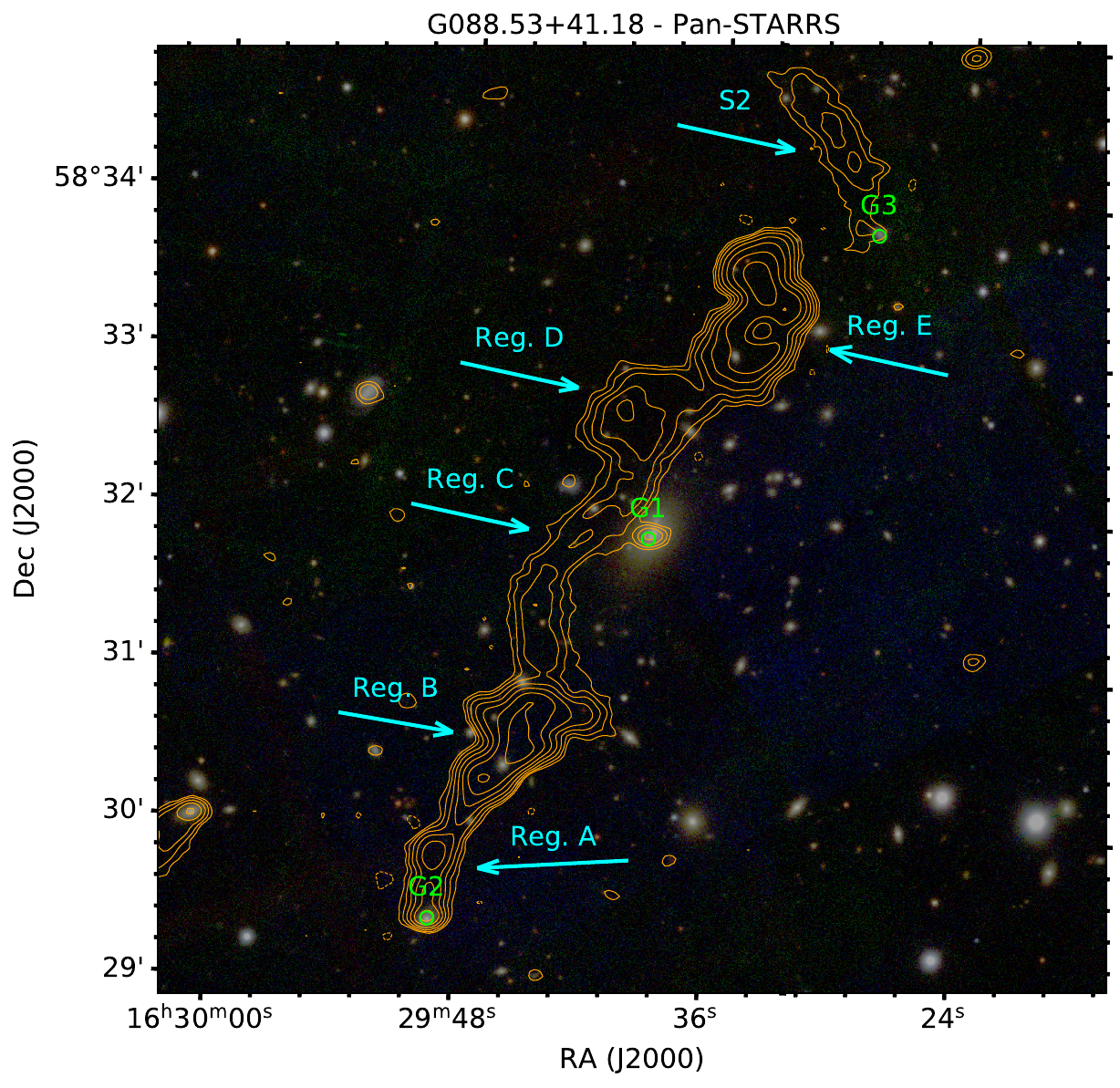}
\includegraphics[width=0.33\textwidth]{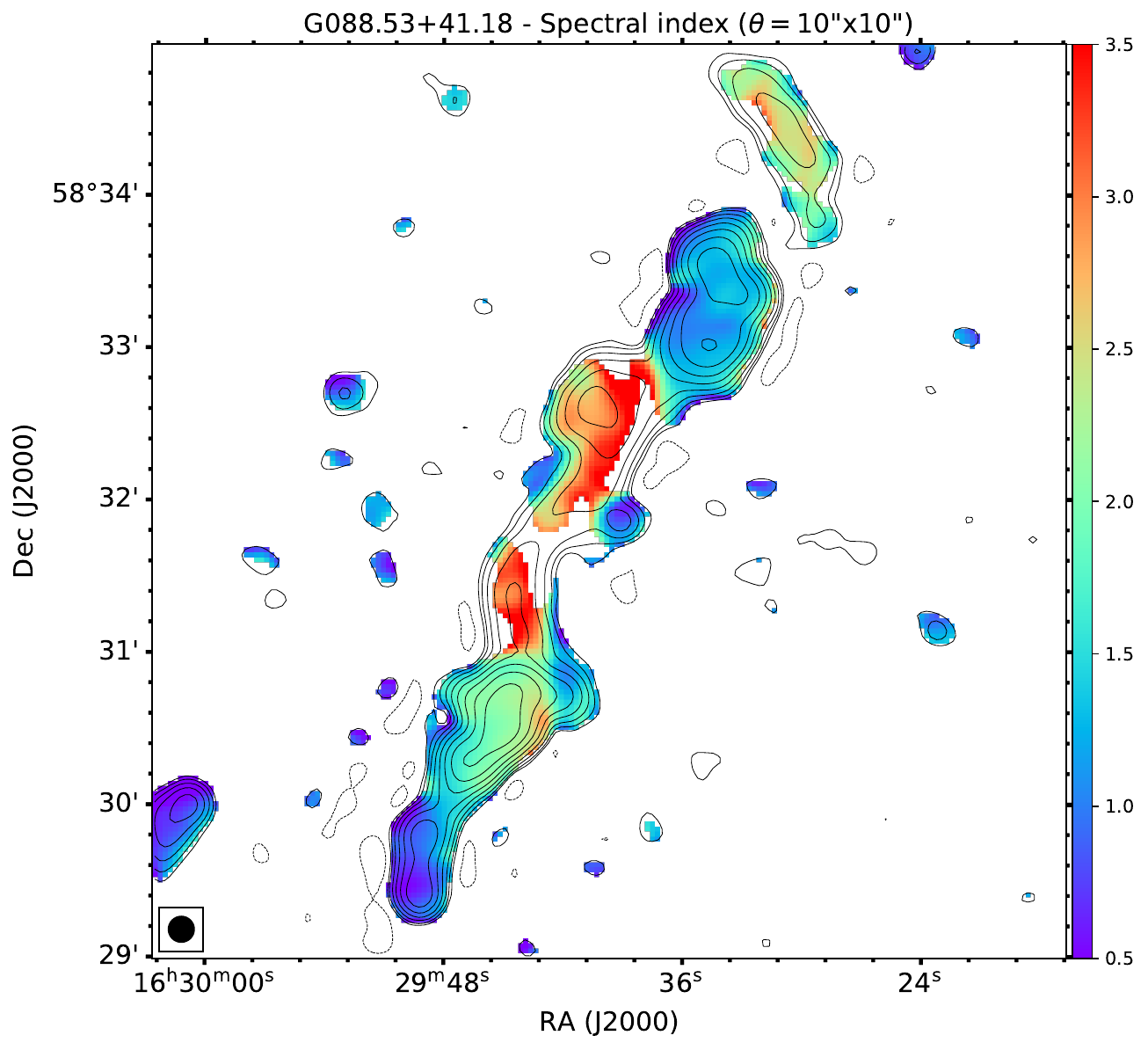} 
\includegraphics[width=0.35\textwidth]{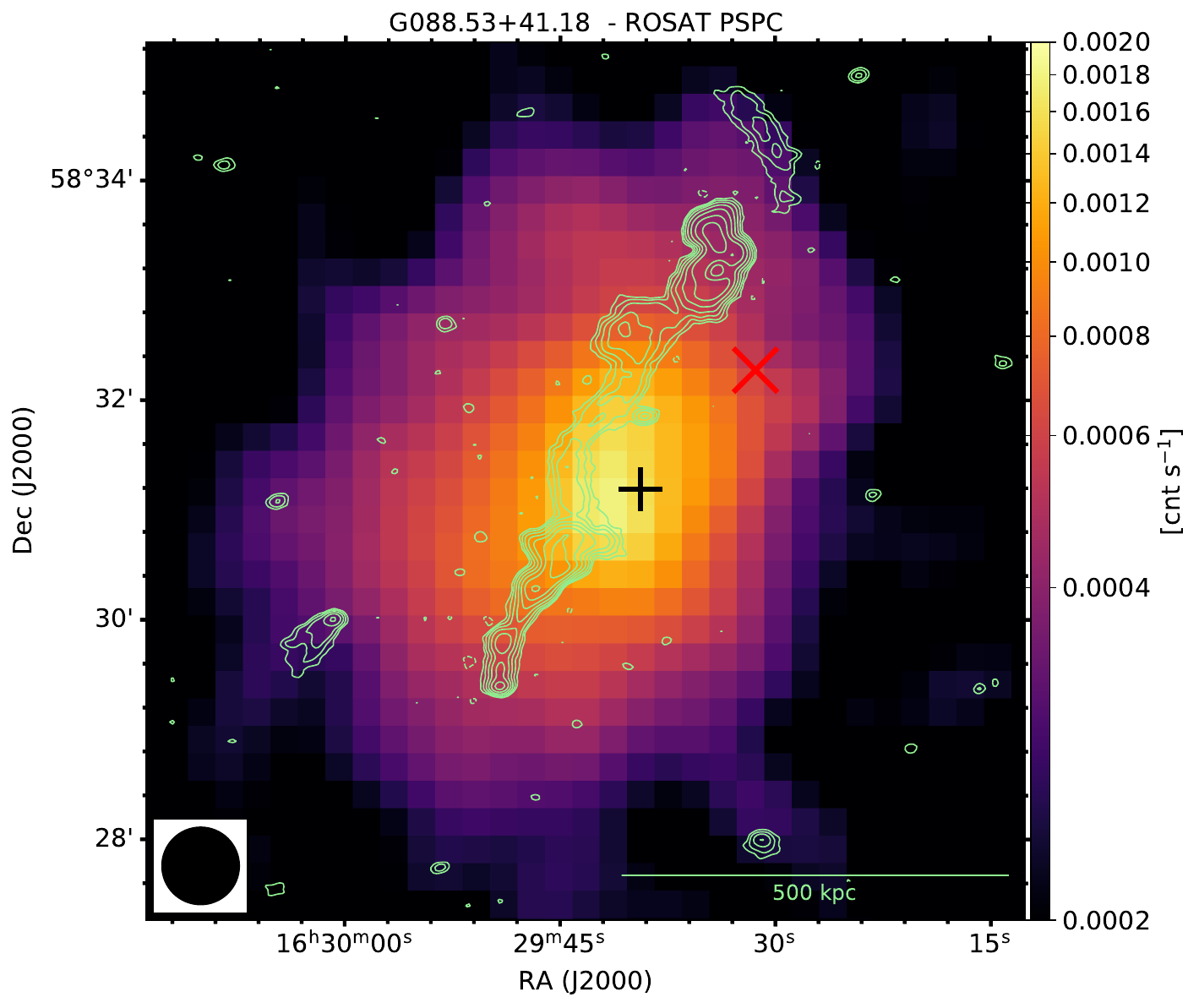}

        \caption{Images of G088.53+41.18. The description of each panel is the same as for Fig. \ref{fig: mappefullres1}.}
        \label{fig: mappefullres2}
\end{figure*}

The radio source in G088 (Fig. \ref{fig: mappefullres2}) is elongated, extending along the NW-SE axis for $\sim 730$ kpc. Several regions (A-E) and a detached interesting radio source (S2) can be distinguished.

Along regions A ($LLS\sim 120$ kpc) and B ($LLS\sim 155$ kpc) we observe a clear spectral gradient (Fig. \ref{fig: profiles G088}), as the spectral index progressively steepens from $\alpha\sim 0.5$ to $\alpha\sim 2$. The brightness peak of region A coincides with the cluster member galaxy G2. These features and overall morphology lead us to conclude that regions A and B are parts of a HT radio galaxy. The surface brightness profile (Fig. \ref{fig: profiles G088}) shows a transition separating regions A and B and an enhancement of the brightness peaking in the middle of the tail; this is unusual for tailed galaxies, but not unique (for instance, a similar structure has been observed for the HT radio galaxy T2 in A2142; \citealt{bruno24b}), and may be suggestive of interplay between the tail and the ICM.

Regions C ($LLS\sim 200$ kpc) and D ($LLS\sim 100$ kpc) are the faintest regions the target. Region C is elongated and characterised by a remarkably steep spectrum of $\alpha \gtrsim 3$, as indeed it is only partially detected at 400 MHz, while region D is patchy and brighter, but exhibits a similar $\alpha \sim 3$. Interestingly, the surface brightness and spectral index are roughly constant along these regions (Fig. \ref{fig: profiles G088}). A compact radio source is located slightly offset with respect to regions C and D, but it is apparently connected to them. It is hosted by G1, which is the brightest cluster galaxy (BCG) in G088, and its radio emission is consistent with AGN activity. Region E ($LLS\sim 215$ kpc) appears to be connected to region D, but the surface brightness abruptly increases and spectral index flattens from $\alpha\sim 3.5$ to $\alpha\sim 1.2$. Such spectral index is uniformly distributed across the region E. The resolution of our radio images is sufficient to reveal an inner ring-like structure, which is however resolved out in the spectral index map.

Finally, we report on the radio source S2, located north-west of our main target. This is roughly orthogonal to the target and has an extent of $LLS\sim 200$ kpc at the cluster redshift. A likely background galaxy, G3 ($z=0.172\pm 0.027$), could be possibly the host of S2. The spectral index is notably steep ($\alpha\sim 2$) and it is approximately constant across the source.

In the X-rays, the ICM shows a moderate core and the morphological parameters (see Appendix \ref{sect: Xray data for G088}) hint at some level of disturbance. The X-ray peak is offset from G1 by  $\sim 100$ kpc, and the eastern boundary of the core appears spatially coincident with the radio emission of regions C and D, suggesting a possible connection between thermal and non-thermal components. Overall, the target extends along the cluster major axis, while S2 is located in the outskirts.

As mentioned, we interpret regions A and B as components of a HT radio galaxy. The measured ultra-steep spectrum confirms that regions C and D consist of fossil plasma. If physically associated with the AGN of the BCG, regions C and D may trace remnant plasma that has drifted apart due to buoyant forces or cluster weather. However, the relatively flat spectral index ($\alpha \sim 0.7$) of the AGN is in contrast with an aged source. Therefore, either the fossil components originated from a previous AGN outburst phase or the association is only seeming. We notice that the spectral index profile is discontinuous at the transition between regions B and C (Fig. \ref{fig: profiles G088}). Such a spectral profile shows similarities to that of the prototypical GReET in A1033 \citep{degasperin17,edler22}. Specifically, in A1033, a spectral flattening was reported, whereas in regions C and D of our target we do not observe a clear flattening, but rather a very slow steepening with the distance, resulting in a plateau around $\alpha \sim 3-3.5$. Therefore, an alternative explanation is that regions C and D are powered by re-energised fossil electrons released by the HT radio galaxy. The constant brightness profile (Fig. \ref{fig: profiles G088}) supports this interpretation, and, considering the spatial coincidence, the gentle re-acceleration might be induced by processes involving the X-ray core region, likely due to weak turbulence, whose lower efficiency compared to A1033 would explain the spectral plateau instead of a flattening. The ring-like morphology of region E is reminiscent of a WAT radio galaxy, but the absence of a spectral gradient and obvious host challenges this interpretation. If S2 is hosted by G3, it may be a HT radio galaxy; if not, its location and orientation suggest that S2 may be a RR. However, both hypotheses are strongly disfavoured by the observed uniformly-distributed ultra-steep spectrum. While the nature of region E and source S2 remains to be understood, we confidently claim that they are unrelated to the HT and candidate GReET emission.

\subsection{G113.29-29.69}
\label{sect: G113}

\begin{figure*}
        \centering

\includegraphics[width=0.33\textwidth]{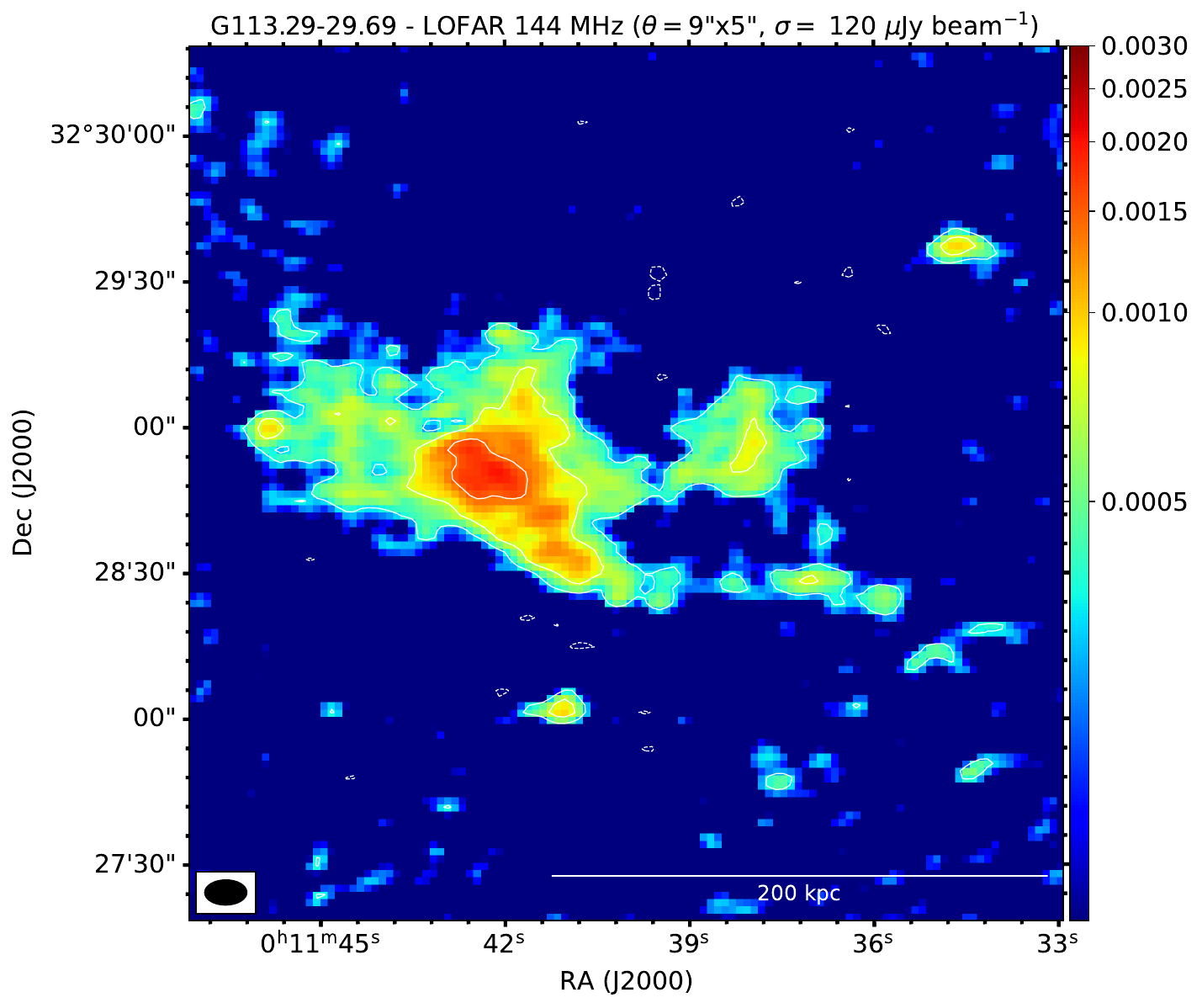}
\includegraphics[width=0.33\textwidth]{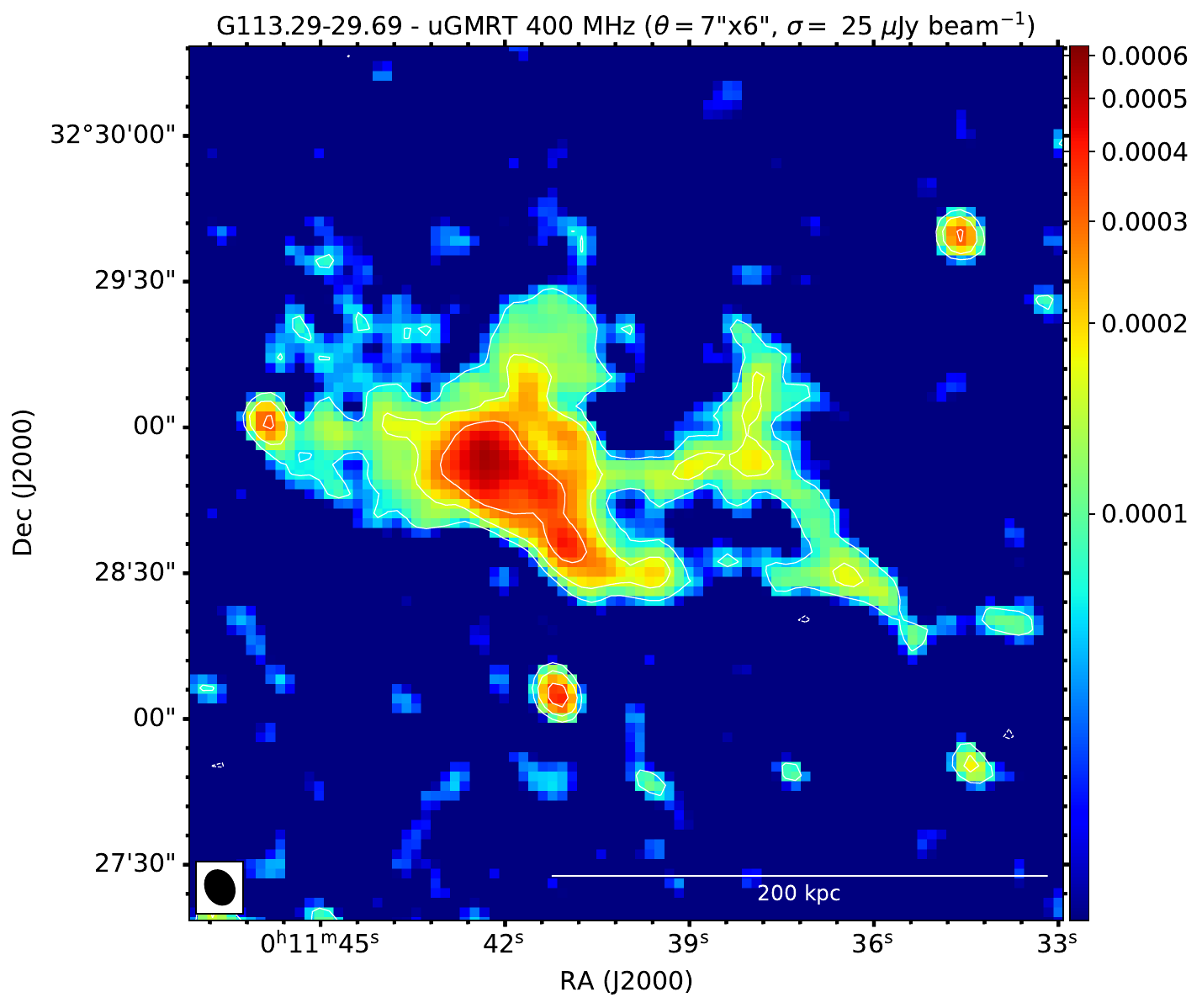} \\
\includegraphics[width=0.31\textwidth]{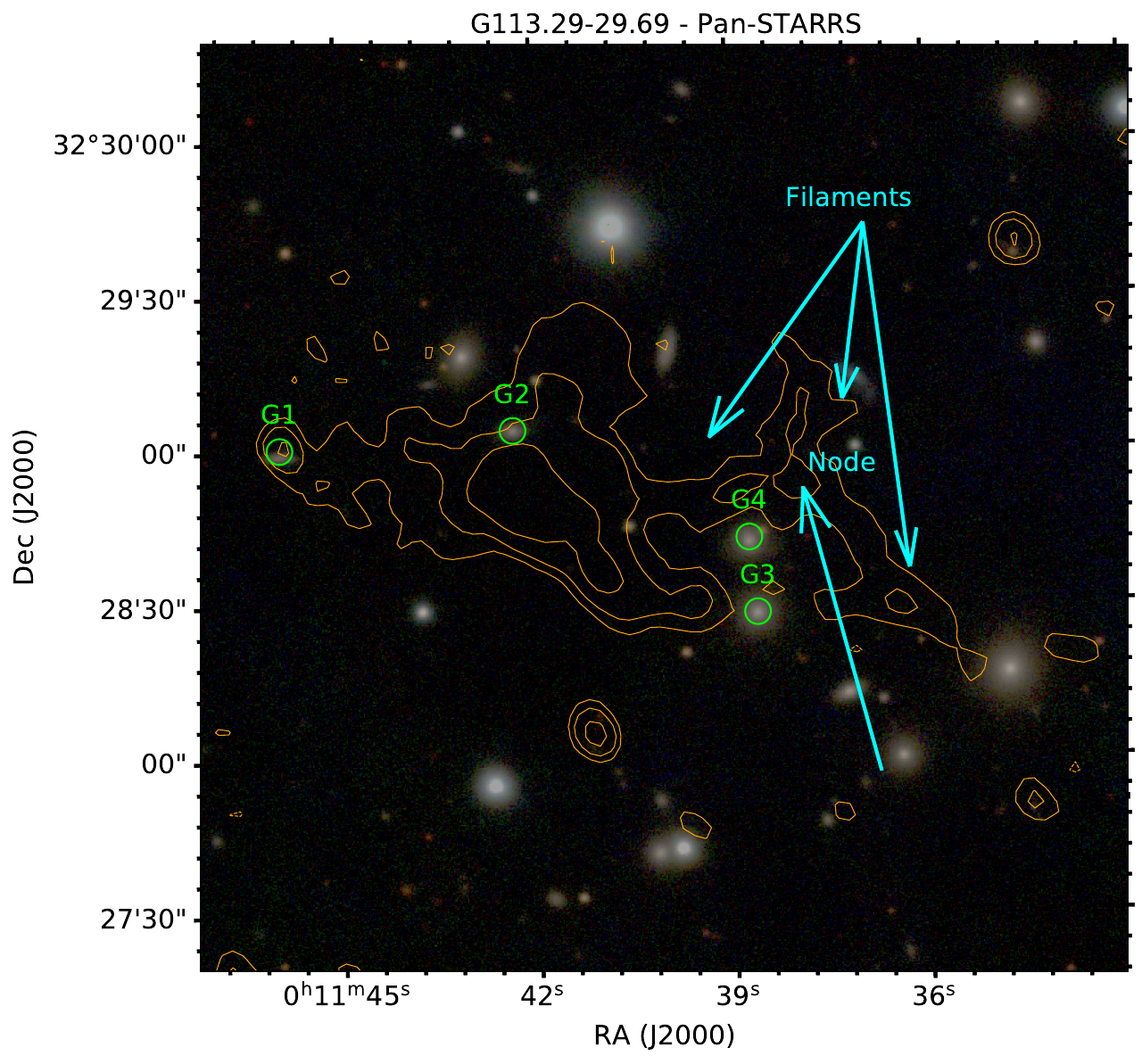}
\includegraphics[width=0.33\textwidth]{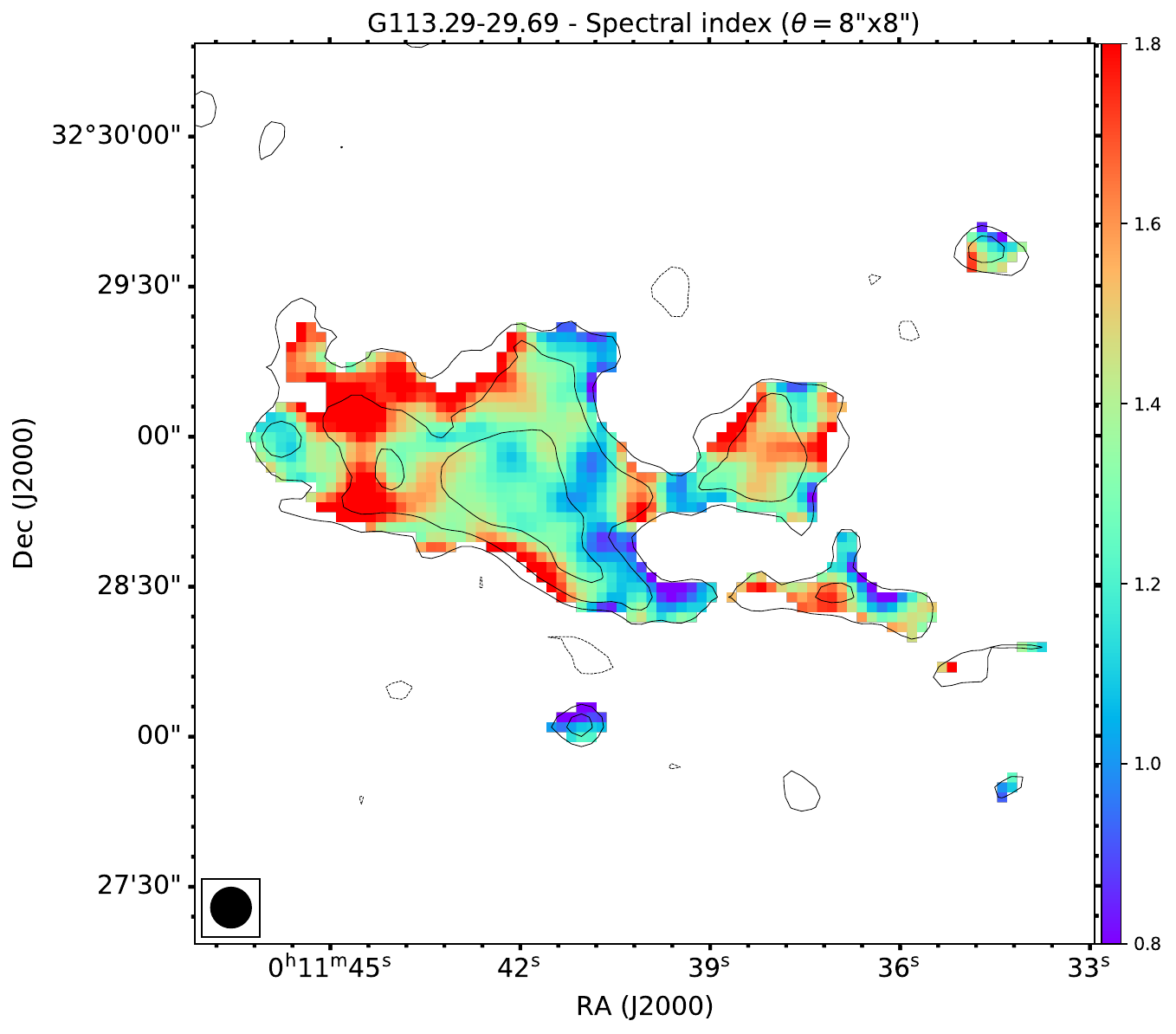}\includegraphics[width=0.33\textwidth]{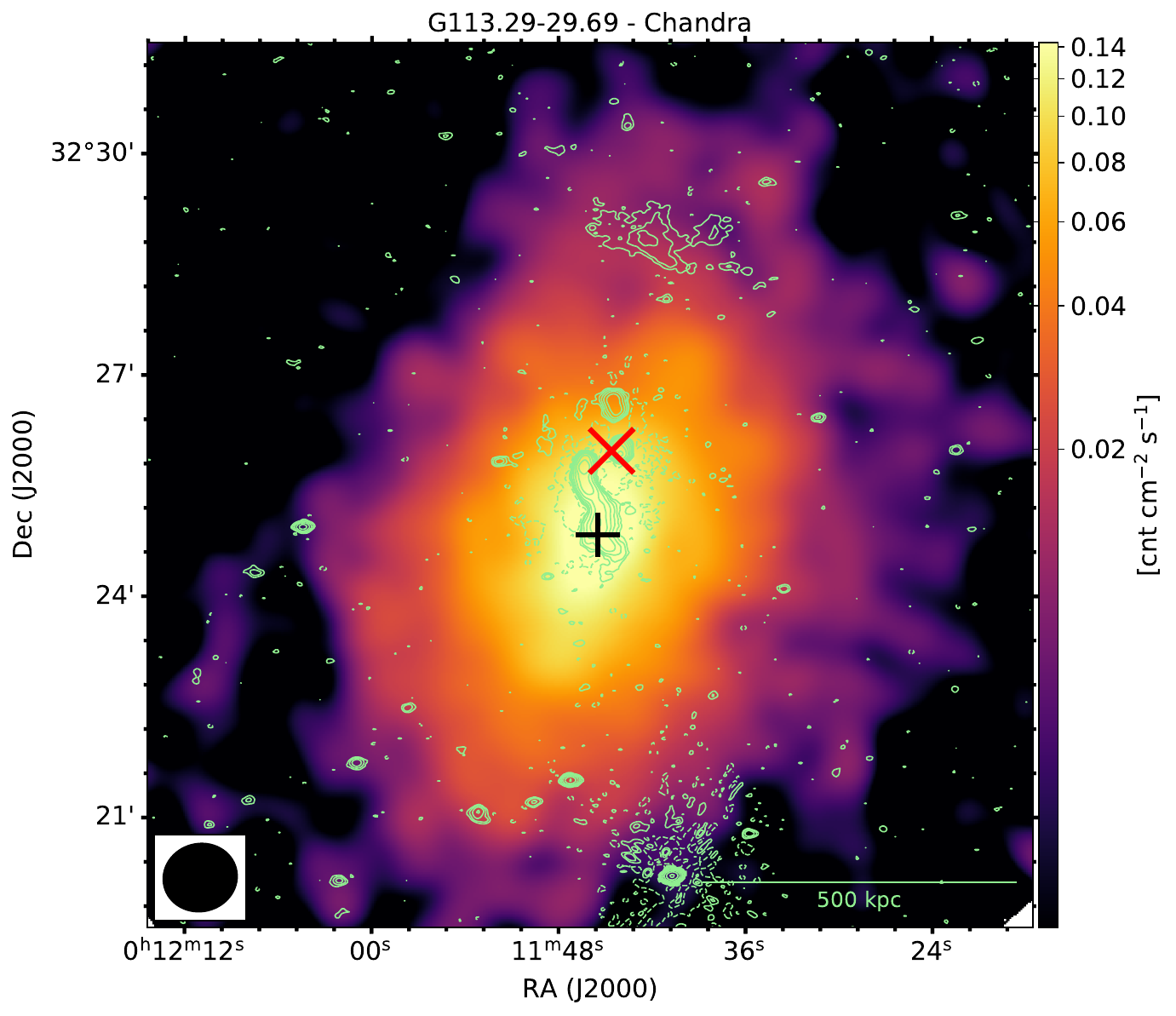}

        \caption{Images of G113.29-29.69. The description of each panel is the same as for Fig. \ref{fig: mappefullres1}. }
        \label{fig: mappefullres3}
\end{figure*}

The radio source in G113 (Fig. \ref{fig: mappefullres3}) has a roughly triangular shape with a length of $\sim 150$ kpc. In the LOFAR image, we detect a secondary blob west of the main structure, which is revealed to be a system of multiple filaments of similar length ($\sim 60$ kpc) intersecting at the position of a local radio peak (the `node') in the uGMRT image. As the 144 and 400 MHz images have a similar resolution, the apparent discrepancy of the morphology of low-surface brightness regions is likely due to a lower quality of the LOFAR data in the direction of the target.

The spectral index map shows patches of steeper and flatter regions ranging in  $\alpha\sim 1-1.8$. However, within errors, the spectral index distribution is roughly uniform, with a mean value of $\alpha\sim 1.3$. The average spectrum of the filaments is consistent with this mean value, but the current data quality prevents us from determining accurate measurements at high resolution.

Several compact radio peaks, best imaged at 400 MHz, are detected across the main region and filaments. Among them, only the brightest peak is associated with a galaxy (G1) in the Pan-STARRS image, but its high spectroscopic redshift ($z=0.125$ against $z=0.107$ for G113) suggests that it may not be a cluster member. The other galaxies (G2, G3, G4) that we identified are not co-located with specific regions of the radio source.

In the X-rays, the ICM is elongated, but it exhibits a bright core. The morphological parameters locate the cluster in between relaxed and merging systems. The radio source lies in the northern cluster outskirts at a projected distance of $\sim 500$ kpc from the X-ray core.

The radio source in G113 exhibits properties similar to those observed for the candidate RP in G071 (Sect. \ref{sect: G071}): overall morphology, location in the cluster outskirts, uncertain association with a host galaxy, and steep-spectrum fossil emission. However, there is no clear evidence of re-energising (such as constant trends in brightness or spectral index), and the average  spectral index is not as ultra-steep as typically observed in RPs. Therefore, we consider it as a remnant, although the possibility of a RP cannot be completely ruled out considering the overall similarities with RPs.

\subsection{G137.74-27.08}
\label{sect: G137}

\begin{figure*}
        \centering

\includegraphics[width=0.33\textwidth]{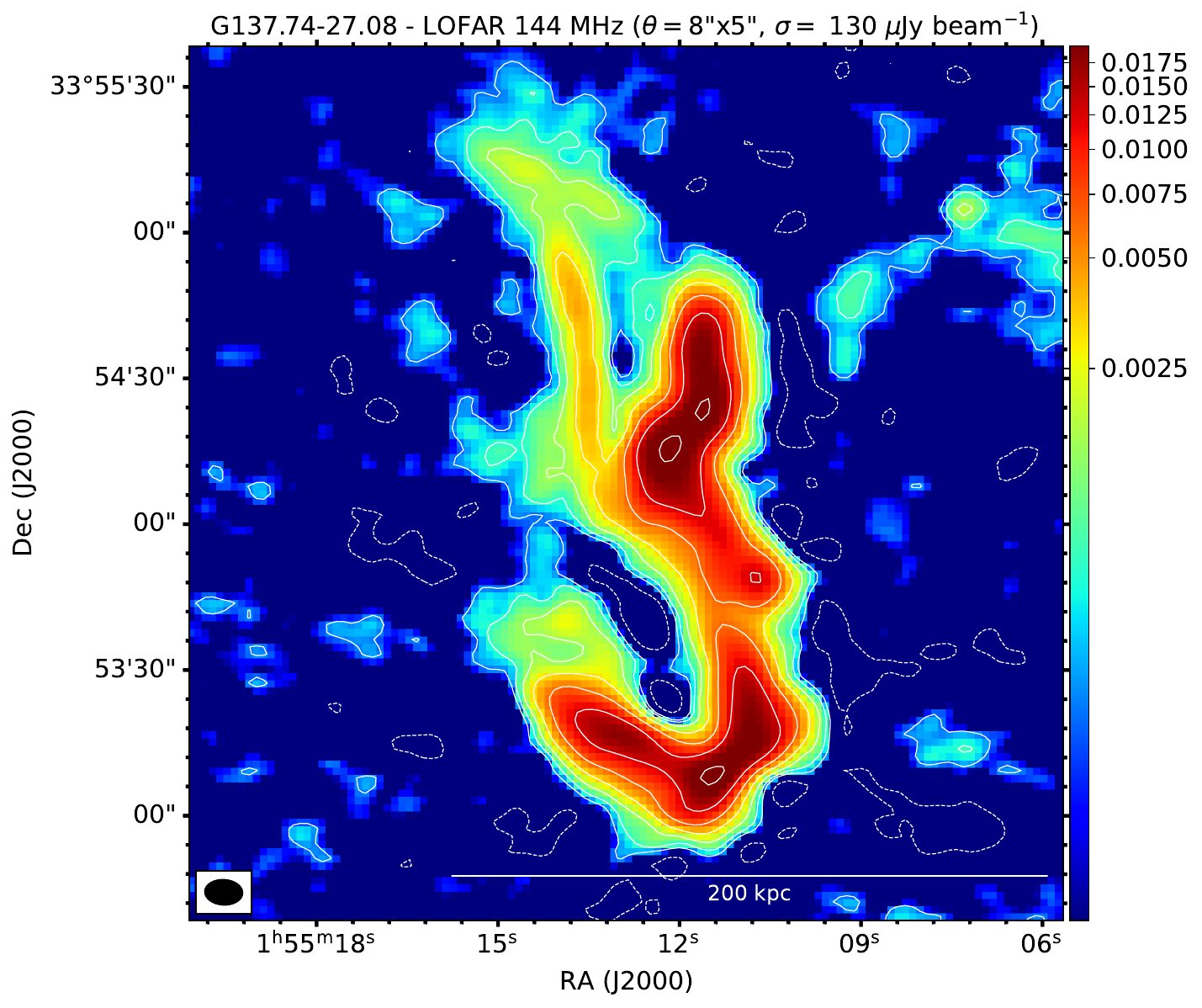}
\includegraphics[width=0.33\textwidth]{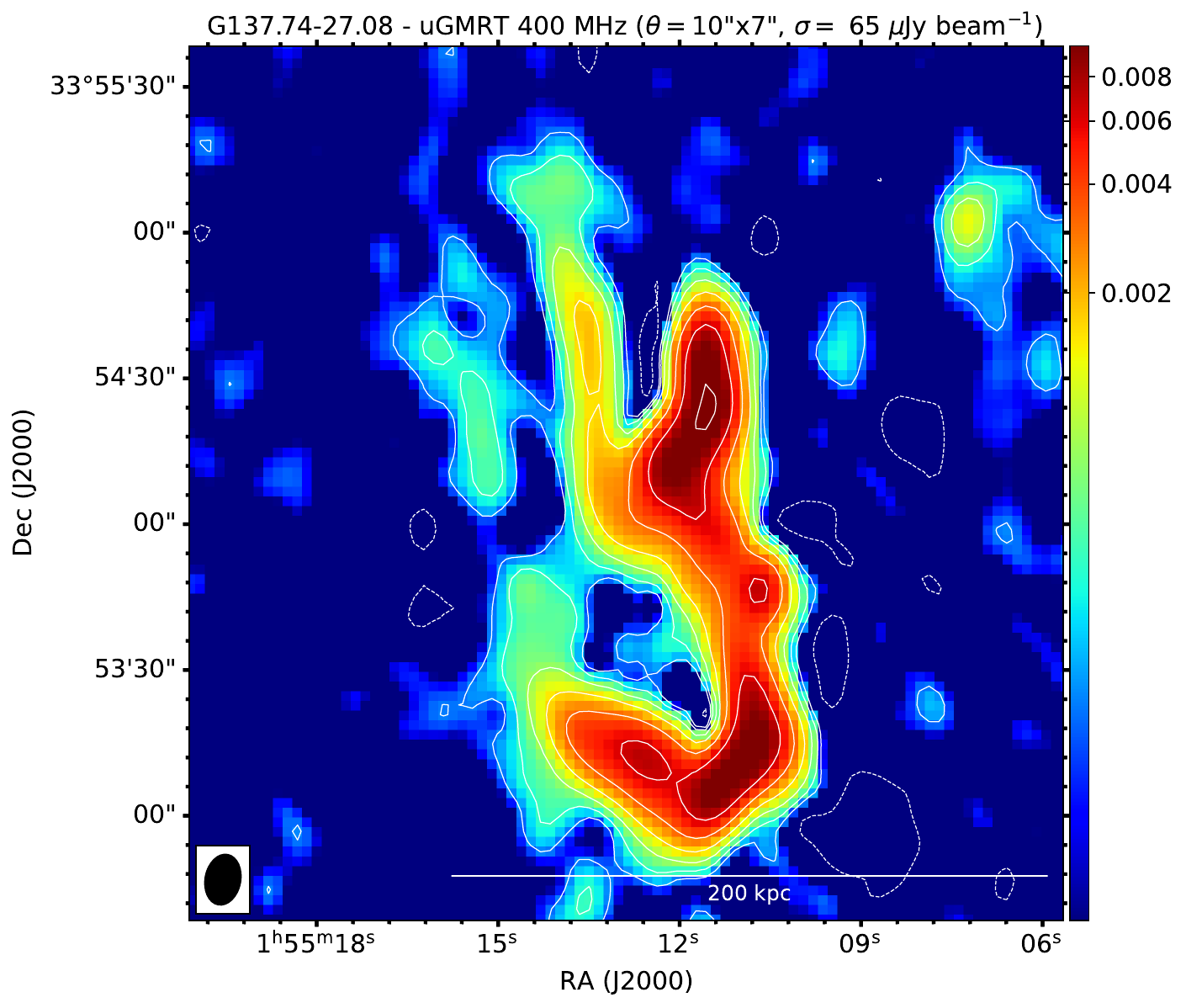}
\includegraphics[width=0.33\textwidth]{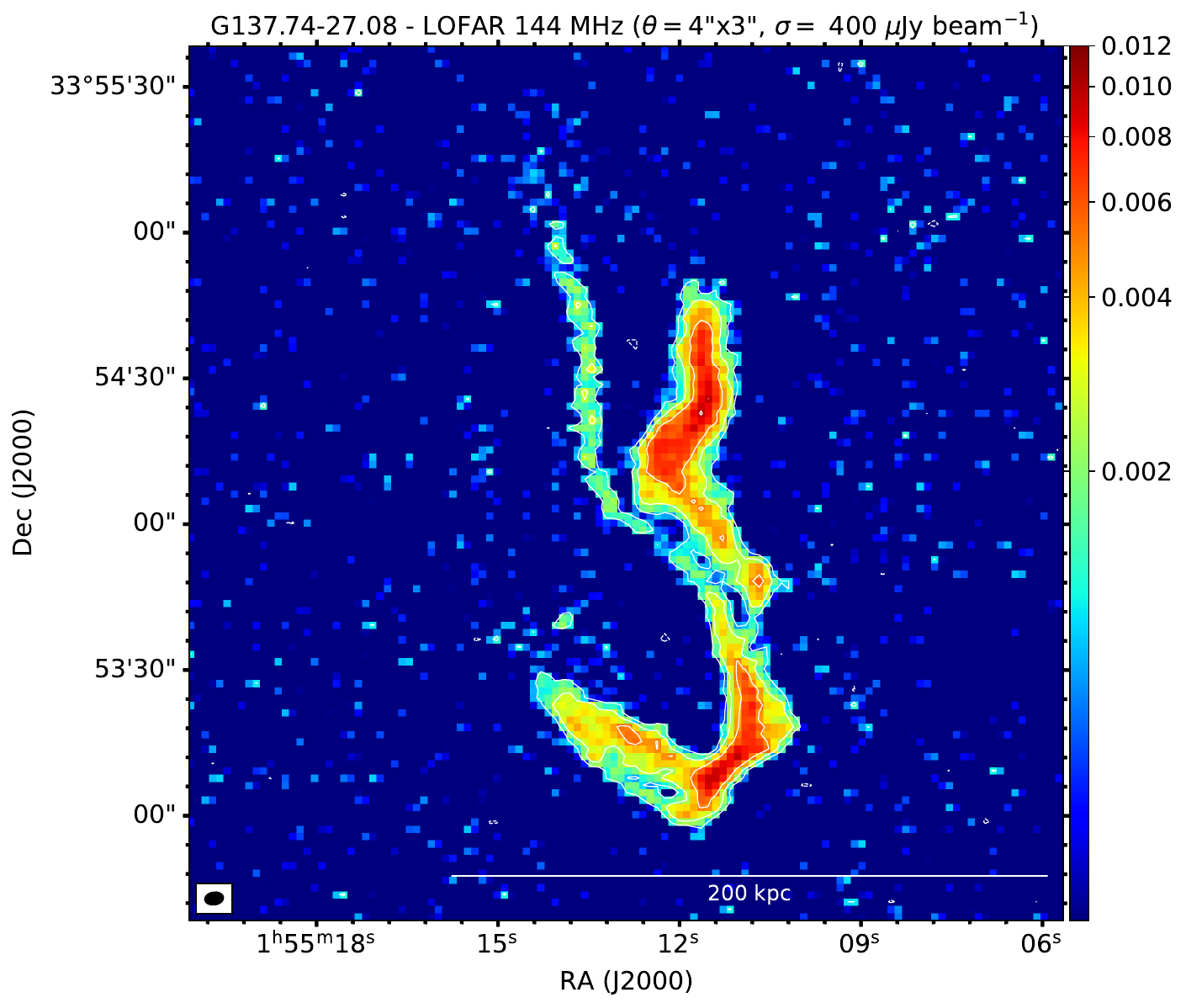}
\includegraphics[width=0.31\textwidth]{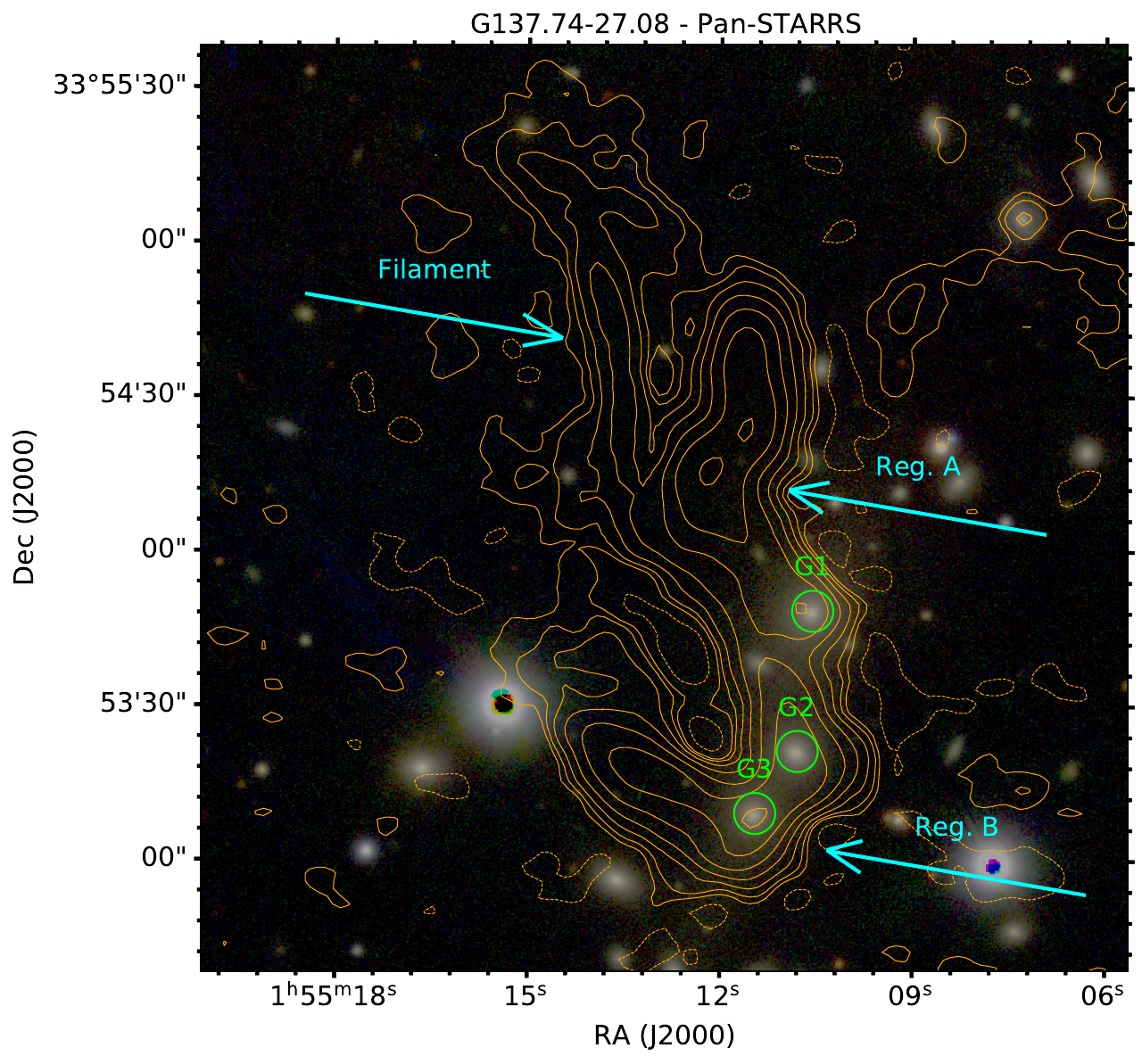}
\includegraphics[width=0.33\textwidth]{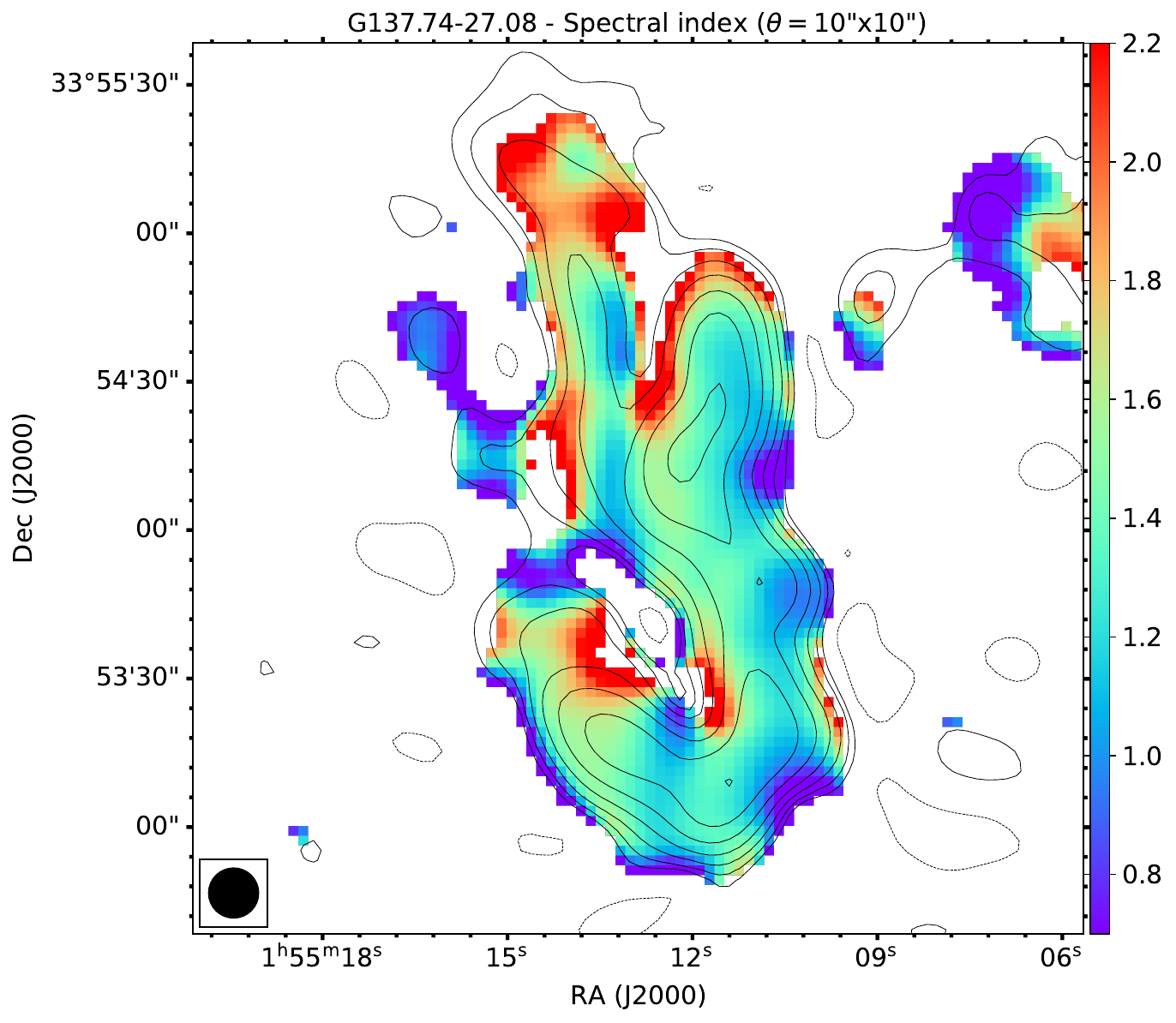}
\includegraphics[width=0.33\textwidth]{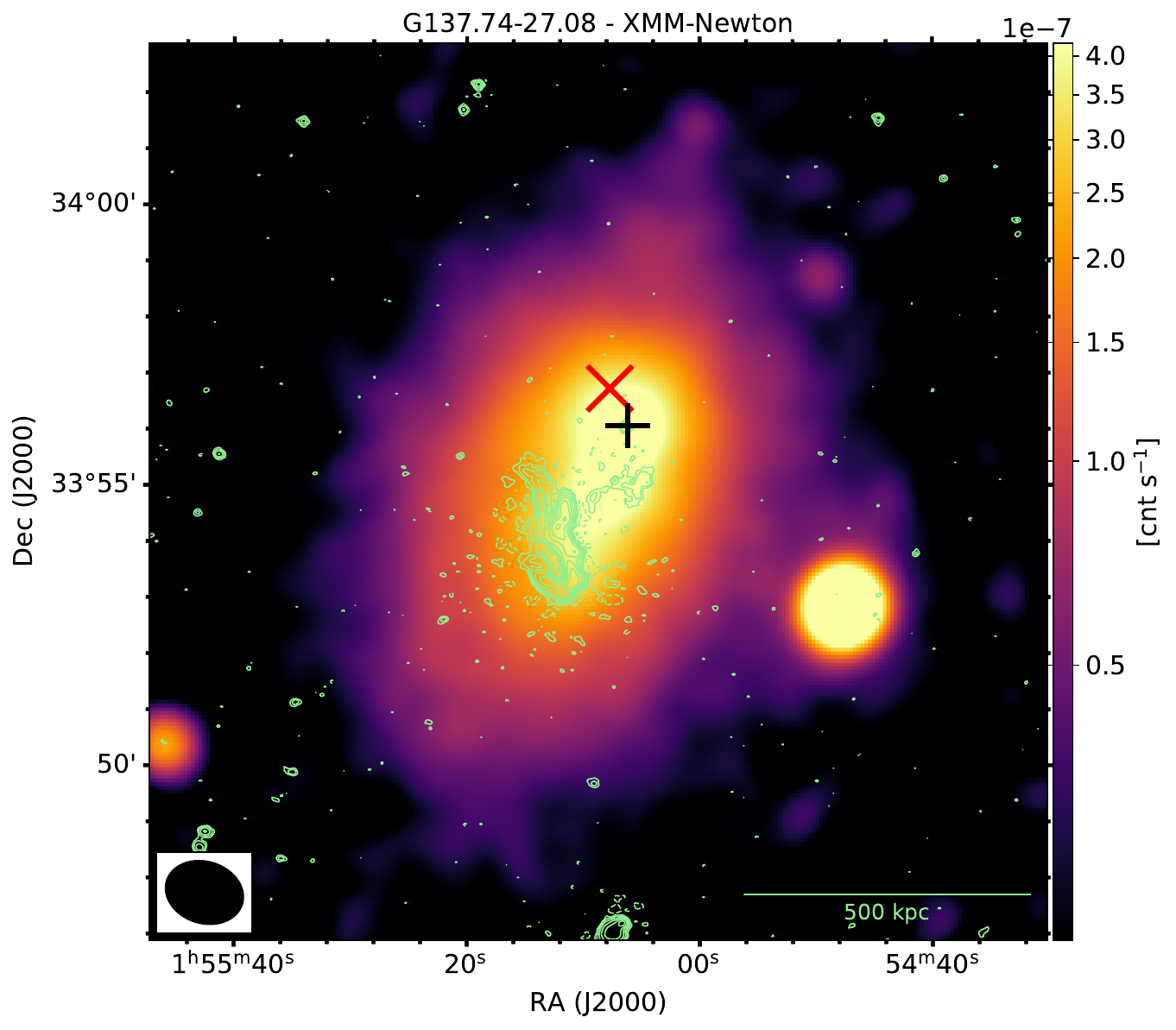}

        \caption{Images of G137.74-27.08. The description of each panel is the same as for Fig. \ref{fig: mappefullres1}. }
        \label{fig: mappefullres4}
\end{figure*}

A multi-wavelength analysis of the galaxy cluster G137.74-27.08 has been recently carried out by \citealt{whyley25}. Specifically, these authors included deep narrow-band GMRT observations at 325 MHz. Our wide-band uGMRT data, with a similar on-source observing time, allowed us to improve the rms noise close to the target by a factor of $\sim 1.4$.

The candidate revived source in G137 (Fig. \ref{fig: mappefullres4}) has a complex morphology. The main structure recalls a Fanaroff-Riley I (FRI; \citealt{fanaroff&riley74}) radio galaxy with S-shaped jets, labelled as regions A and B, departing from a central compact radio source hosted by the galaxy G1 (`G5' in  \citealt{whyley25}). A filament appears to be connected to the putative jets. The spectral index is approximately constant across the whole source, being $\alpha= 1.4 \pm 0.1$ consistent with the integrated spectrum reported by \cite{whyley25}.

The Pan-STARRS image shows that G1 shares a stellar envelope with two bright galaxies, G2 and G3 (`G4' and `G3', respectively, in \citealt{whyley25}), suggesting merging activity. The galaxies G2 and G3 are co-spatial with the local radio peaks of region B, but the resolution is not sufficient to distinguish whether only one or both of them are radio emitting. In this respect, we notice that in the Karl G. Jansky Very Large Array Sky Survey (VLASS; \citealt{lacyVLASS20}) there is no detection of radio activity from G1, G2, and G3 at a threshold of $3\sigma_{\rm VLASS}=210 \; {\rm \mu Jy \; beam^{-1}}$ (see Fig. \ref{fig: vlass1} in Appendix \ref{sect: VLASS images}). The merging nature of the cluster is confirmed by the X-ray data. Interestingly, \cite{whyley25} reported on the detection of a shock towards the region of our target, at a distance of $\sim 200$ kpc from the cluster centre. These authors thus interpreted the radio source as resulting from adiabatic compression of its lobes triggered by the shock passage. In the top right panel of Fig. \ref{fig: mappefullres4} we show a higher-resolution LOFAR image of G137, and in Fig. \ref{fig: profiles G137} the corresponding surface brightness profile. These images suggest alternative scenarios, which we discuss below.

In the first scenario, regions A and B are radio lobes of a single radio galaxy, with G1 being its host. This would be supported by the similar integrated spectra and radio powers of the two regions (see Table \ref{table: flux}). The S-shaped morphology can be driven by jet precession induced by G2 and G3. However, the absence of a global spectral gradient and the ultra-steep spectrum cannot be easily reconciled with a single radio galaxy.

In the second scenario, regions A and B are independent sources, specifically a HT (hosted by G1) and a WAT (hosted by either G2 or G3) radio galaxy, respectively. Interestingly, in the high-resolution image, the filament appears to be the terminal part of the western jet of the WAT rather than being associated with region A. The surface brightness and spectral index profiles (Fig. \ref{fig: profiles G137}) support the WAT classification and suggest that G3 is its host: the surface brightness peaks at the position of G3 and decreases along the two putative bent jets, and the spectral index steepens departing from G3. The filament can be explained in terms of drifting of the WAT towards south, which leaves electrons behind (see e.g. the WAT 111 in Ophiuchus for a similar case; \citealt{botteon25}, \citealt{cotton25}). On the other hand, the brightness overall increases and the spectral index is roughly constant along region A, challenging the simple HT scenario.

In the third scenario, the merging event involving G1, G2, and G3 is responsible for the observed radio emission. The collision of these galaxies induced re-energising of fossil electrons and reshaped the radio source via ICM shocks and/or turbulence.

Confirming a preferential scenario is challenging, but current data favour the WAT plus filament interpretation for region B. We notice that the surface brightness profile shows fluctuations around a constant value over a length of $\sim 130$ kpc along the filament, whereas its spectrum gradually steepens (although spectral measurements can be biased by the lower resolution of the spectral index map). Constant and enhanced brightness, as observed for both region A and the filament, and flat spectral trend, as observed for region A, are indicative of ongoing re-energising of aged electrons and/or amplification of the magnetic field. However, the nature of the radio source in G137 remains uncertain, and this is a clear case warning against a simple classification as RP or GReET based on limited information.

\subsection{G155.80+70.40} 
\label{sect: G155}

\begin{figure*}
        \centering

\includegraphics[width=0.33\textwidth]{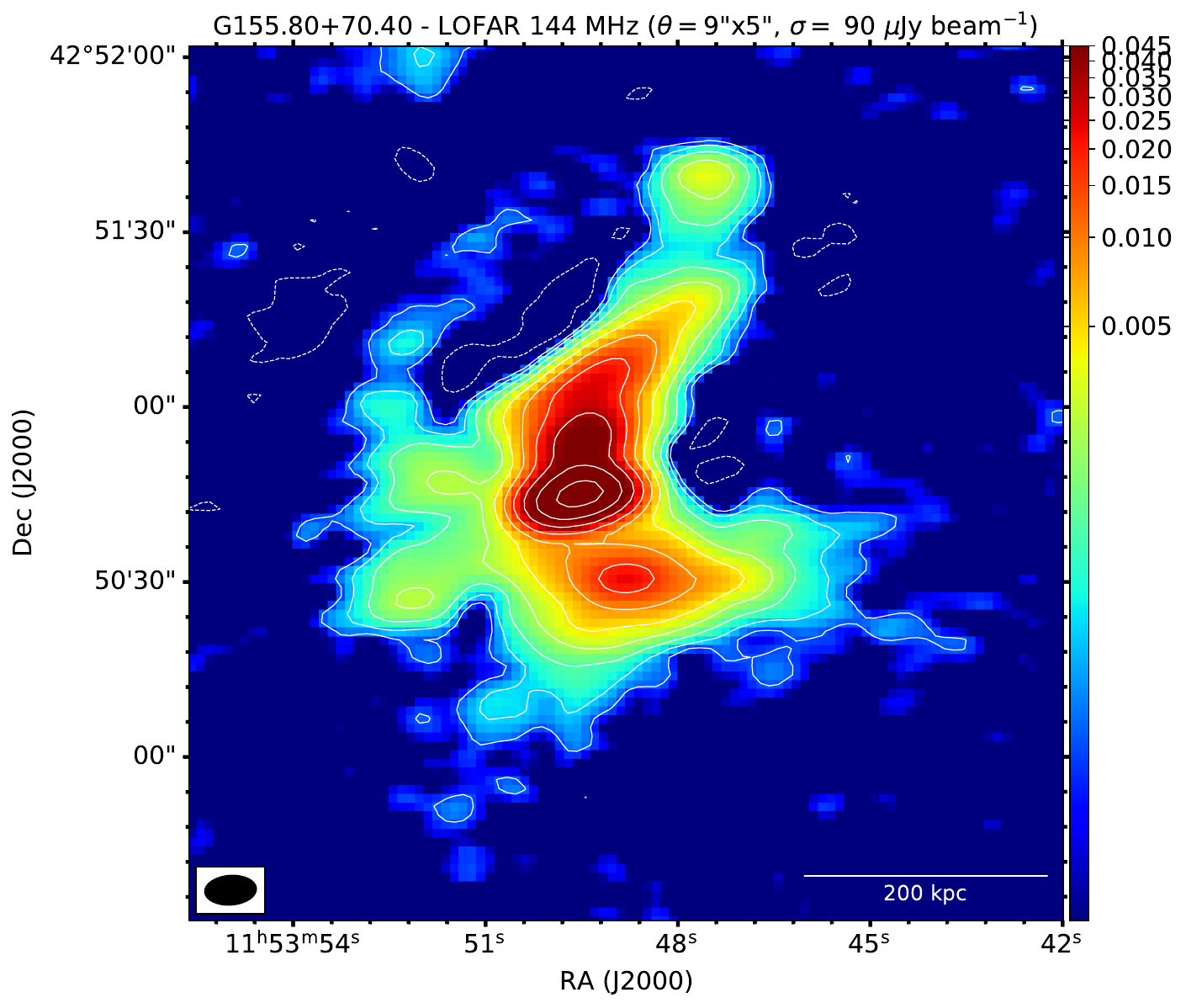}
\includegraphics[width=0.33\textwidth]{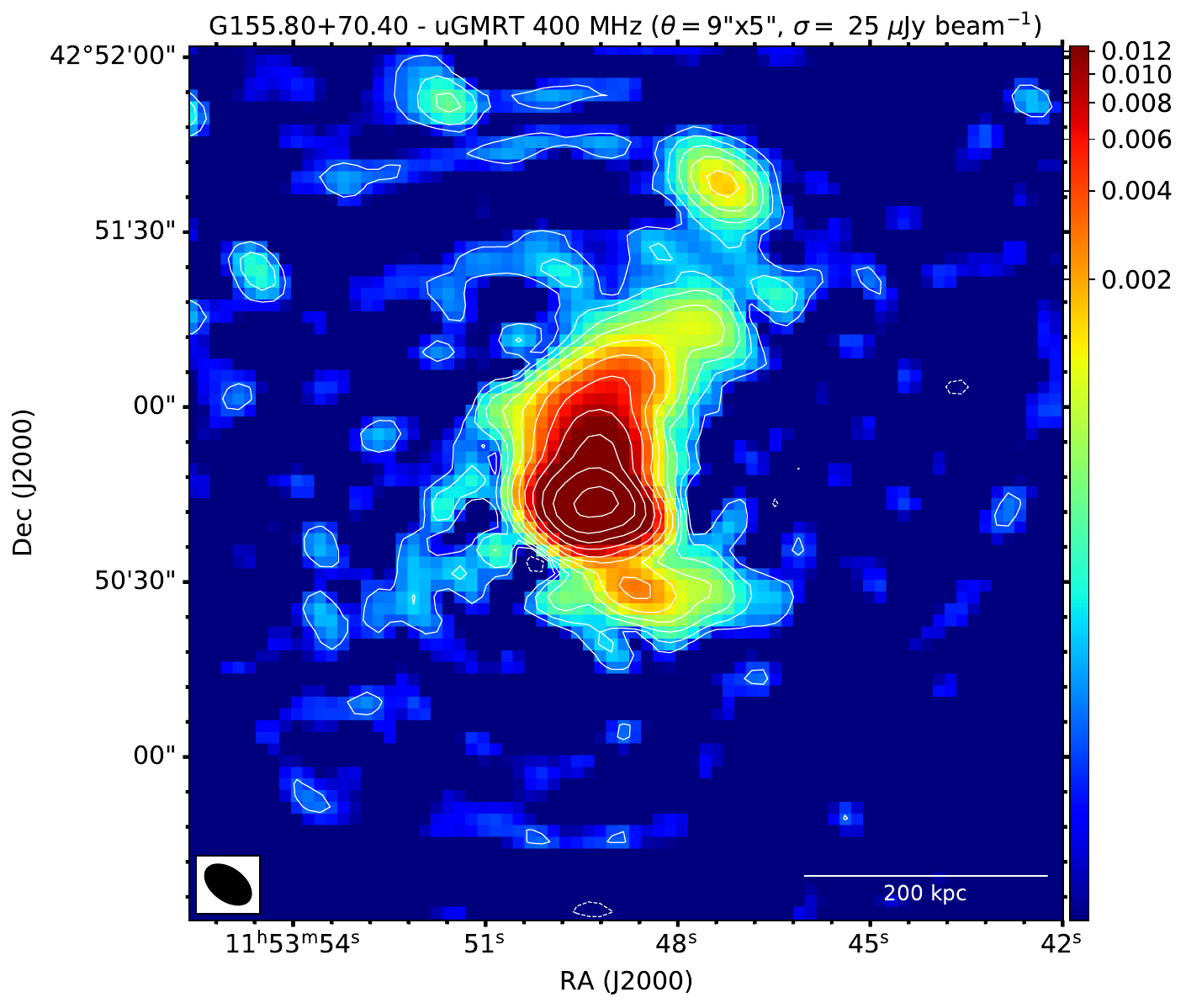}
\includegraphics[width=0.33\textwidth]{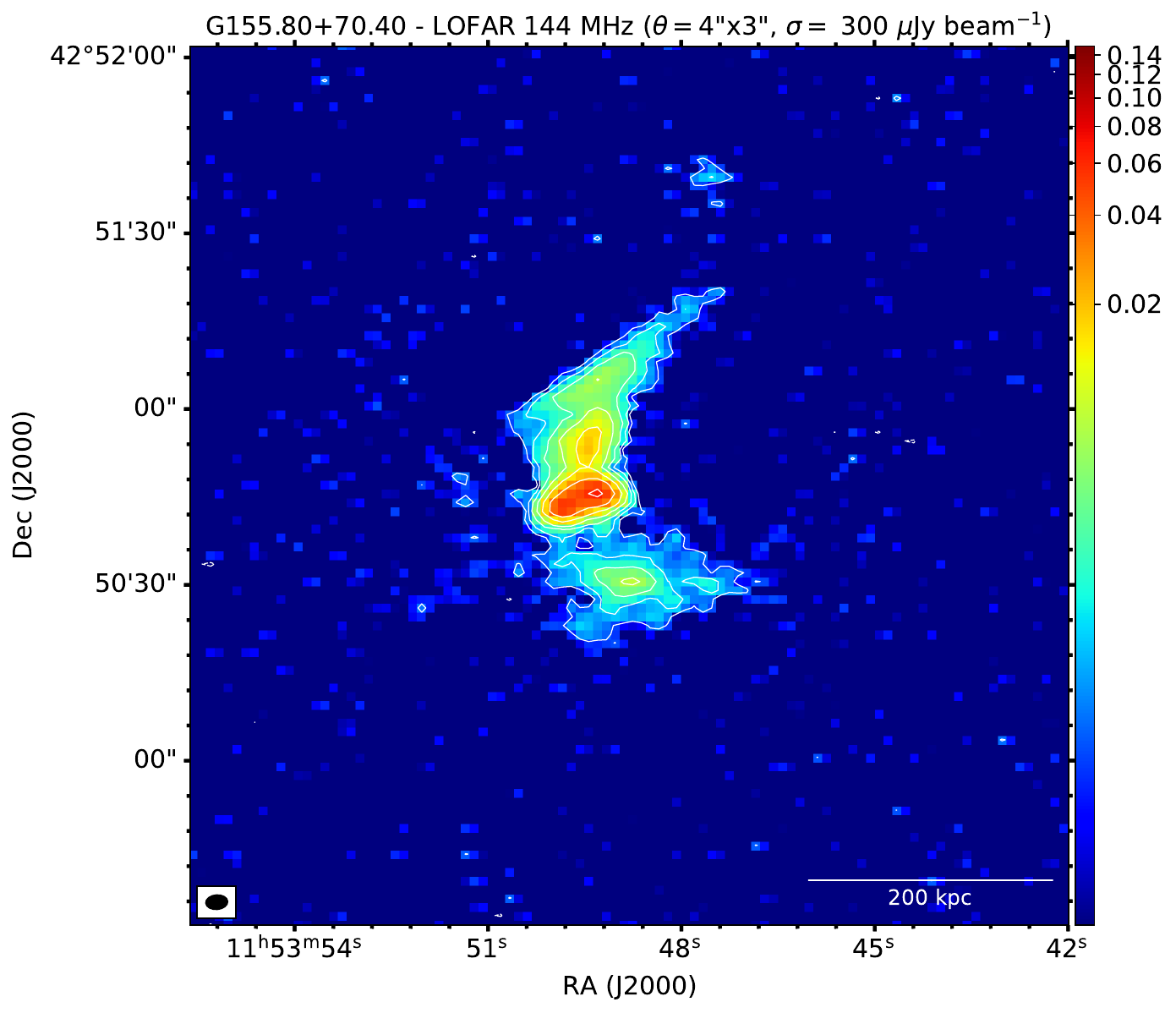}
\includegraphics[width=0.31\textwidth]{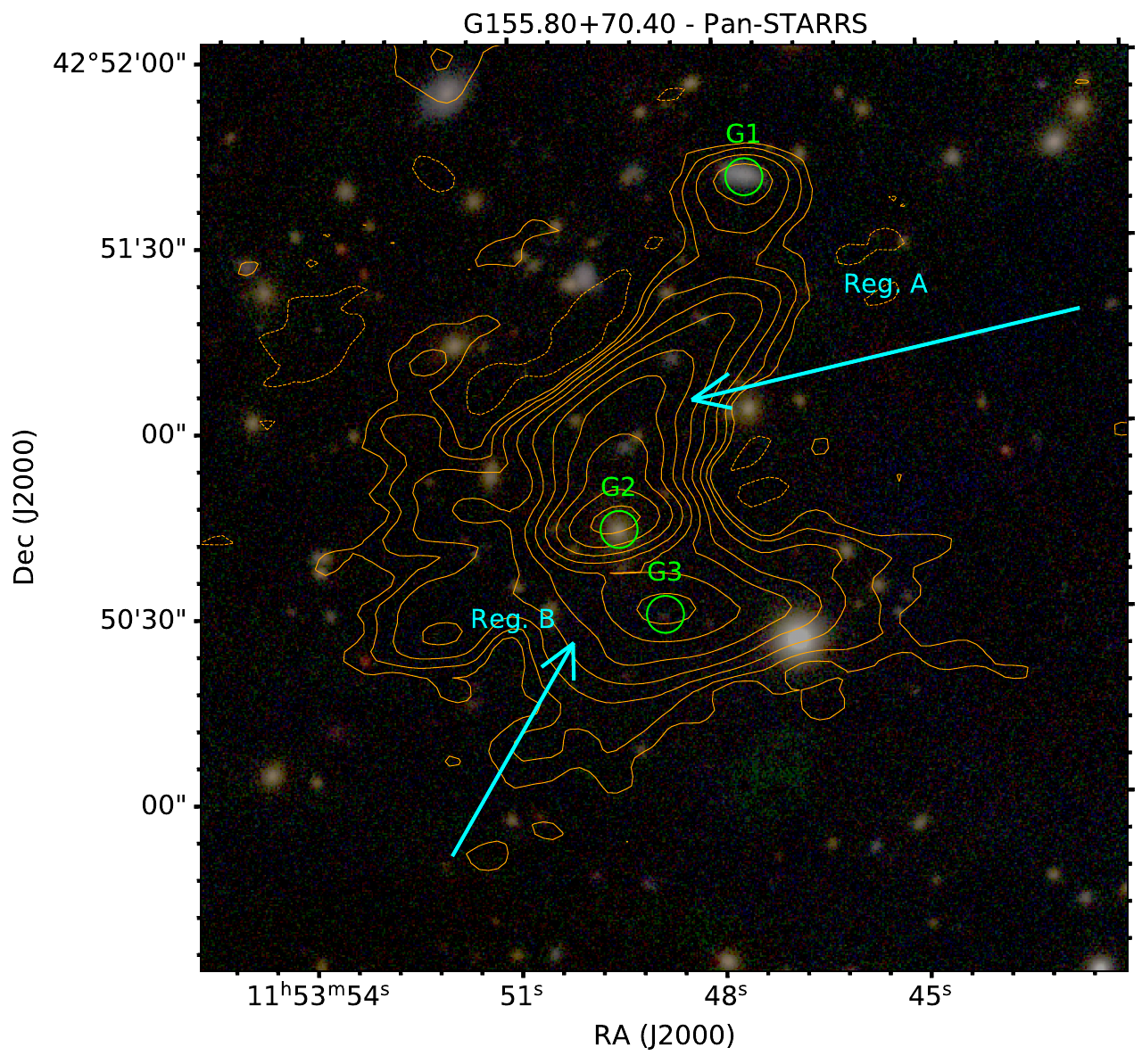}
\includegraphics[width=0.33\textwidth]{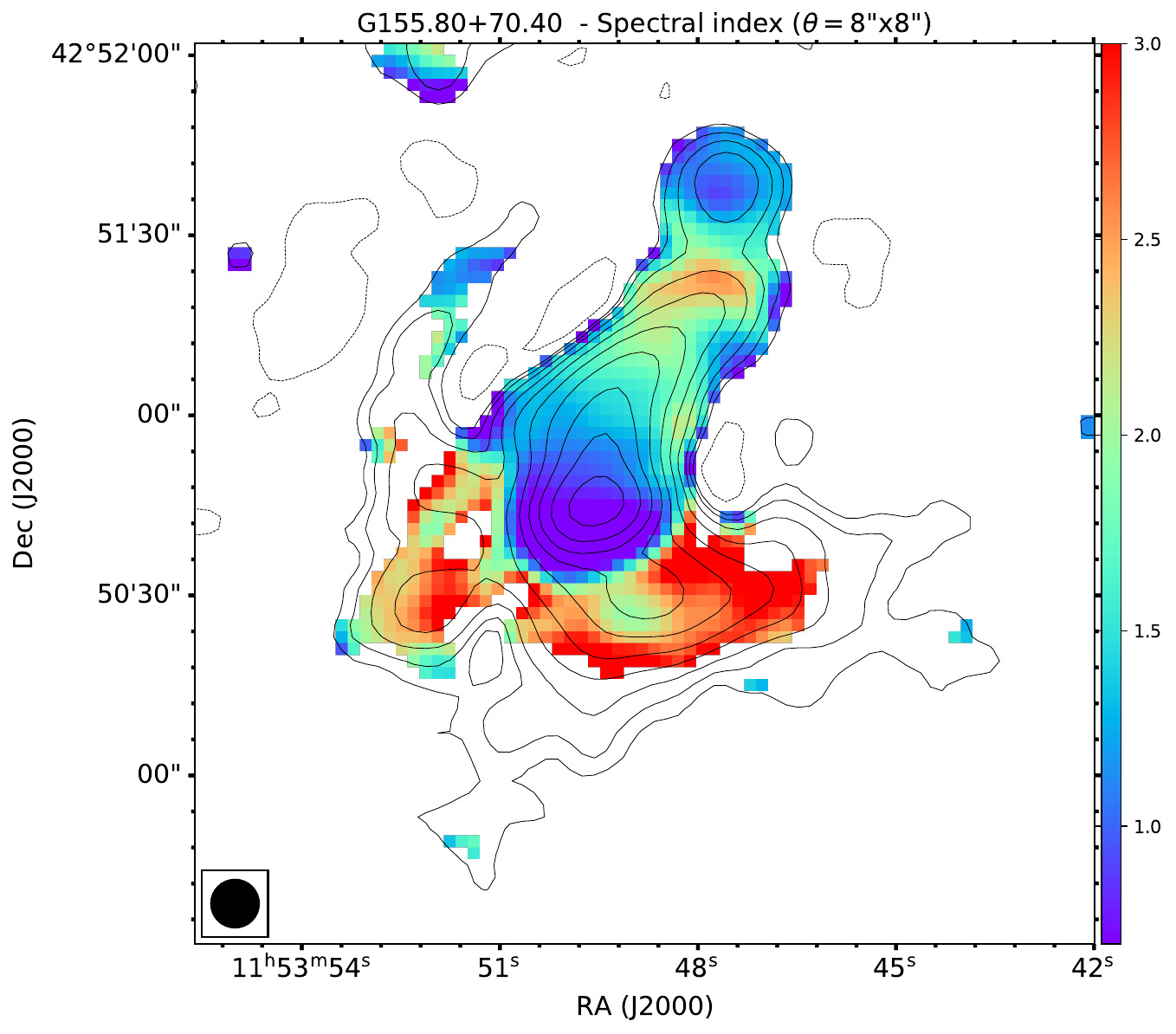} 

        \caption{Images of G155.80+70.40. The description of each panel is the same as for Fig. \ref{fig: mappefullres1}. }
        \label{fig: mappefullres5}
\end{figure*}

X-ray data of G155 are not available and  optical images do not reveal the presence of a dominant member galaxy. Therefore, determining the position of the radio source in G155 (Fig. \ref{fig: mappefullres5}) with respect to the cluster centre is challenging; assuming the \textit{Planck} centre, the target lies at a projected distance of $\sim 400$ kpc. The radio source in G155 is diffuse and exhibits multiple radio peaks. We distinguished two regions within the source.

Within region A, the two radio peaks are co-spatial with the cluster member galaxies G1 and G2, which have a projected separation of $\sim 300$ kpc. The spectral indices at the location of the radio peaks are $\alpha\sim 1$ and $\alpha\sim 0.6$ for G1 and G2, respectively. As shown by the spectral index profile (Fig. \ref{fig: profiles G155}), we measured a gradual steepening from G2 towards G1.

Region B refers to the diffuse southern emission of the target. It has a size of $\sim 450$ kpc, as measured along E-W. The radio peak is co-spatial with G3, a non-cluster member galaxy at high redshift ($z=0.687$, Table \ref{table: galaxies}). Our uGMRT image recovers only the brightest emission of this region (inner $\sim 150$ kpc), where we measured a remarkably ultra-steep spectral index of $\alpha\sim 2.7$. The undetected emission at 400 MHz is expected to have an even steeper spectrum.

The spectral information that we obtained provides crucial hints on the nature of the target. The spectral profile suggests that G2 is the host of a HT radio galaxy moving southwards. This scenario is confirmed by the LOFAR image at higher resolution, which reveals unresolved radio jets departing from the northern region of G2 towards east and west directions, in line with the expected deflection towards north. Notably, this image shows a sharp transition of the tail, which is abruptly deflected from N-S to NW-SE. Similar transitions are becoming common in tailed radio galaxies, with NGC 4869 in Coma \citep{lal20} being a prototypical example, and plausibly originate from the interaction of the tail with the ICM. As is typical for tailed galaxies, the surface brightness decreases with the distance from the core (Fig. \ref{fig: profiles G155}), but this decline is slowed down along the transition.

Region B is amorphous and has an ultra-steep spectrum, confirming its fossil nature. It is unlikely that G3 is the host of such emission, as it would require a chance alignment between our target and an high-$z$ remnant source. A plausible explanation is that region B traces remnant emission from a previous AGN activity phase (possibly of G1, considering its steep spectrum).

\subsection{G165.68+44.01} 
\label{sect: G165}

\begin{figure*}
        \centering

\includegraphics[width=0.33\textwidth]{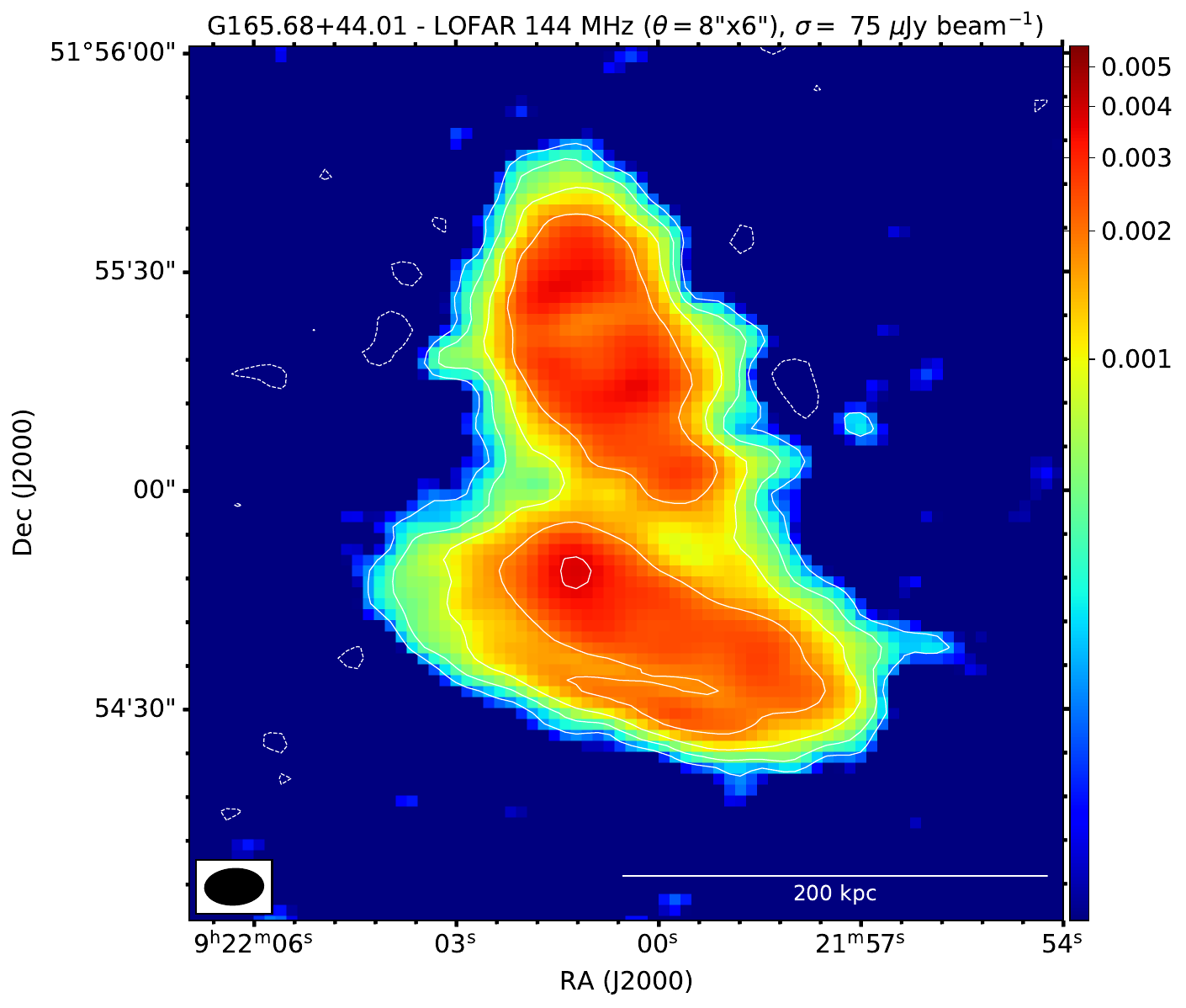}
\includegraphics[width=0.33\textwidth]{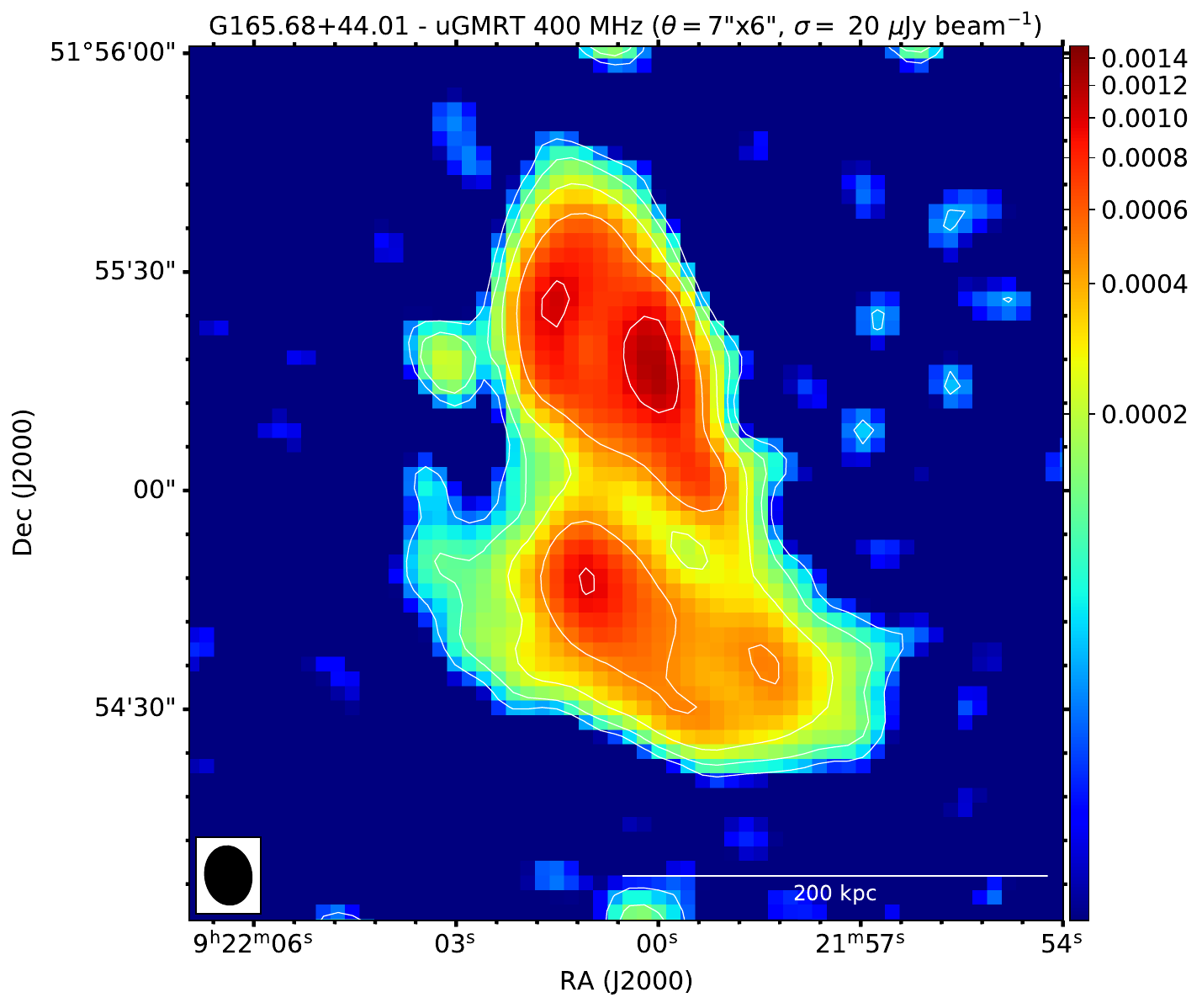}
\includegraphics[width=0.33\textwidth]{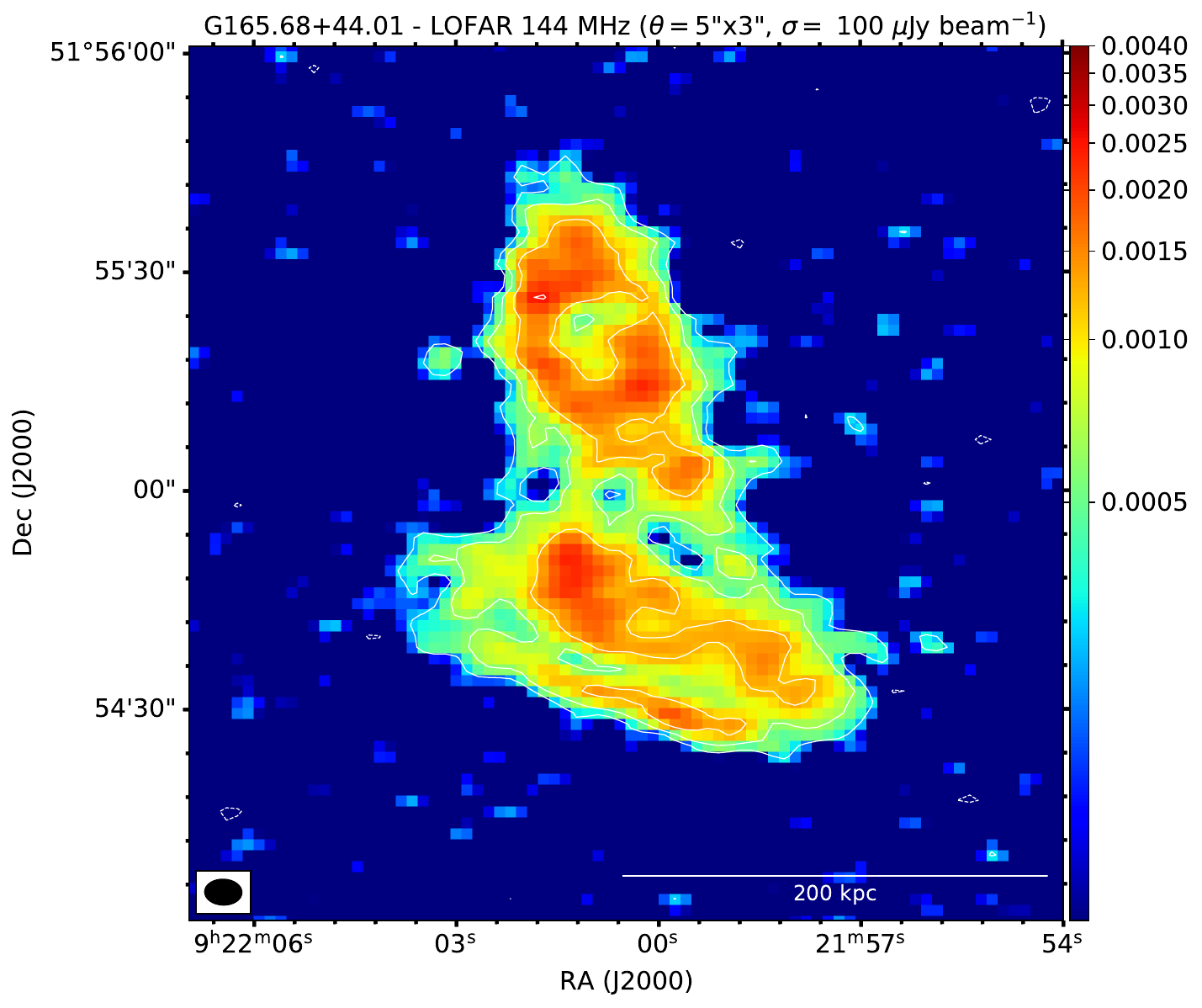}
\includegraphics[width=0.31\textwidth]{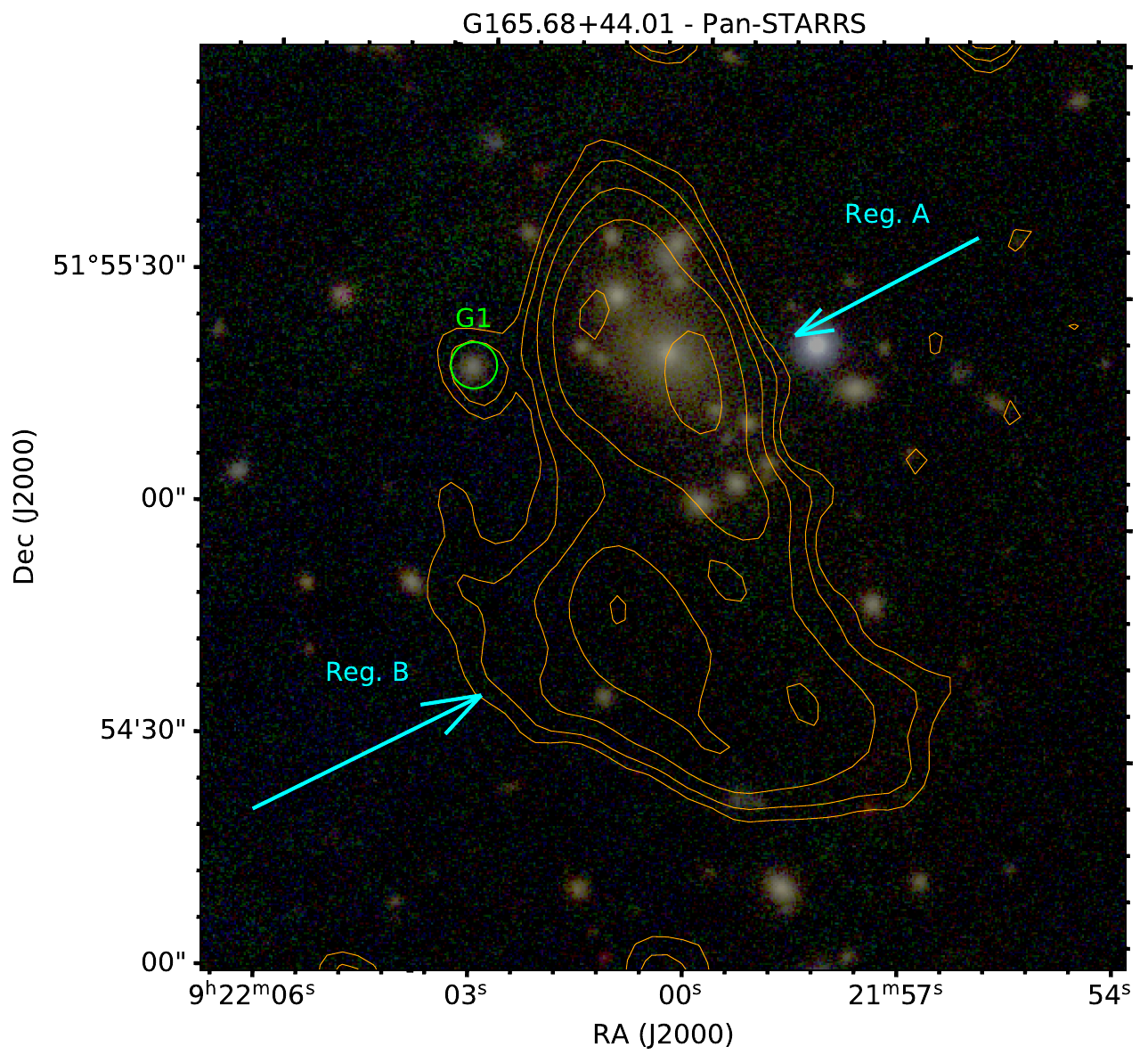}
\includegraphics[width=0.33\textwidth]{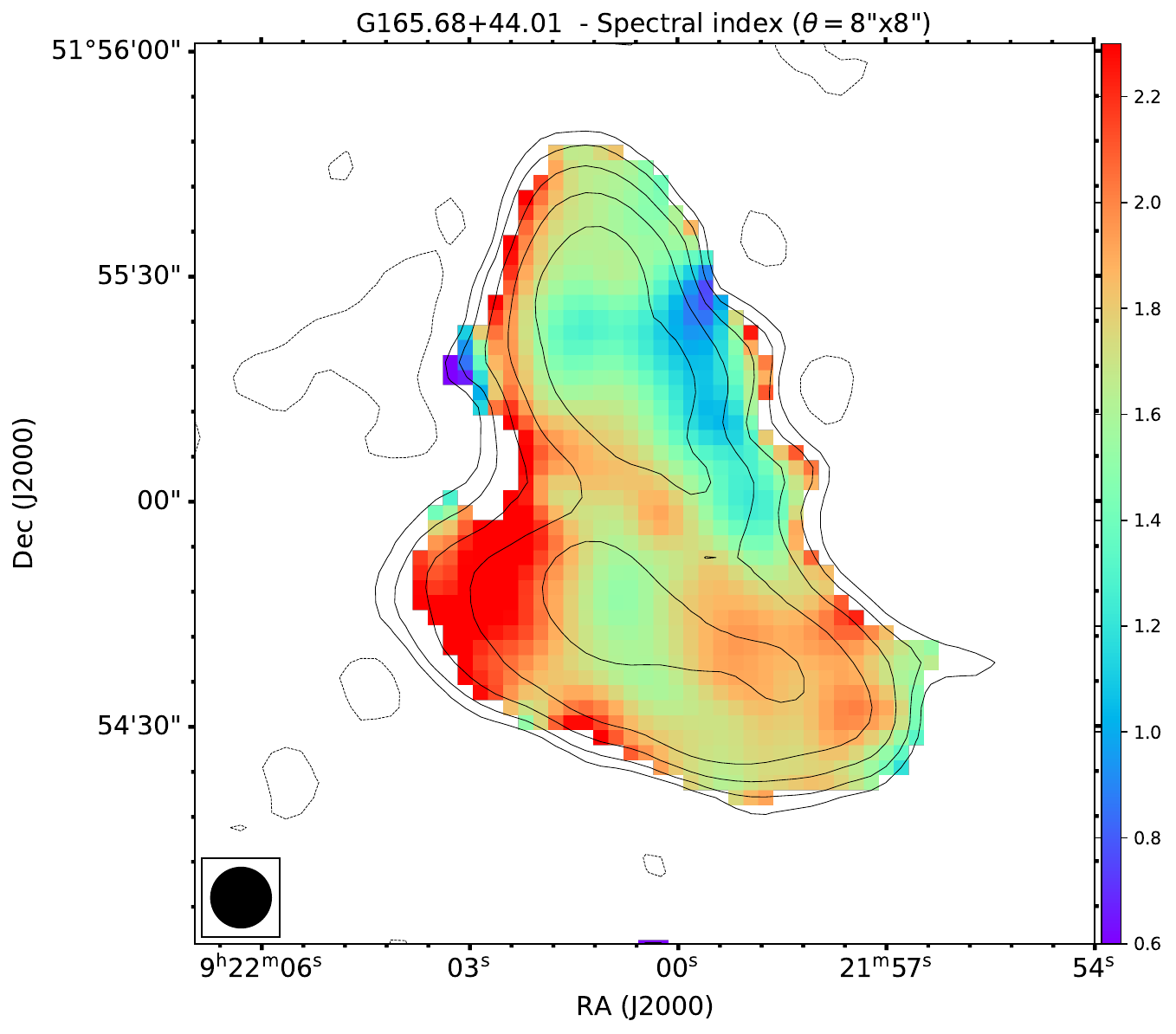}
\includegraphics[width=0.33\textwidth]{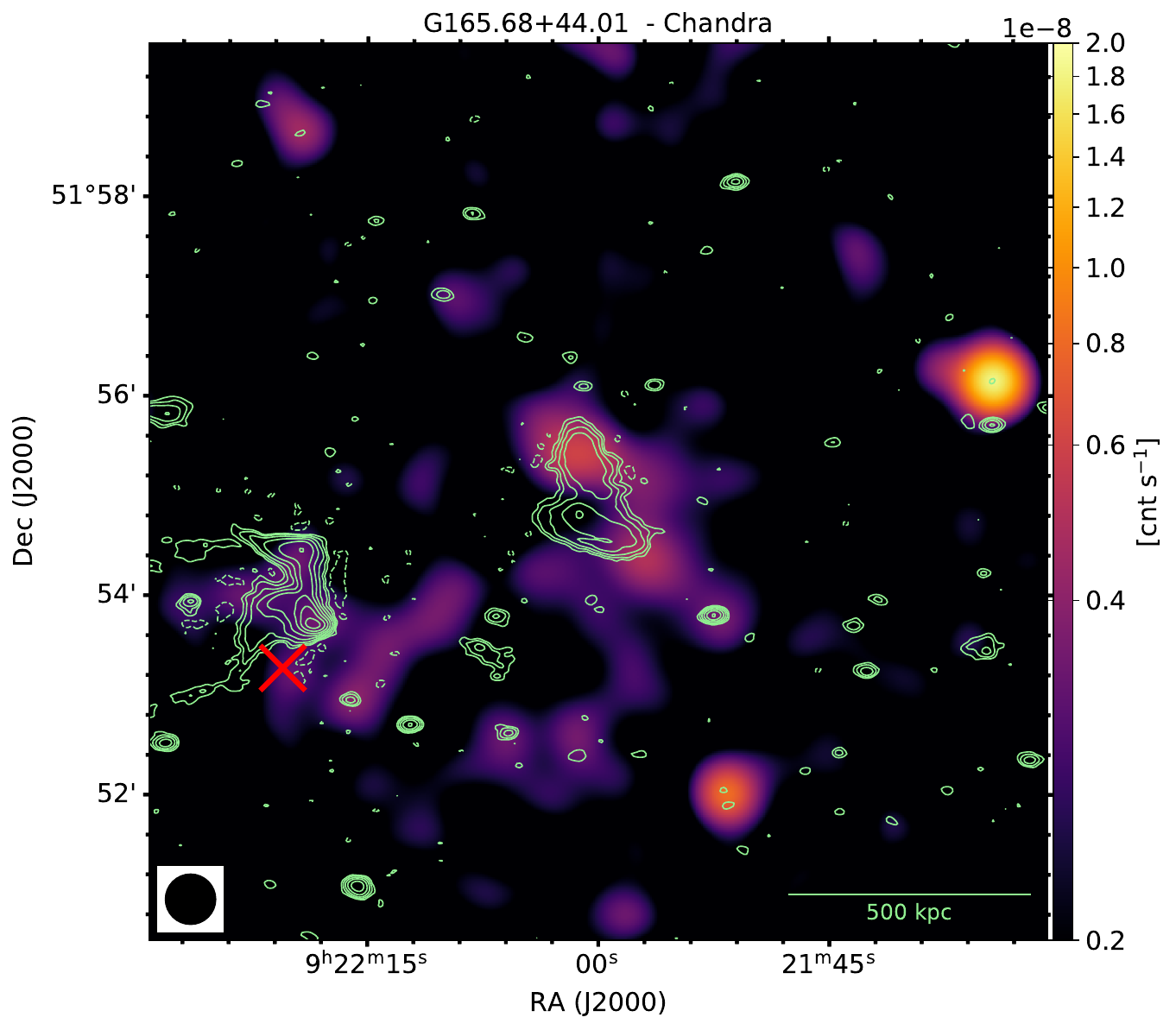}

        \caption{Images of G165.68+44.01. The description of each panel is the same as for Fig. \ref{fig: mappefullres1}. }
        \label{fig: mappefullres6}
\end{figure*}

The radio source in G165 (Fig. \ref{fig: mappefullres6}) consists of two apparently connected regions of similar ellipsoidal shape and size ($\sim 250$ kpc). However, the LOFAR image at $\theta=5''\times3''$ resolution reveals different internal structures, with region A (in the north) exhibiting a clear C-shaped structure and region B (in the south) consisting of filamentary components. 

C-shaped morphologies are typical of WAT radio galaxies, but there is no evidence of a radio core within region A that can confirm this interpretation. Indeed, we identified only a compact radio source, associated with the galaxy G1, but it is largely offset from the putative jet bending point and its photometric redshift ($z=0.302\pm0.047$) suggests that it is a background source. Similarly, the filamentary components of region B might be interpreted as bent radio jets from a tailed galaxy seen almost edge-on, but we do not observe any potential radio core. The spectral index map shows that both regions A and B consist of ultra-steep spectrum emission and, although the distribution is not uniform, there are no gradients, which would be expected if they were tailed radio galaxies. The integrated spectral indices are $\alpha \sim 1.5$ and $\alpha \sim 1.8$ for regions A and B, respectively, but this difference is not particularly significant within errors, and it is moslty driven by the presence of a flatter-spectrum patch ($\alpha \sim 1.1$) and a steeper-spectrum patch ($\alpha \sim 2.3$) at the eastern and western edges of the two regions, respectively.

The location of the target within the cluster is unclear. The reported \textit{Chandra} X-ray image (see also Appendix \ref{sect: Xray data for G088}) is not sufficiently deep, but it shows tentative detection of diffuse emission both close to the \textit{Planck} centre, where a WAT radio galaxy (see overlaid contours) is located, and co-spatial with the target radio source. In the optical, the same regions are occupied by several member galaxies gathered around dominant galaxies of similar magnitudes. The projected distance of the two dominant galaxies of the systems, and analogously of the X-ray patches, is $\sim 670$ kpc. Therefore, the cluster centre is largely uncertain, and it is possible that G165 actually consists of two main groups rather than being a single system.

In summary, our analysis is not conclusive on the nature of the radio emission in G165. Both regions A and B consist of fossil plasma, but it is unclear whether this is associated with a single remnant or two distinct remnants. None of them are clearly associated with an optical host and X-ray information is inadequate. Deeper X-ray data are necessary to investigate the thermal emission in G165 and its dynamical state, with important implications on the interpretation of the target. In this respect, we cannot rule out possible interplay with the ambient medium; for instance, the complex internal structures may arise from reprocessing of fossil emission by the cluster (or group) weather (see a similar case in the group NGC 507; \citealt{brienza22}).

\subsection{G172.74+65.30} 
\label{sect: G172}

\begin{figure*}
        \centering

\includegraphics[width=0.33\textwidth]{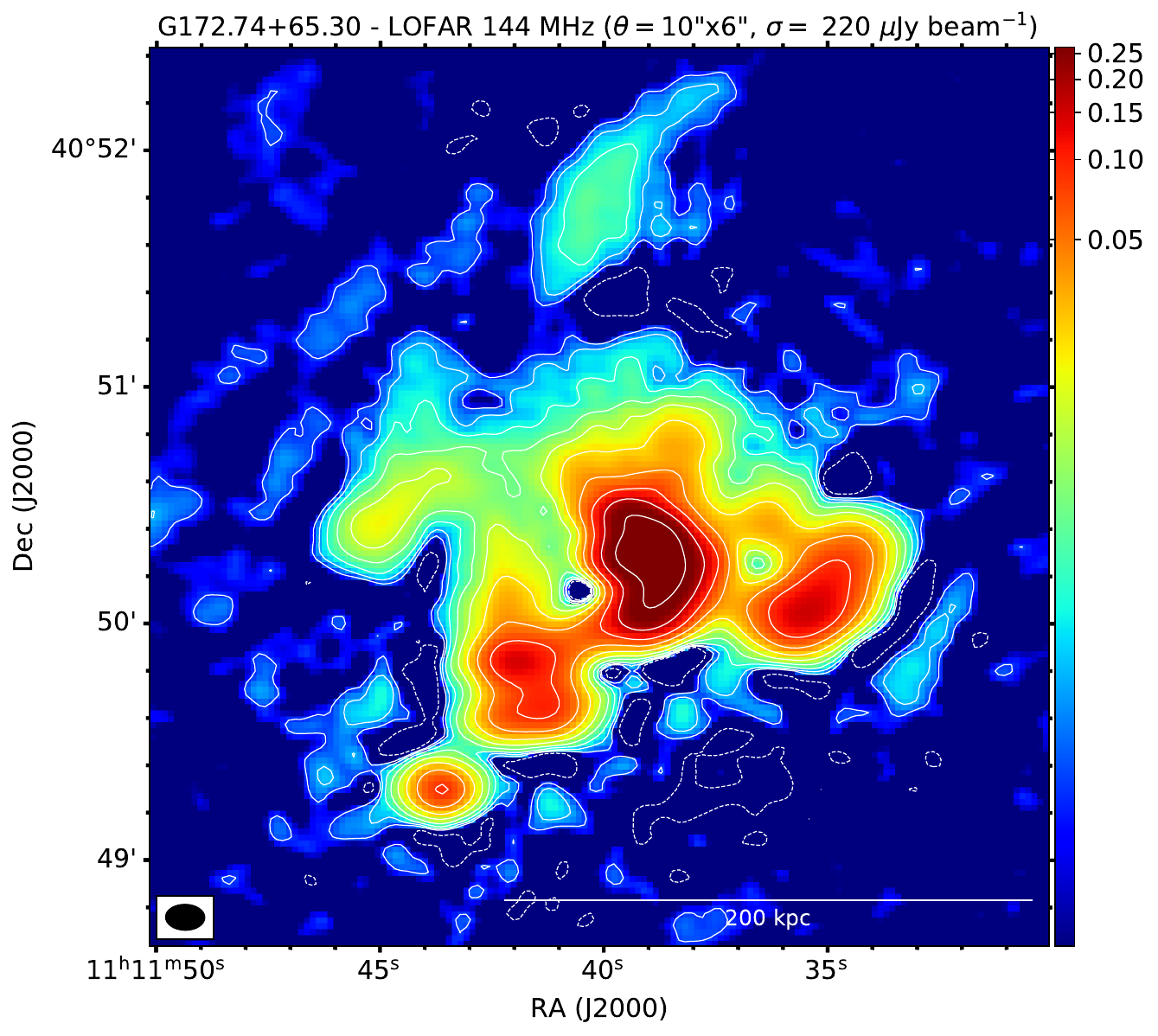}
\includegraphics[width=0.33\textwidth]{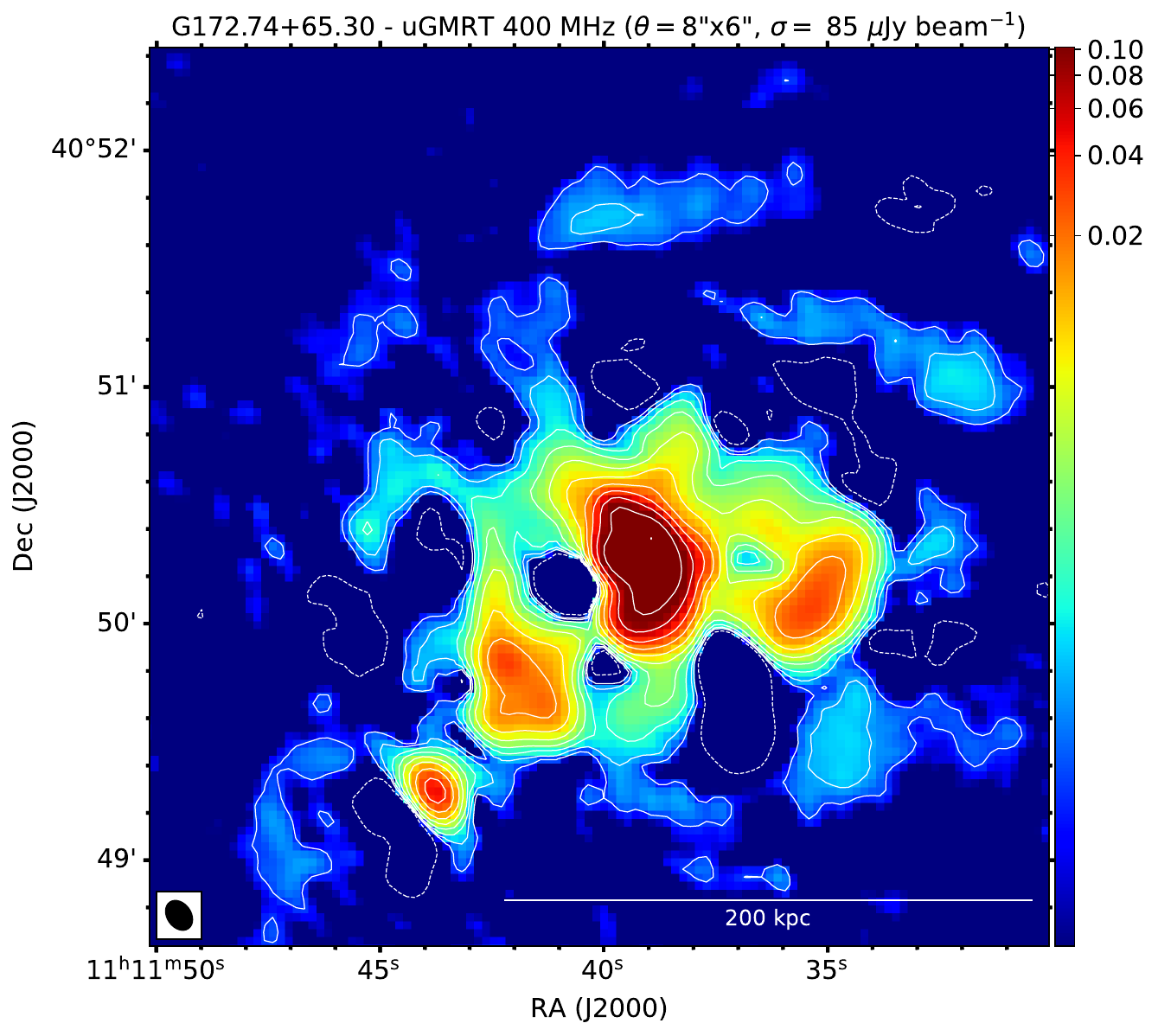}
\includegraphics[width=0.33\textwidth]{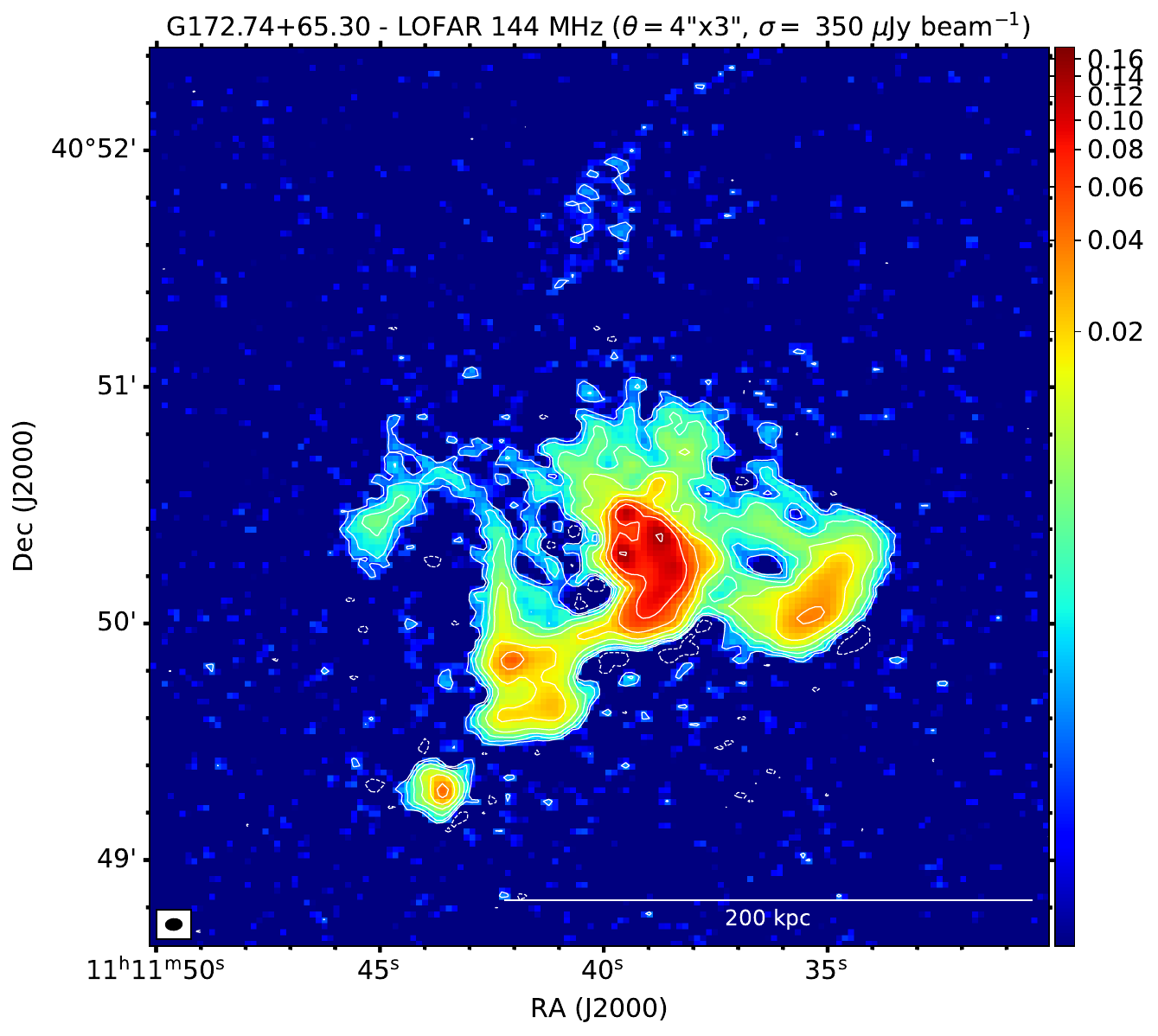}
\includegraphics[width=0.31\textwidth]{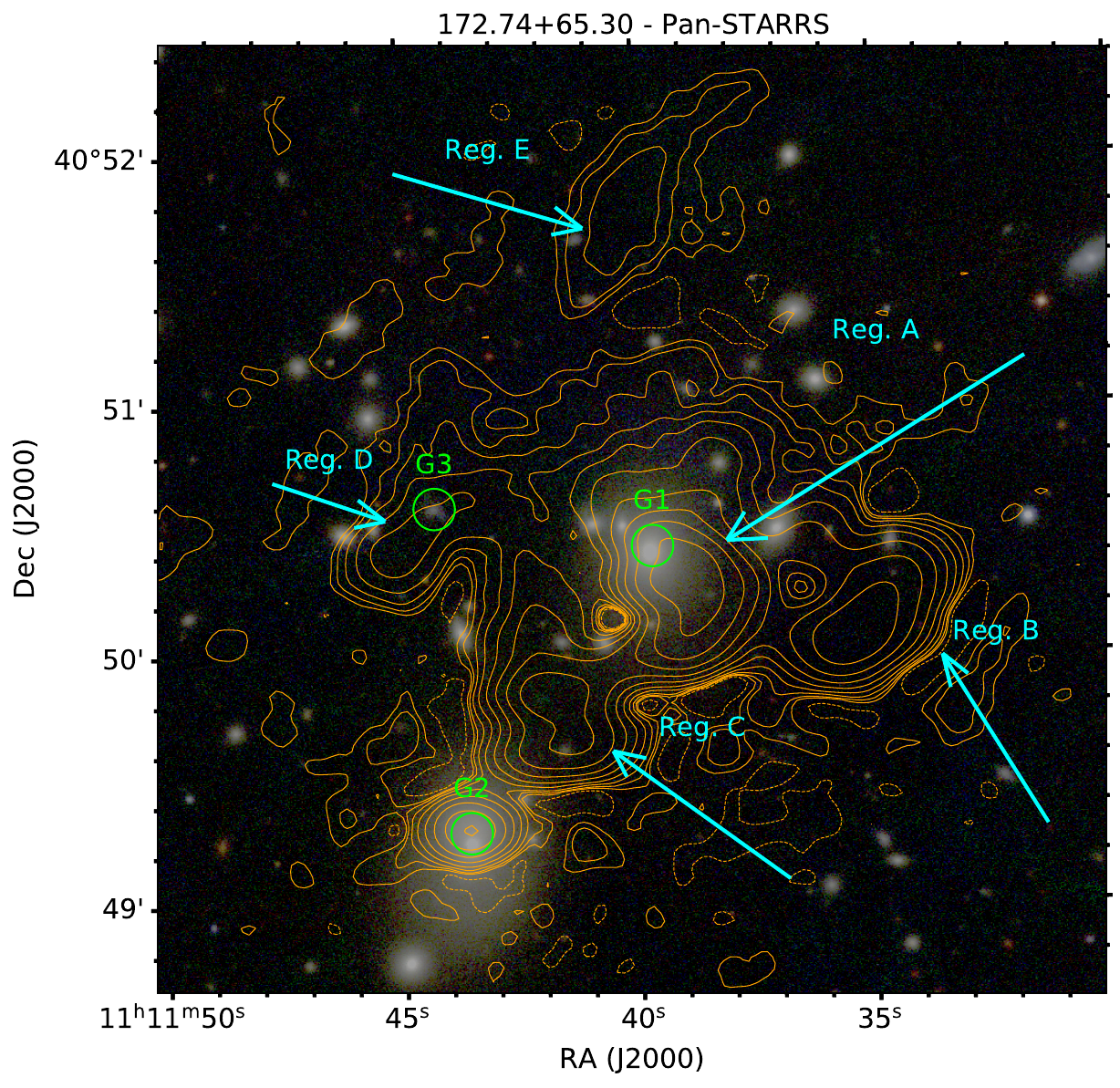}
\includegraphics[width=0.33\textwidth]{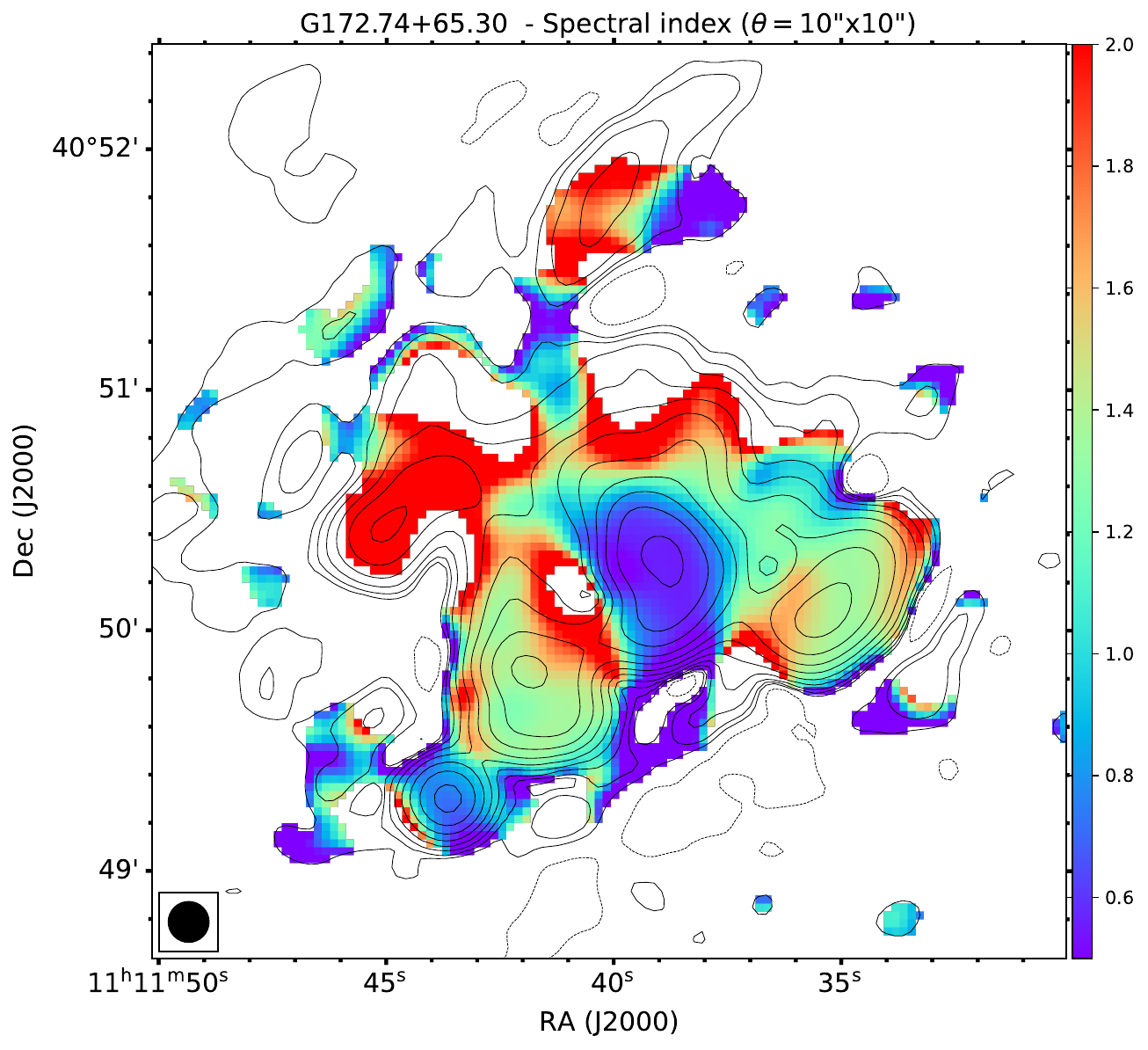}
\includegraphics[width=0.34\textwidth]{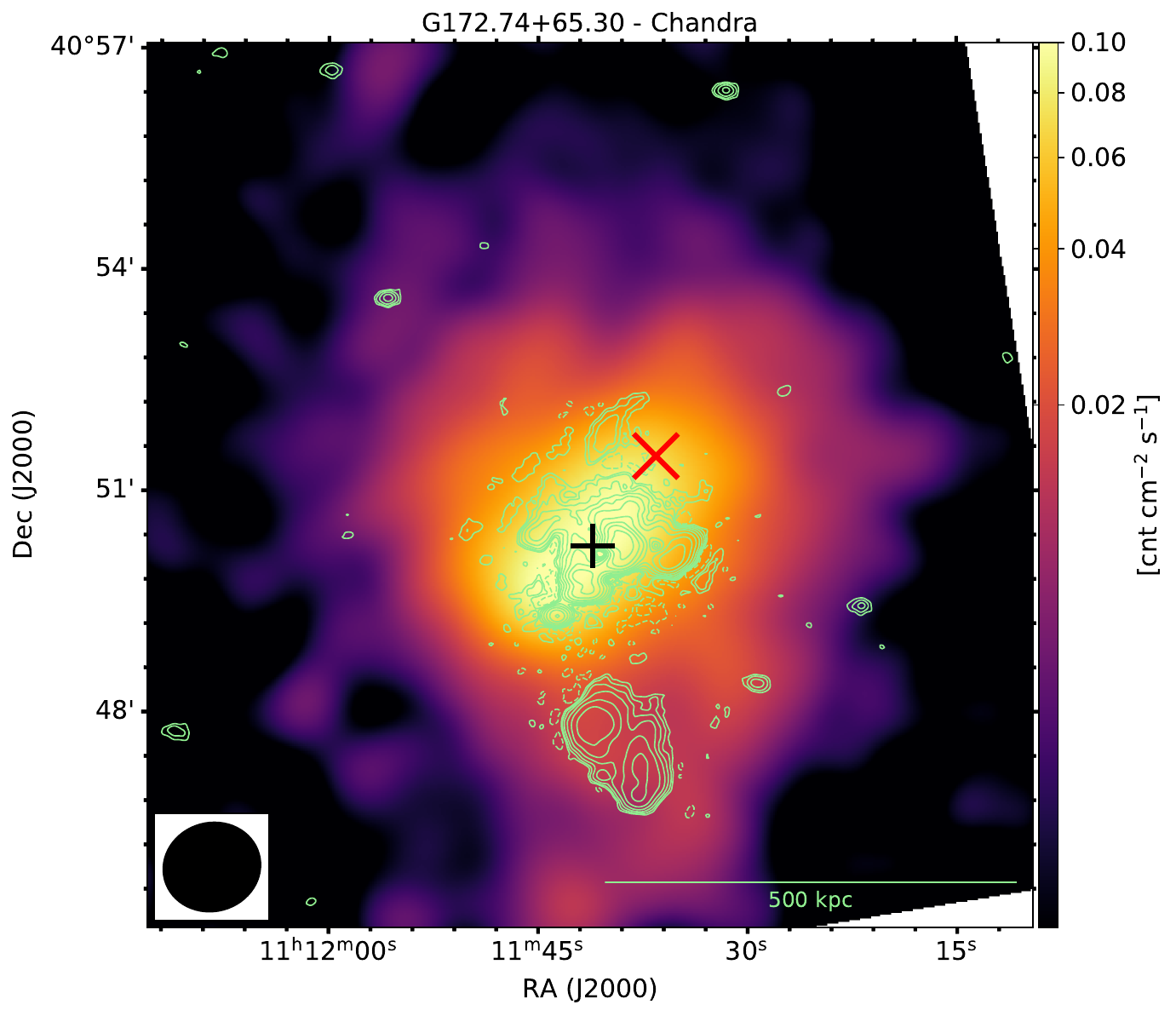}

        \caption{Images of G172.74+65.30. The description of each panel is the same as for Fig. \ref{fig: mappefullres1}. }
        \label{fig: mappefullres7}
\end{figure*}

The bright target at the centre of G172 (Fig. \ref{fig: mappefullres7}) has an impressively complex diffuse morphology with multiple brightness peaks and gaps, inner filaments, and arc-shaped regions. The limited dynamic range is responsible for artefacts around the source that are not easily distinguishable from real (low surface brightness) emission; for instance, region E could be an artefact.

The X-ray image reveals an elongated and disturbed system lacking an X-ray core. The merging nature of the cluster is confirmed by the presence of the two brightest galaxies G1 and G2, separated by a projected distance of $\sim 120$ kpc and both active in the radio band. The radio emission associated with G2 has a roundish morphology extending for $\sim 30$ kpc, whereas G1 is associated with the peak emission of region A. In the higher-resolution LOFAR image, the structure of region A is resolved, revealing that it is a NAT radio galaxy with distinguishable core and bent jets (see also the VLASS image in Fig. \ref{fig: vlass2} in Appendix \ref{sect: VLASS images}). The average  spectral indices of both emission associated with G1 and G2 are $\alpha \sim 0.7$, as typical of radio galaxies.

Our LOFAR and uGMRT images at $\sim 6''$ suggest a possible physical connection between the NAT (region A) and regions B and C via filamentary structures. The high-resolution LOFAR image supports this hypothesis. Indeed, the jets of the NAT appear to be twisted, with one of the two jets (`jet 1' in Fig. \ref{fig: vlass2}) first bent towards west and then towards east, in the direction of region C. In this scenario, region B could represent the western part of the tail, produced by the other jet (`jet 2', the one first bent towards east), which is partly hidden in projection due to the twisting motion. The consistent spectral indices $\alpha\sim 1.4$ of regions B and C could indicate that these have a common evolutionary history, but it is not trivial to account for their discrepant morphology observed at higher resolution. 

Region D, exhibiting an ultra-steep spectrum of $\alpha \sim 2.5$, appears to be connected to region C via a filament. A cluster member (G3) lies close the intersection of the two regions, but there is no  evidence for a clear association with the radio emission.

All regions discussed above are embedded in diffuse emission with a steep spectral index $\alpha\sim 1.3$. Such diffuse emission extends on scales of at least $\sim 200$ kpc, and it is interestingly more concentrated towards the northern regions of the target. Considering the merging environment and the spectral and morphological radio properties, it is possible that the diffuse emission is produced by particle re-acceleration from ICM turbulence, similarly to the formation of RHs (in particular, we notice morphological similarities with the candidate mini RH in A3582S reported by \citealt{digennaro25}). Moreover, the diffuse emission may consist of aged electrons spread by the NAT during its motion and reprocessed by the cluster weather.

Regardless of the nature of the diffuse emission, re-acceleration processes are plausibly ongoing. However, the dynamics of the NAT and projection effects hamper any interpretation. Except for the NAT emission, we keep the classification by \cite{botteon22} of the other regions of the target in G172  as uncertain.

\section{Discussion and conclusions}
\label{sect: Discussion}

\begin{table*}[!h]
 \fontsize{7.5}{7.5}\selectfont
\centering
	\caption[]{Properties and (tentative) classification of the analysed radio sources.  }
	\label{table: flux}   
	\begin{tabular}{ccccccccccc}
	\hline
	\noalign{\smallskip}
Host & Reg. & Class. & $D_{\rm c}$ & $LLS$ & $A$ & $S_{\rm 144}$ & $S_{\rm 400}$ & $\alpha$ & $P_{\rm 150}$ \\
&  & & (kpc) & (kpc) & ($10^3$ kpc$^2$) & (mJy) & (mJy) & & ($10^{24}\; {\rm W \; Hz^{-1}}$)  \\
\hline
\noalign{\smallskip}
 G071 & - & RP$^{(\rm c, *)}$ & 920$^{\rm +}$ &  580 & 78 & $197.8 \pm 19.8$ & $19.5 \pm 1.2$ & $2.3 \pm 0.1$ & $14.6\pm 1.5$ \\
G088 & A, B & HT$^{(\rm *)}$ & 290$^{\rm +}$ &  275 & 29 & $431.1 \pm 43.1$ & $74.7 \pm 4.5$ & $1.7 \pm 0.1$ & $20.6\pm 2.1$ \\
G088 & C, D & GReET$^{(\rm c, *)}$ & 155$^{\rm +}$ &  315 & 28 & $123.1 \pm 12.3$ & $6.1 \pm 0.4$ & $2.9 \pm 0.1$ & $6.5\pm 0.6$ \\
G088 & E & Uncertain & 390$^{\rm +}$ &  215 & 21 & $432.1 \pm 43.2$ & $121.6 \pm 7.3$ & $1.2 \pm 0.1$ & $19.8\pm 2.0$ \\
G088 & S2 & Uncertain & 535$^{\rm +}$ &  210 & 14 & $24.6 \pm 2.5$ & $2.4 \pm 0.2$ & $2.3 \pm 0.1$ & $1.3\pm 0.2$ \\
G113 & - & Remnant  & 580$^{\rm +}$ &  $210$ & 21 & $48.4 \pm 5.0$ & $13.1 \pm 0.8$ & $1.3 \pm 0.1$ & $1.4\pm 0.1$ \\
G137 & A & Uncertain$^{(\rm *)}$ & 240$^{\rm +}$ &  130 & 6 & $342.0 \pm 34.2$ & $86.0 \pm 5.2$ & $1.4 \pm 0.1$ & $6.3\pm 0.6$ \\
G137 & B, Fil. & WAT$^{(\rm *)}$ & 300$^{\rm +}$ &  280 & 17 & $403.7 \pm 40.4$ & $99.7 \pm 6.0$ & $1.4 \pm 0.1$ & $7.4\pm 0.7$ \\
G155 & A & HT & 435$^{\rm x}$ &  285 & 41 & $581.5 \pm 58.2$ & $254.0 \pm 15.2 $ & $0.8 \pm 0.1$ & $195.2\pm 20.3$ \\
G155 & B & Remnant & 520$^{\rm x}$ &  430 & 53 & $85.8 \pm 8.6$ & $5.6 \pm 0.4$ & $2.7 \pm 0.1$ & $46.0\pm 4.8$ \\
G165 & A & Remnant & 750$^{\rm x}$ &  240 & 23 & $56.7 \pm 5.7$ & $12.1 \pm 0.7$ & $1.5 \pm 0.1$ & $7.6\pm 0.8$ \\
G165 & B & Remnant & 720$^{\rm x}$ &  260 & 25 & $61.6 \pm 6.2$ & $9.5 \pm 0.6$ & $1.8 \pm 0.1$ & $8.6\pm 0.8$ \\
G172 & A & NAT & 40$^{\rm +}$ &  90 & 7 & $4104.4 \pm 410.4$ & $2020.9 \pm 121.3$ & $0.7 \pm 0.1$ & $60.0\pm 6.0$ \\
G172 & B, C, D & Uncertain & 40$^{\rm +}$ &  220 & 21 & $2116.1 \pm 211.6$ & $460.1 \pm 27.6$ & $1.5 \pm 0.1$ & $32.8\pm 3.2$ \\
\noalign{\smallskip}
	\hline
	\end{tabular}  
	\begin{tablenotes}
\item    {\small \textbf{Notes}. Cols. 1-2: host cluster and considered region of the radio source. Col. 3: (tentative) classification; `c' stands for `candidate' and `*' indicates evidence of re-energising based on radio data only (not considering X-rays). Cols. 4-6: projected distance of target from the \textit{Planck} centre ($^{\rm x}$) or X-ray peak ($^{\rm +}$), largest linear size, and area. Cols. 7-10: flux densities measured within regions encompassing the $3\sigma$ level of the 144 MHz image, integrated spectral index, and $k$-corrected radio power at 150 MHz.} 
 \end{tablenotes}	
\end{table*}

In this work, we reported on the spectral study at $144-400$ MHz of a sample of 7 candidate revived fossil radio sources. These have been selected by visual inspection of amorphous and filamentary sources detected among nearby and low-mass galaxy clusters in LoTSS-DR2 (Sect. \ref{sect: Sample selection}, Fig. \ref{fig: collection}) and then followed-up with the uGMRT. The results reported in Sect. \ref{sect: Morphological and spectral properties} demonstrated that our morphology-based selection is effective to identify steep-spectrum sources, as all of the targets exhibit regions having spectral indices $\alpha\gg 1$. We showed that sensitive high-resolution images are essential to avoid misclassification. Our analysis highlights that there is not a common dominant physical process underlying all the presented sources. However, interpreting their nature and reconstructing their history remain challenging in many cases. In this section, we discuss our findings and future prospects. The (tentative) classification based on morphology, surface brightness profiles, spectral properties, and optical associations, and measured quantities  (distance from the cluster centre, size, flux densities, integrated spectral index, and radio power) are reported in Table \ref{table: flux}.

The fossil radio source in G071 (Sect. \ref{sect: G071}) shows signs of re-acceleration along the filaments and we classify it as a radio phoenix. Although the radio source in G113 (Sect. \ref{sect: G113}) shares properties with the RP in G071, we have no evidence of possible re-acceleration, and we thus consider it as a remnant. In G088 (Sect. \ref{sect: G088}) we identified a HT radio galaxy terminating with a Gently Re-Energised Tail and two unrelated sources of unclear nature. The radio emission in G137 (Sect. \ref{sect: G137}) reasonably consists of two distinct sources, namely a WAT radio galaxy ending with a long filament and a radio source of uncertain nature, both of which showing evidence of re-acceleration. In G155 (Sect. \ref{sect: G155}) we identified a HT radio galaxy and an unrelated remnant component. In G165 (Sect. \ref{sect: G165}) we detected either a single or two distinct fossil components. In G172 (Sect. \ref{sect: G172}) we revealed the presence of a central NAT radio galaxy embedded in diffuse emission of unclear origin, which is likely the result of re-acceleration processes. The targets in G137, G155, and G172 are clear examples presenting complex regions that have been revealed to be components of tailed galaxies, which could be easily misclassified or over-interpreted in the absence of sensitive, high-resolution data. The confirmation of ongoing re-energising processes and definitive classification should be supported by a detailed study of the local conditions of the ICM with X-ray data. Specifically, detecting discontinuities in the thermodynamic properties of the ICM co-spatial with the radio emission would be solid evidence of the thermal and non-thermal interplay. Furthermore, the various  scenarios that we proposed should be investigated with additional radio observations; our campaign at 400 MHz provided vital flux density and spectral index measurements (Table \ref{table: flux}) to plan follow-up observations.

Among the original goals of our work, we aimed to provide a first look on the occurrence of revived sources in galaxy clusters and search for possible correlations similar to those found for radio halos and relics that link the radio emission with the host cluster properties \citep[e.g.][]{cassano10A,cuciti23,jones23}. In the light of our findings and necessity of deeper investigation for a genuine classification, no quantitative conclusions can be drawn. Qualitatively, our analysis is in line with results from the literature showing that revived sources can inhabit low-mass systems with hints of large-scale disturbance \citep[e.g.][]{mandal20,edler22}; specifically, all our targets with available X-ray data are found in (mildly or highly) disturbed systems, as confirmed by the morphological parameter distribution shown in Fig. \ref{fig: cw} in Appendix \ref{sect: Xray data for G088}. Moreover, our images show that fossil electrons can be spread across a large fraction of the cluster volume, thus serving as reservoirs of seed cosmic rays in the ICM, in agreement with predictions from simulations \citep[e.g.][]{zuhone21,vazza&botteon24}.

Revived fossils are also particularly interesting in the framework of physics of radio filaments, a topic that is becoming increasingly appealing with the advent of the current generation of interferometers \citep[e.g.][]{ramatsoku20,condon21,yusef-zadeh22,rudnick22,candini23,brienza25,churazov25,derubeis25,giacintucci25}. Whether such filaments originate from radio galaxies or trace local conditions of the ICM magnetic fields is unclear. It is also not understood if all filaments are produced and powered by the same mechanisms. In this context, we found different behaviours for filaments in our radio sources. For G071, the surface brightness increases and decreases along different parts of filaments A and B, and the ultra-steep spectral index ($\alpha \sim 2.3$) of filament A is constant across $\sim 100$ kpc (Fig. \ref{fig: profiles G071}). Along the filament in G137, the surface brightness oscillates around an approximately constant value across $\sim 130$ kpc, but the spectral index gradually steepens (Fig. \ref{fig: profiles G137}). Intricate intersected filaments are found for G113 and G172 (Figs. \ref{fig: mappefullres2}, \ref{fig: mappefullres7}).

In conclusion, our work needs to be complemented by further multi-wavelength analysis to disentangle the mechanisms responsible for the observed radio emission. Nevertheless, it represents a step towards developing observing strategies and methodologies for identifying and characterising the diversity of fossil and revived fossil sources in radio surveys, such as carrying out explorative campaigns to constrain the spatial spectral distribution serving for further follow-ups, and pushing for the resolution of the instruments. While our sample is limited to nearby low-mass clusters, \cite{botteon22} reported possible revived fossils in wider redshift and mass ranges, suggesting that these apparently rare sources may be more common than generally thought. In this context, the upcoming release of LoTSS-DR3 (Shimwell et al., in prep.) will enable the search for potential revived sources across the entire northern sky. Their search can be automatised through machine learning based algorithms (e.g. \citealt{Mostert21,Gupta22,stuardi24}), as we have validated our visual-based selection method. Moreover, the (sub)-arcsecond resolution provided by the international stations of LOFAR-VLBI will offer new insights into the intricate structures of these sources. Finally, the Square Kilometre Array and its precursors will still play a key role, thanks to their combination of high resolution and sensitivity . We therefore anticipate significant progress in the near future in the study of these elusive and complex targets, which have so far been largely overlooked.

\begin{acknowledgements}
We thank the referee for their constructive comments on the manuscript. MarBr\"u acknowledges support from the Deutsche Forschungsgemeinschaft under Germany's Excellence Strategy - EXC 2121 “Quantum Universe” - 390833306. EDR is supported by the Fondazione ICSC, Spoke 3 Astrophysics and Cosmos Observations. National Recovery and Resilience Plan (Piano Nazionale di Ripresa e Resilienza, PNRR) Project ID CN\_00000013 "Italian Research Center for High-Performance Computing, Big Data and Quantum Computing" funded by MUR Missione 4 Componente 2 Investimento 1.4: Potenziamento strutture di ricerca e creazione di "campioni nazionali di R\&S (M4C2-19)" - Next Generation EU (NGEU). FdG acknowledges support from the ERC Consolidator Grant ULU 101086378. AI acknowledges funding from the European Research Council (ERC) under the European Union's Horizon 2020 research and innovation programme (grant agreement No. 833824).

LOFAR \citep{vanhaarlem13} is the Low Frequency Array designed and constructed by ASTRON. It has observing, data processing, and data storage facilities in several countries, which are owned by various parties (each with their own funding sources), and that are collectively operated by the ILT foundation under a joint scientific policy. The ILT resources have benefited from the following recent major funding sources: CNRS-INSU, Observatoire de Paris and Universit\'e d’Orl\'eans, France; BMBF, MIWF- NRW, MPG, Germany; Science Foundation Ireland (SFI), Department of Business, Enterprise and Innovation (DBEI), Ireland; NWO, The Netherlands; The Science and Technology Facilities Council, UK; Ministry of Science and Higher Education, Poland; The Istituto Nazionale di Astrofisica (INAF), Italy. This research made use of the Dutch national e-infrastructure with support of the SURF Cooperative (e-infra 180169) and the LOFAR e-infra group. The J\"ulich LOFAR Long Term Archive and the German LOFAR network are both coordinated and operated by the J\"ulich Supercomputing Centre (JSC), and computing resources on the supercomputer JUWELS at JSC were provided by the Gauss Centre for Supercomputing e.V. (grant CHTB00) through the John von Neumann Institute for Computing (NIC). This research made use of the University of Hertfordshire high-performance computing facility and the LOFAR-UK computing facility located at the University of Hertfordshire and supported by STFC [ST/P000096/1], and of the Italian LOFAR-IT computing infrastructure supported and operated by INAF, including the resources within the PLEIADI special `LOFAR' project by USC-C of INAF, and by the Physics Department of Turin University (under the agreement with Consorzio Interuniversitario per la Fisica Spaziale) at the C3S Supercomputing Centre, Italy.  We thank the staff of the GMRT that made these observations possible. GMRT is run by the National Centre for Radio Astrophysics of the Tata Institute of Fundamental Research. The National Radio Astronomy Observatory is a facility of the National Science Foundation operated under cooperative agreement by Associated Universities, Inc. The Pan-STARRS1 Surveys (PS1) and the PS1 public science archive have been made possible through contributions by the Institute for Astronomy, the University of Hawaii, the Pan-STARRS Project Office, the Max-Planck Society and its participating institutes, the Max Planck Institute for Astronomy, Heidelberg and the Max Planck Institute for Extraterrestrial Physics, Garching, The Johns Hopkins University, Durham University, the University of Edinburgh, the Queen's University Belfast, the Harvard-Smithsonian Center for Astrophysics, the Las Cumbres Observatory Global Telescope Network Incorporated, the National Central University of Taiwan, the Space Telescope Science Institute, the National Aeronautics and Space Administration under Grant No. NNX08AR22G issued through the Planetary Science Division of the NASA Science Mission Directorate, the National Science Foundation Grant No. AST–1238877, the University of Maryland, Eotvos Lorand University (ELTE), the Los Alamos National Laboratory, and the Gordon and Betty Moore Foundation. This work is based on observations obtained with XMM-Newton, an ESA science mission with instruments and contributions directly funded by ESA Member States and NASA. This research has made use of data obtained from the \textit{Chandra} Data Archive and
software provided by the Chandra X-ray Center (CXC) in the application package CIAO. This research has made use of SAOImageDS9, developed by Smithsonian Astrophysical Observatory \citep{ds9}. This research has made use of the VizieR catalogue access tool, CDS, Strasbourg Astronomical Observatory, France (DOI: 10.26093/cds/vizier). This research made use of APLpy, an open-source plotting package for Python \citep{robitaille&bressert12APLPY}, Astropy, a community-developed core Python package for Astronomy \citep{astropycollaboration13,astropycollaboration18}, Matplotlib \citep{hunter07MATPLOTLIB}, Numpy \citep{harris20NUMPY}, SciPy \citep{scipy}.

\end{acknowledgements}

\bibliographystyle{aa}
\bibliography{bibliografia}

\begin{appendix}
\onecolumn
\FloatBarrier

\section{Spectral index error maps}
\label{sect: errspixmap}

\begin{figure*}[!h]
        \centering

\includegraphics[width=0.33\textwidth]{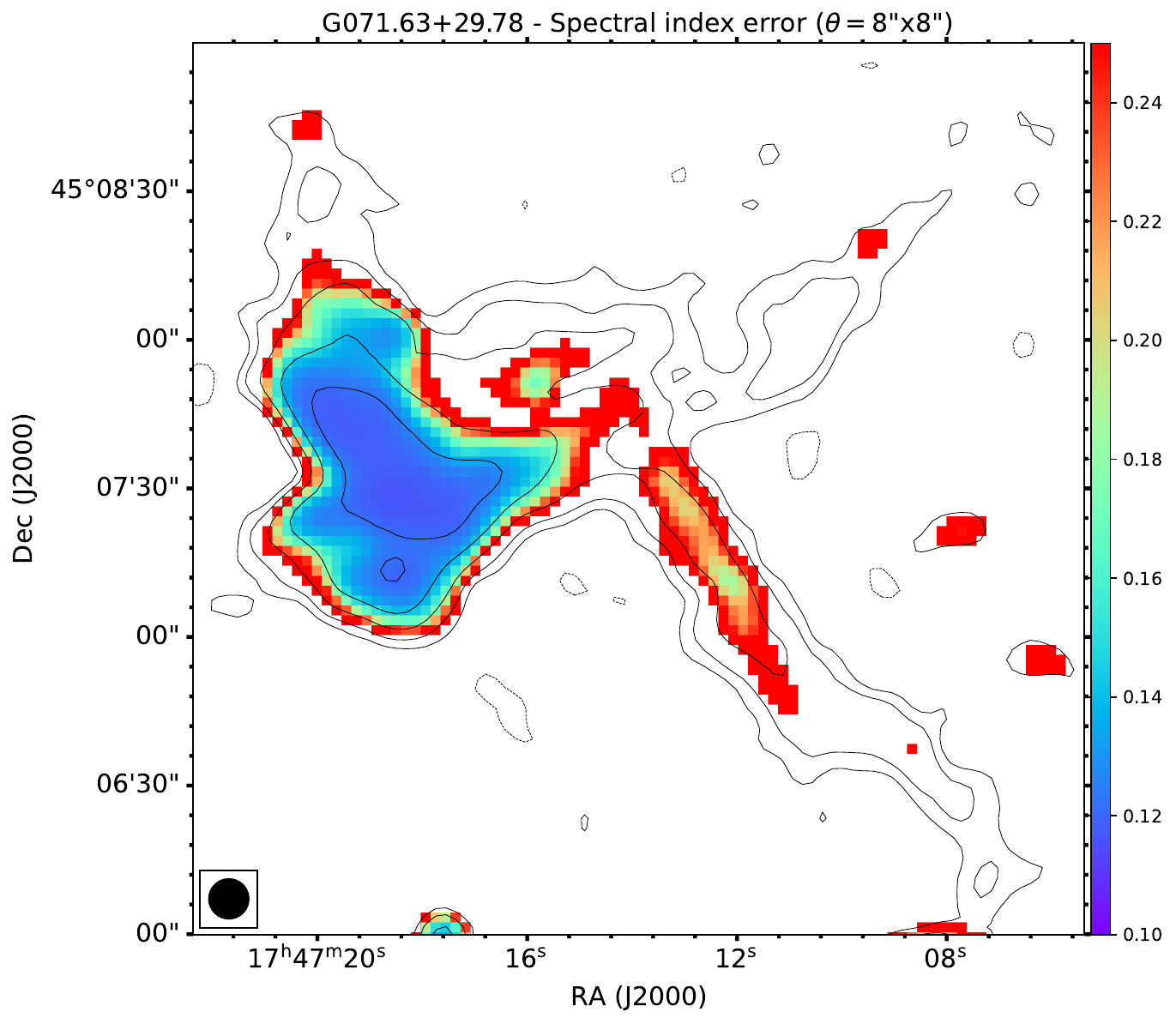}
\includegraphics[width=0.33\textwidth]{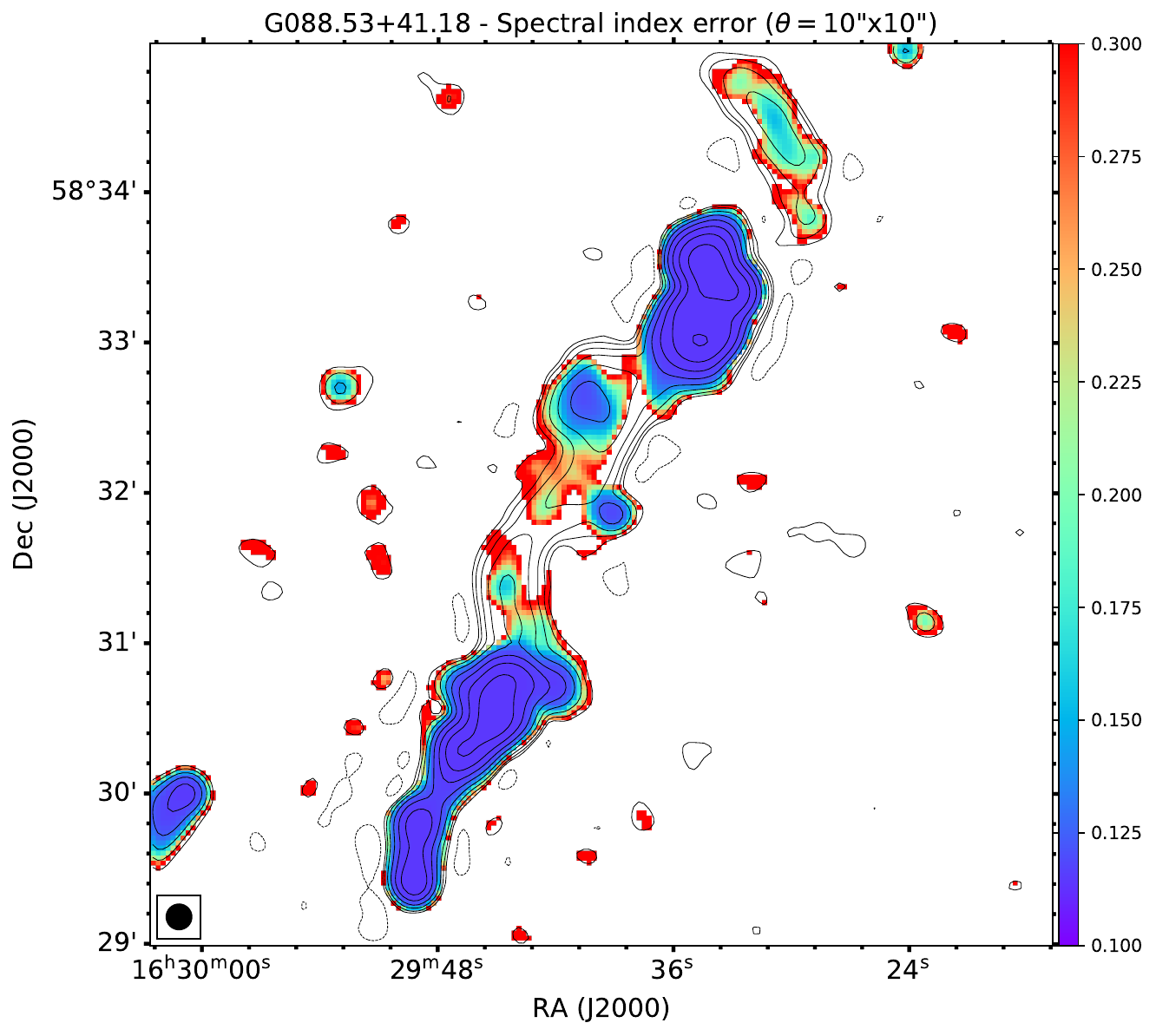}
\includegraphics[width=0.33\textwidth]{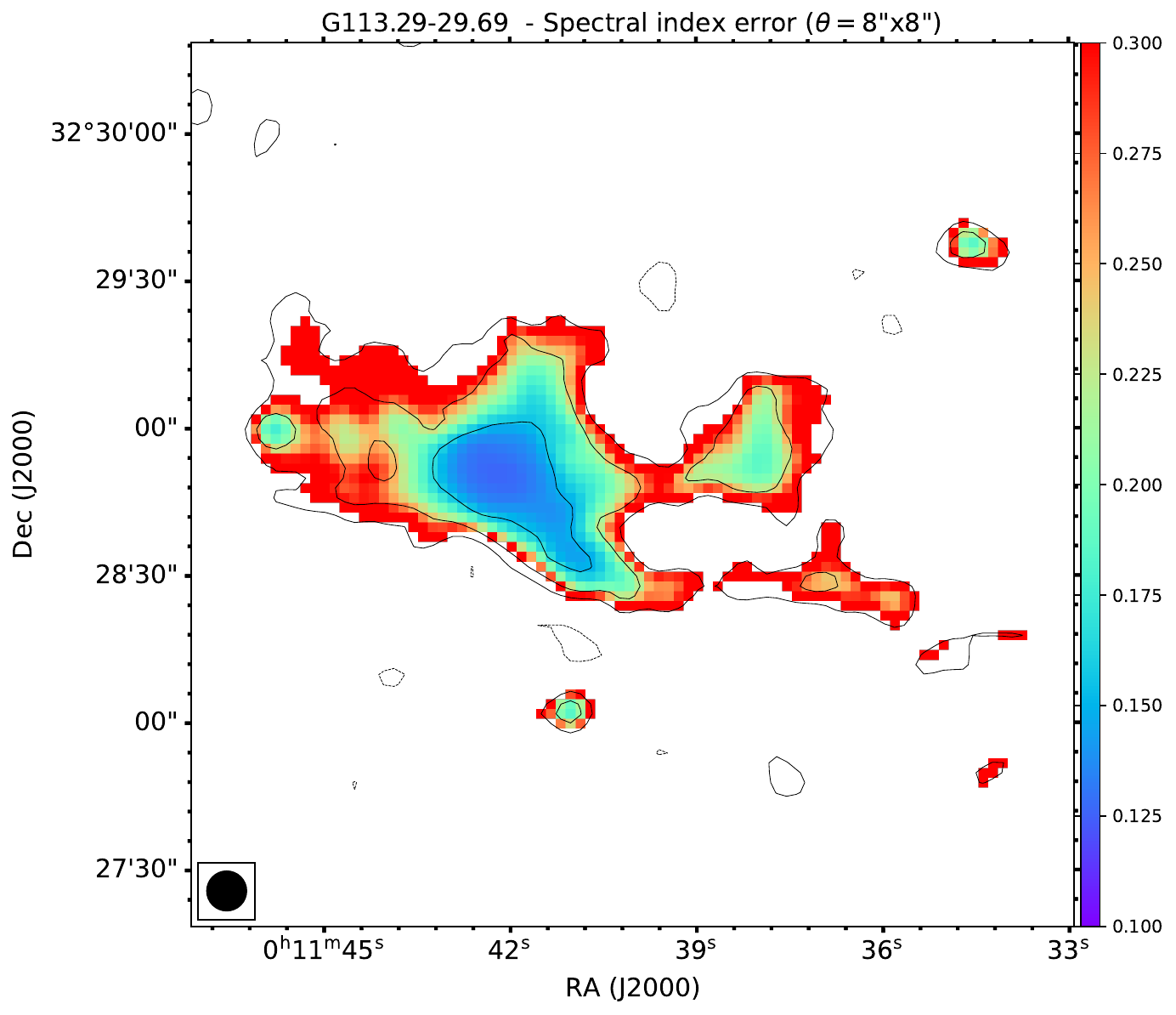}
\includegraphics[width=0.33\textwidth]{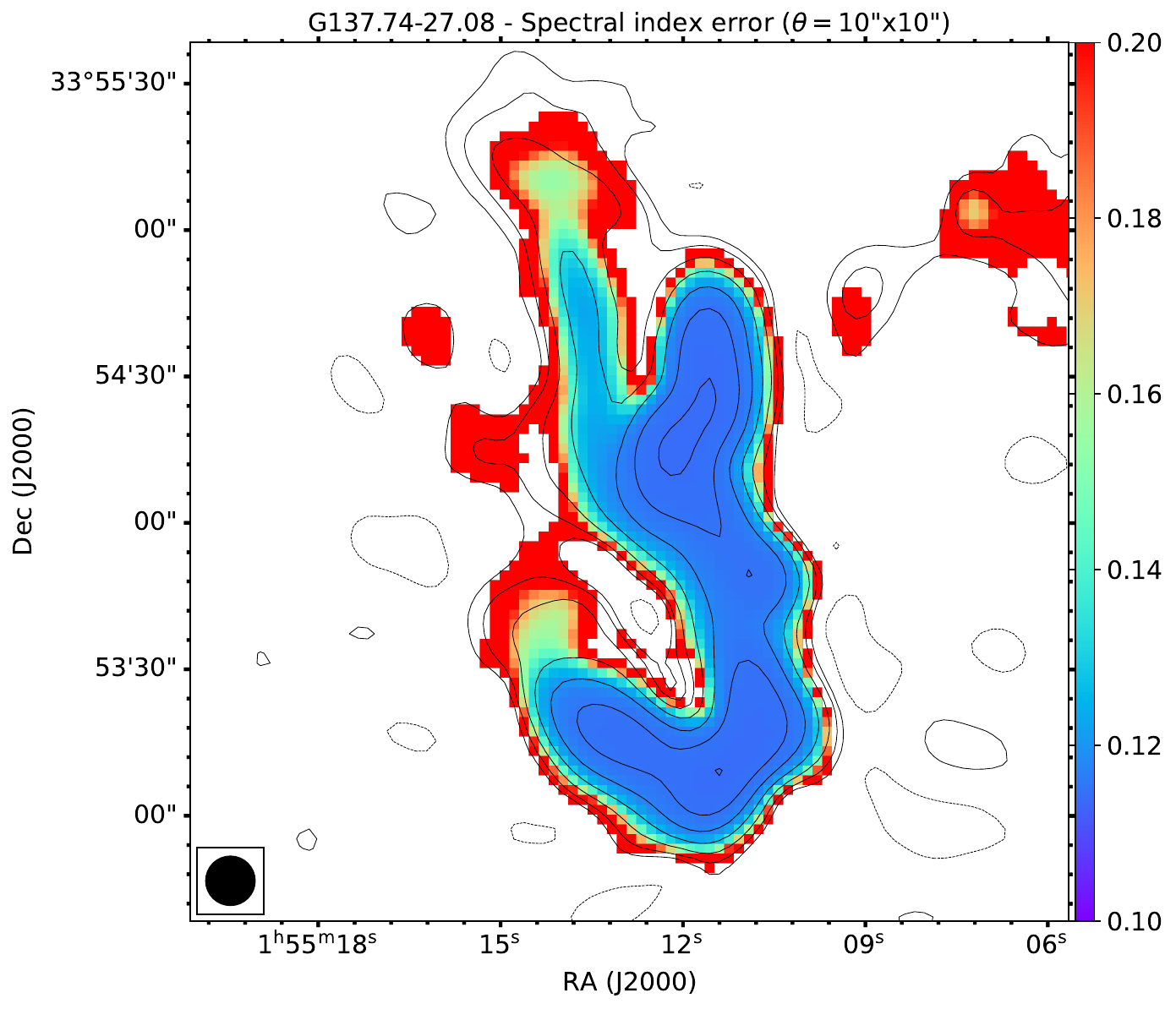}
\includegraphics[width=0.33\textwidth]{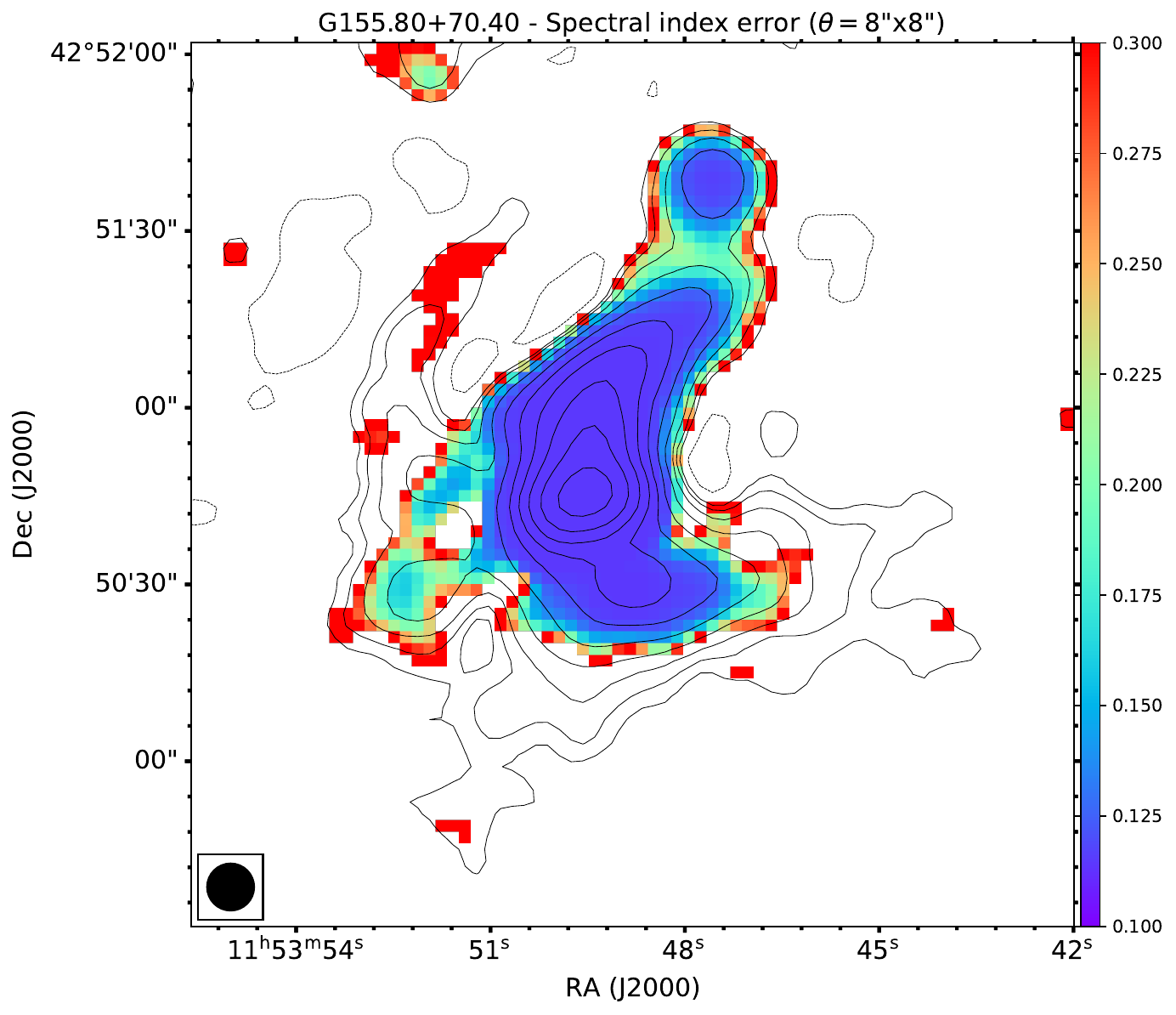}
\includegraphics[width=0.33\textwidth]{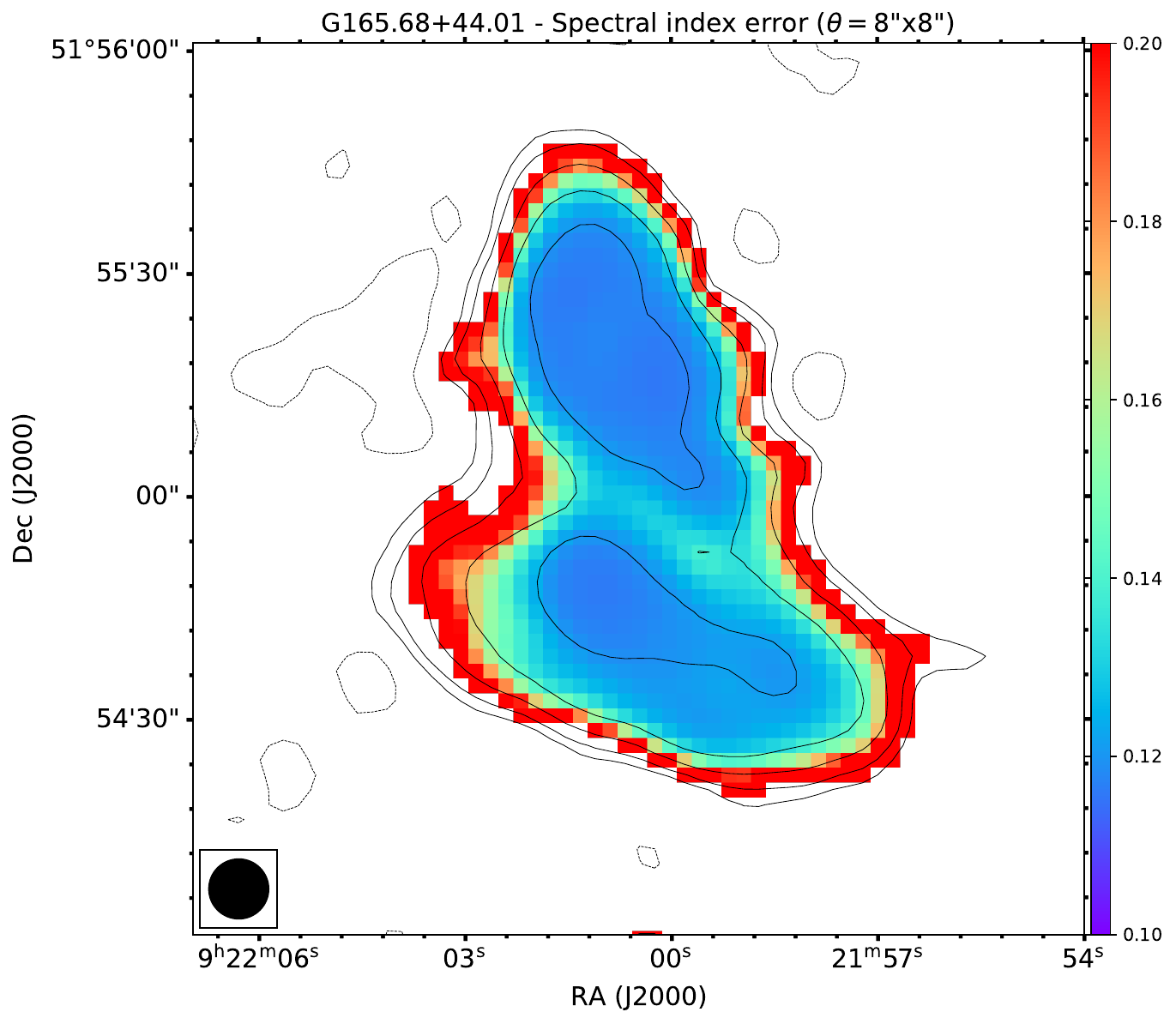}
\includegraphics[width=0.33\textwidth]{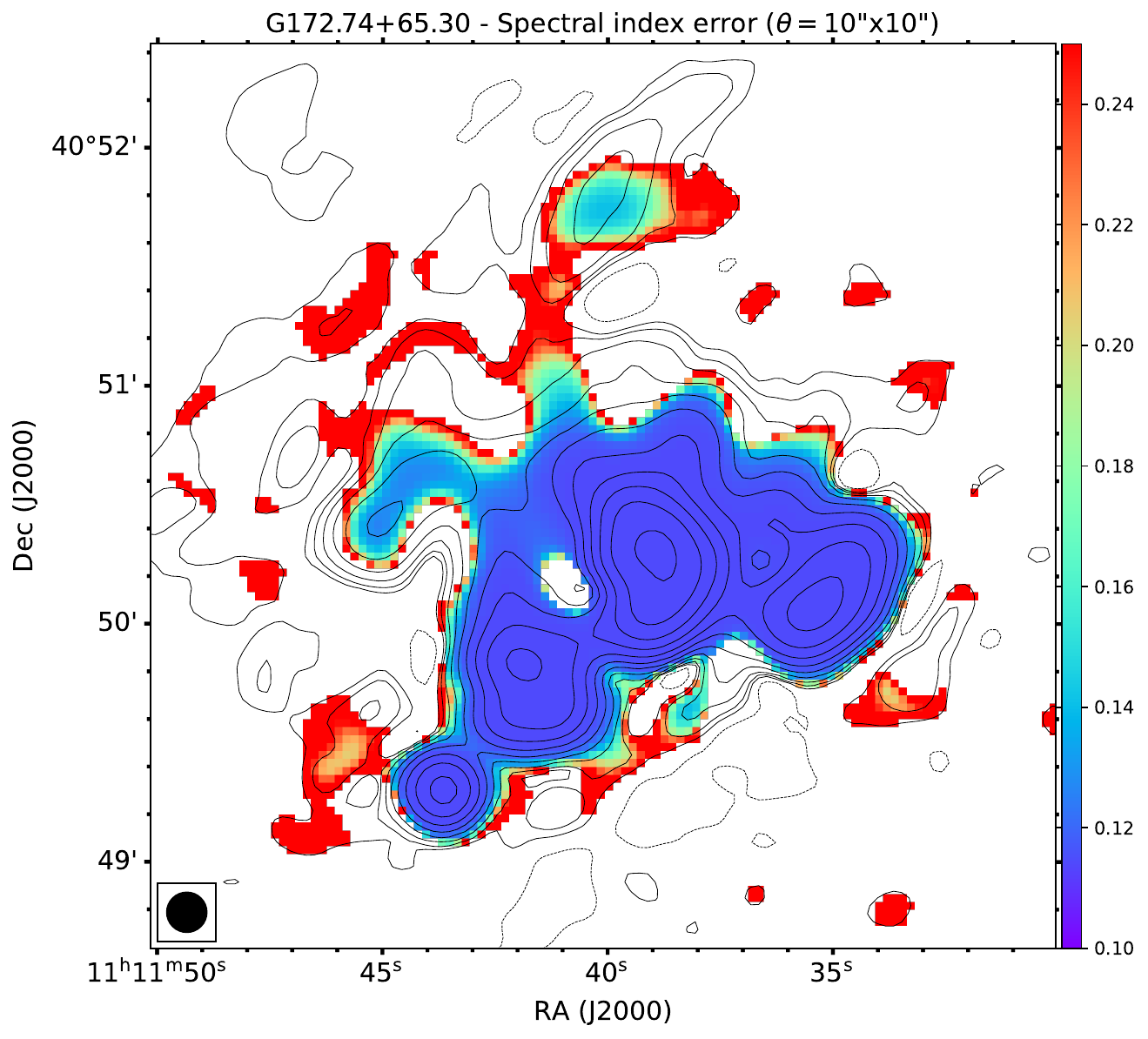}
        \caption{Spectral index error maps corresponding to the maps shown in Figs. \ref{fig: mappefullres1}-\ref{fig: mappefullres7}.}
        \label{fig: errspixmap}
\end{figure*}

In Fig. \ref{fig: errspixmap}, we report the error maps associated with the spectral index maps shown in Figs. \ref{fig: mappefullres1}-\ref{fig: mappefullres7}.

\FloatBarrier

\section{Surface brightness and spectral index profiles}
\label{sect: Surface brightness and spectral index profiles}

\begin{figure*}[!h]
        \centering

\includegraphics[width=0.288\textwidth]{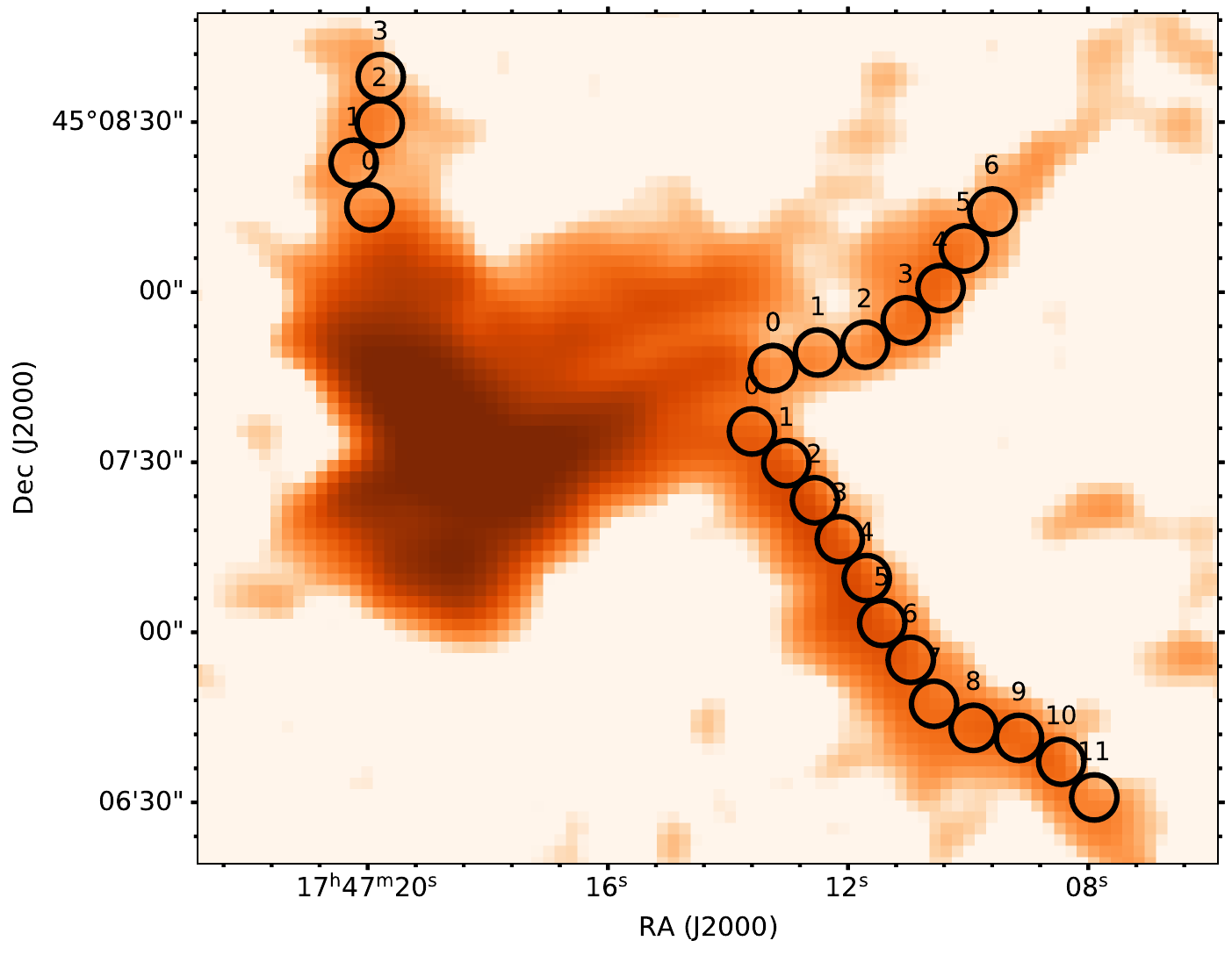}
\includegraphics[width=0.342\textwidth]{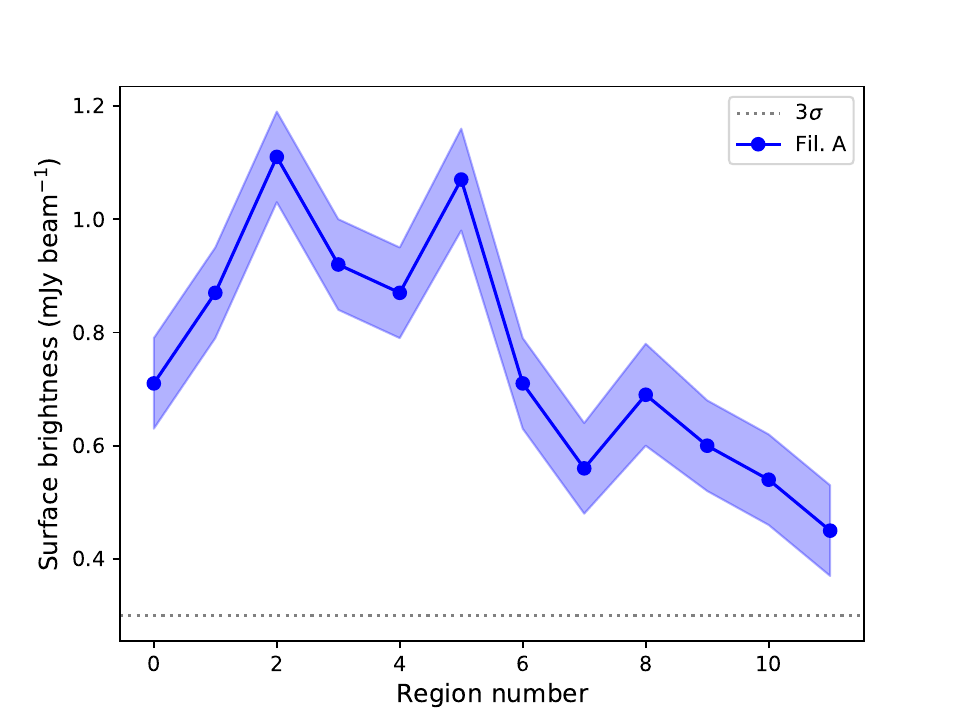}
\includegraphics[width=0.342\textwidth]{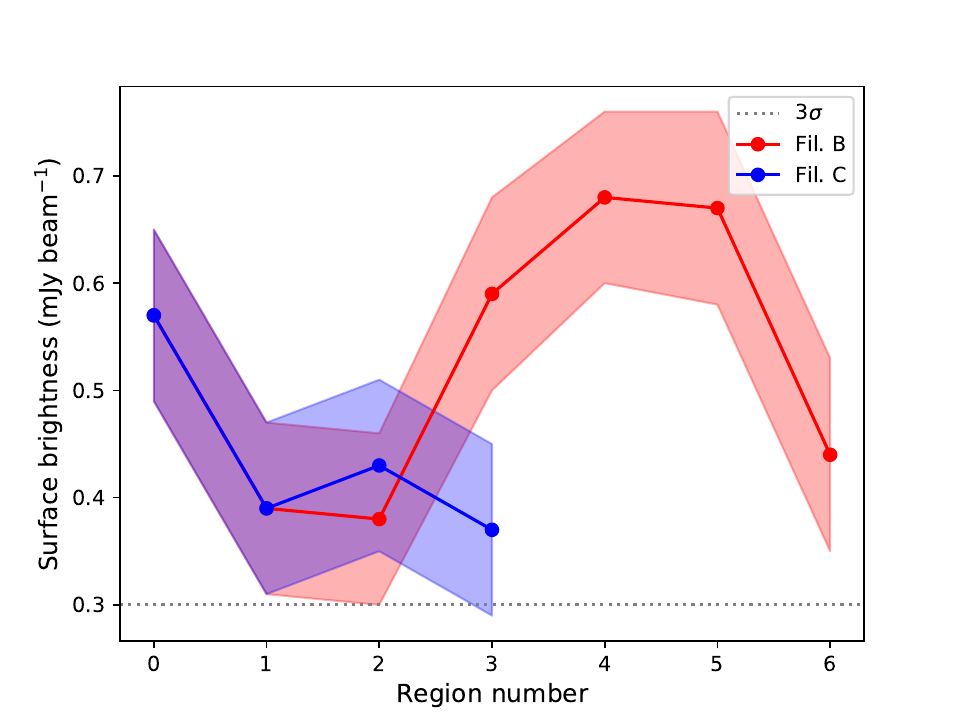}
\includegraphics[width=0.32\textwidth]{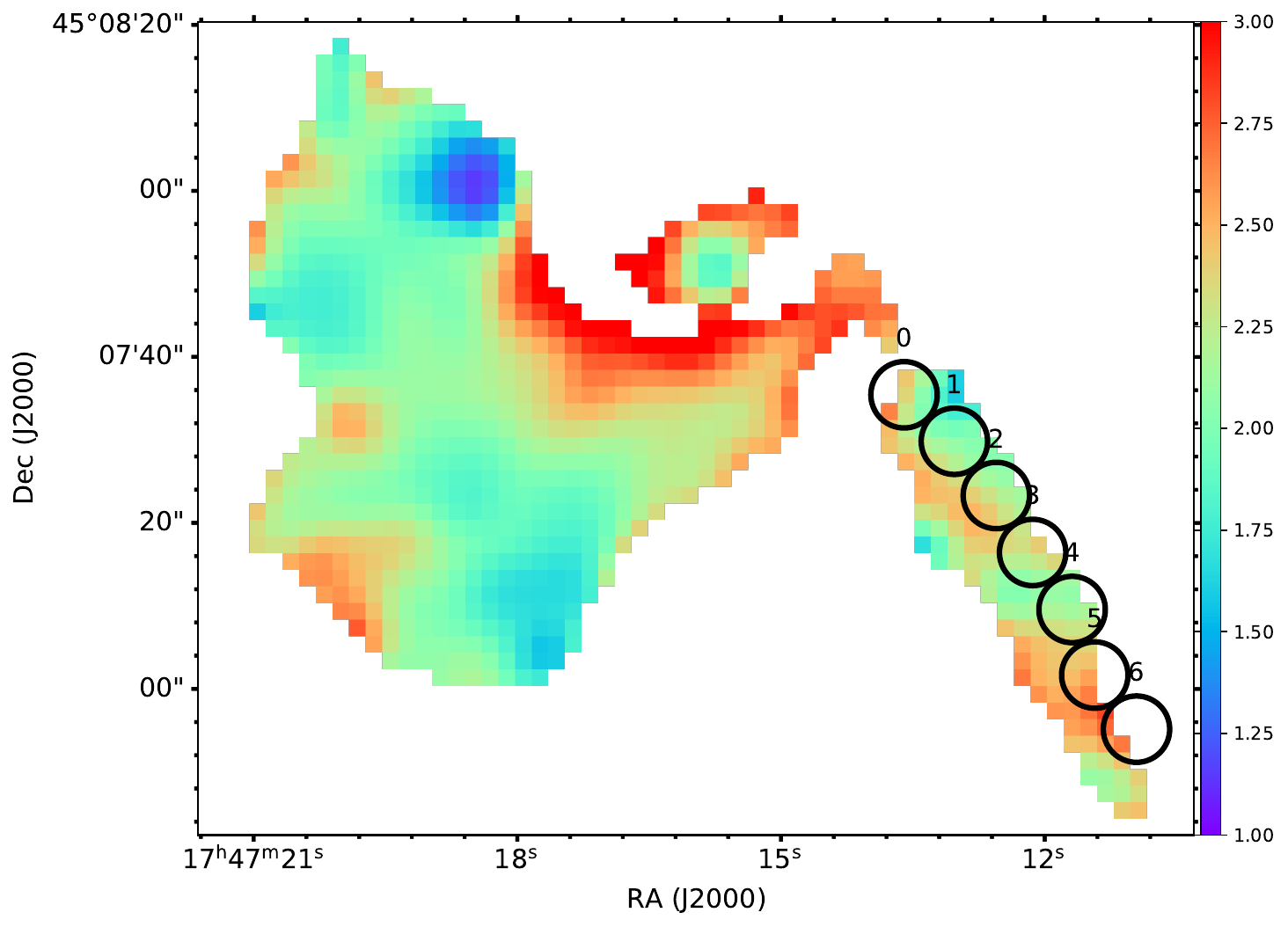}
\includegraphics[width=0.34\textwidth]{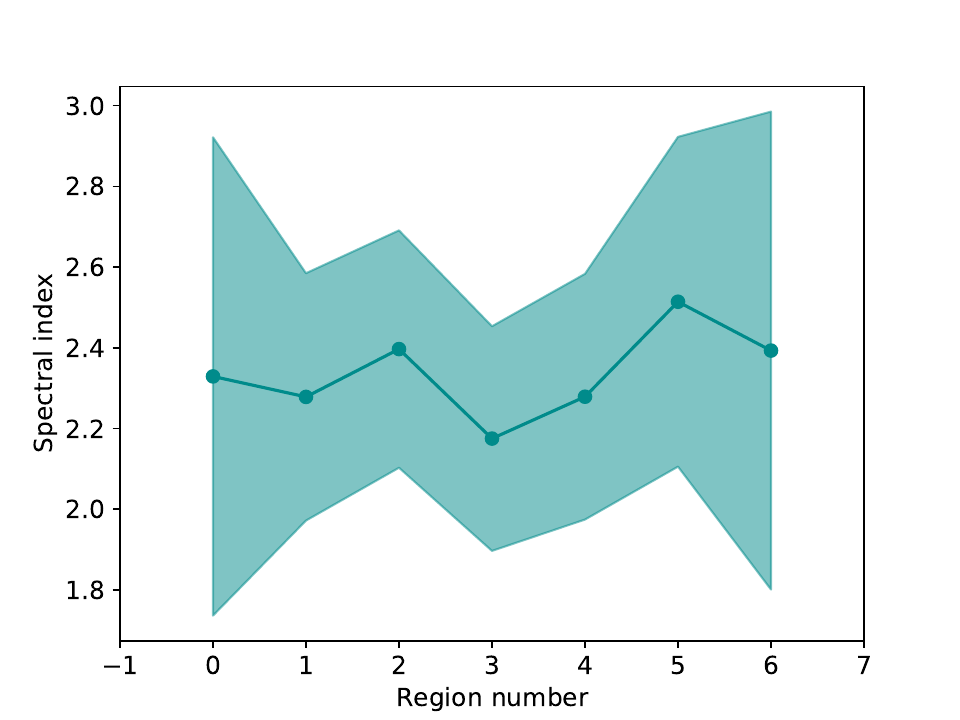}

        \caption{Profiles of the target in G071. \textit{Top left}: 144 MHz image at resolution $\theta=8''\times8''$ sampled with beam-size circles ($8''=22$ kpc at $z=0.157$). \textit{Top centre}: surface brightness profile along filament A. \textit{Top right}: surface brightness profile along filaments B and C. In top centre and right panels, the horizontal dotted line indicates the $3\sigma$ level of the image. \textit{Bottom left}: spectral index map sampled with beam-size ($\theta=8''\times8''$) circles. \textit{Bottom right}: Spectral index profile along filament A. }
        \label{fig: profiles G071}
\end{figure*}

\begin{figure*}[!h]
        \centering

\includegraphics[width=0.2\textwidth]{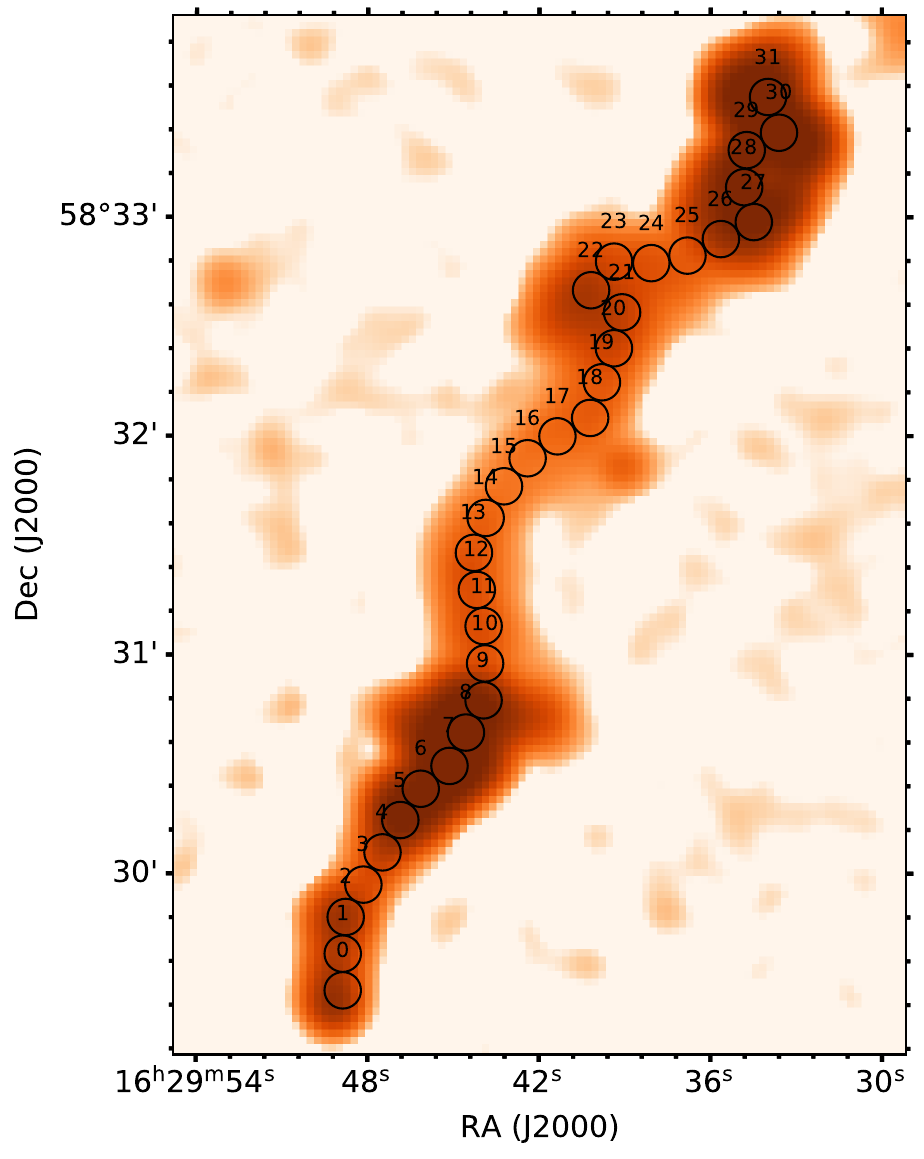}
\includegraphics[width=0.38\textwidth]{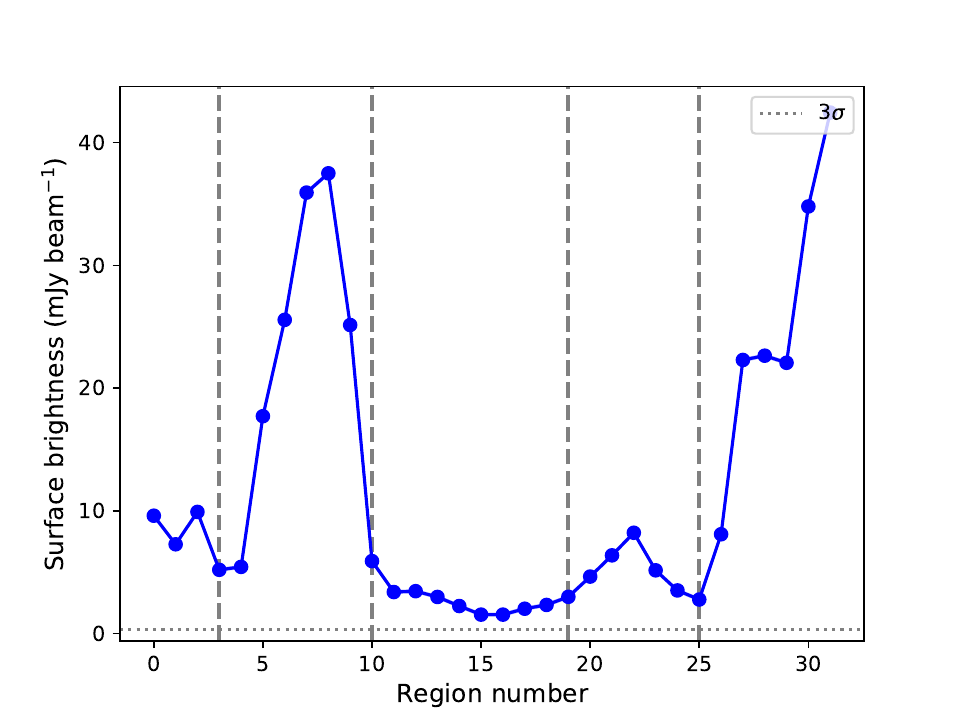} \\
\includegraphics[width=0.22\textwidth]{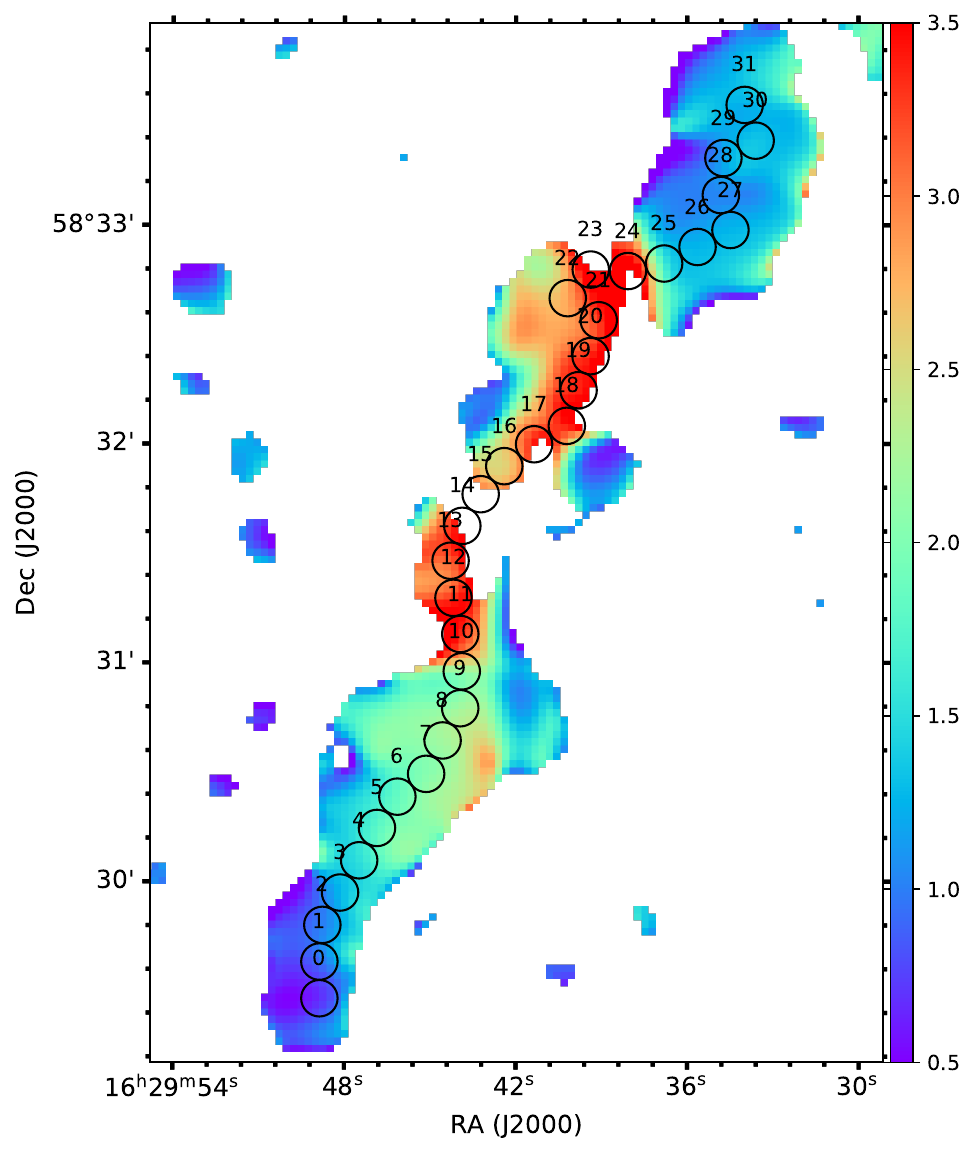}
\includegraphics[width=0.38\textwidth]{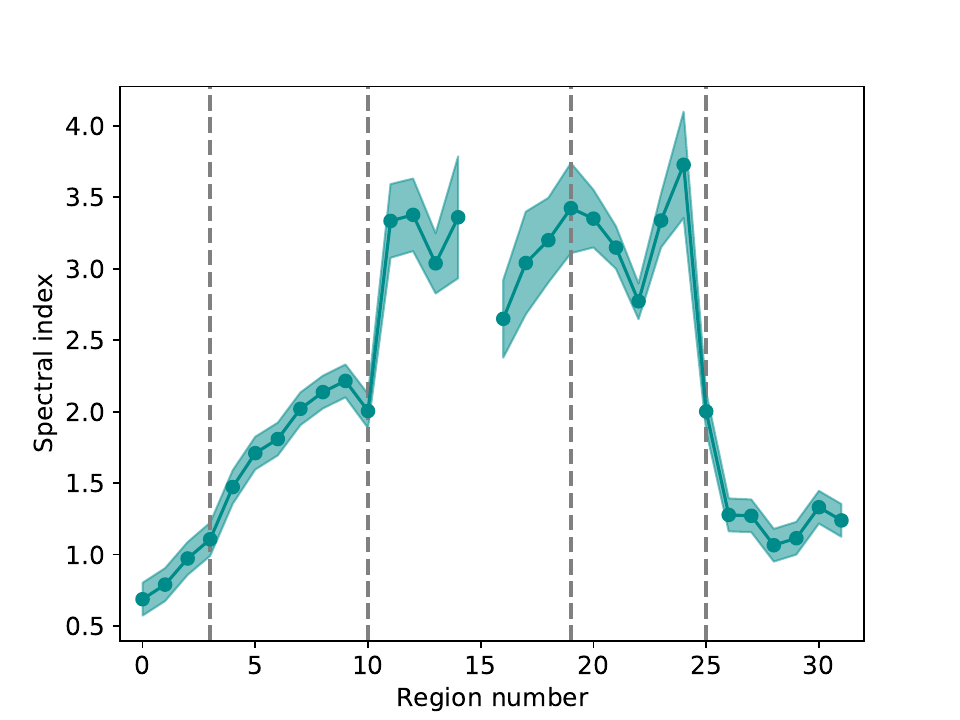}

        \caption{Profiles of the target in G088.  \textit{Top left}: 144 MHz image at resolution $\theta=10''\times10''$ sampled with beam-size circles ($10''=24$ kpc at $z=0.133$). \textit{Top right}: surface brightness profile. The horizontal dotted line indicates the $3\sigma$ level of the image. \textit{Bottom left}: spectral index map sampled with beam-size ($\theta=10''\times10''$) circles. \textit{Bottom right}: Spectral index profile. In both right panels, the transition between the source regions (A to E) are indicated by vertical dashed lines. }
        \label{fig: profiles G088}
\end{figure*}

\begin{figure*}[!h]
        \centering

\includegraphics[width=0.18\textwidth]{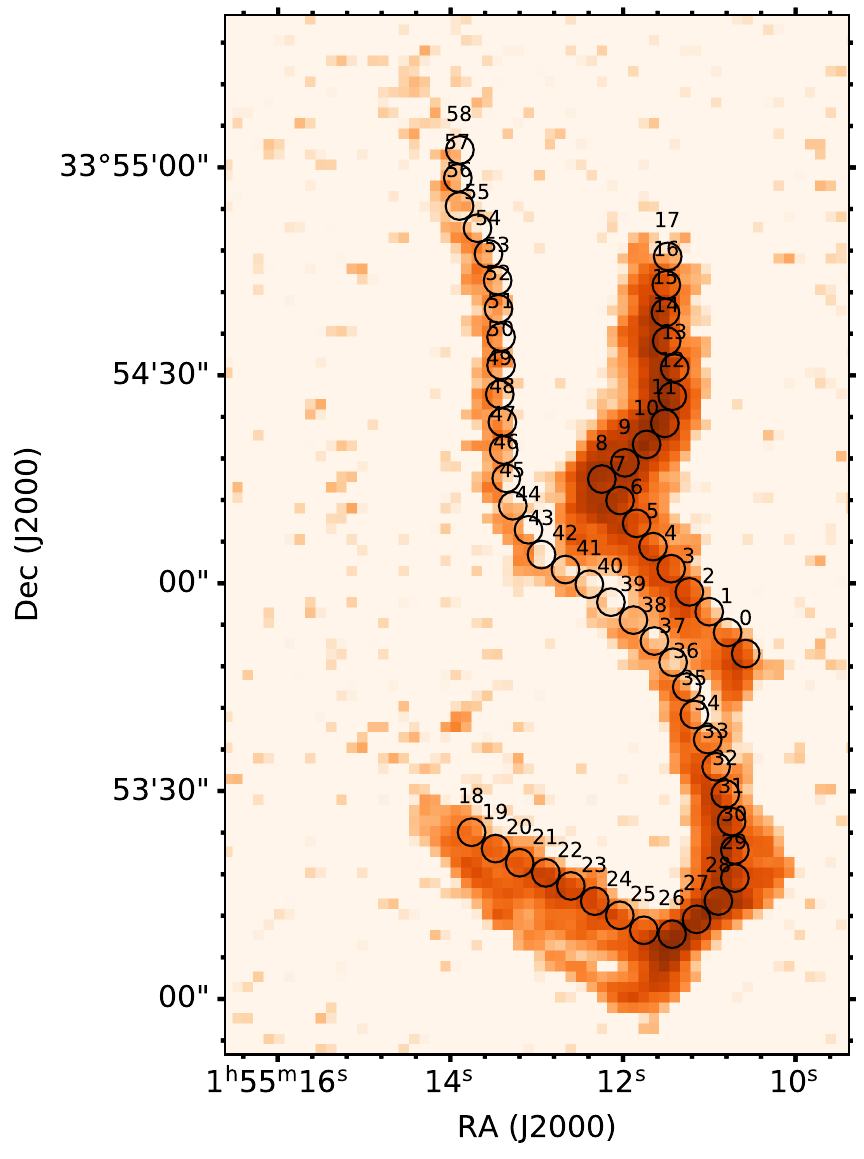}
\includegraphics[width=0.36\textwidth]{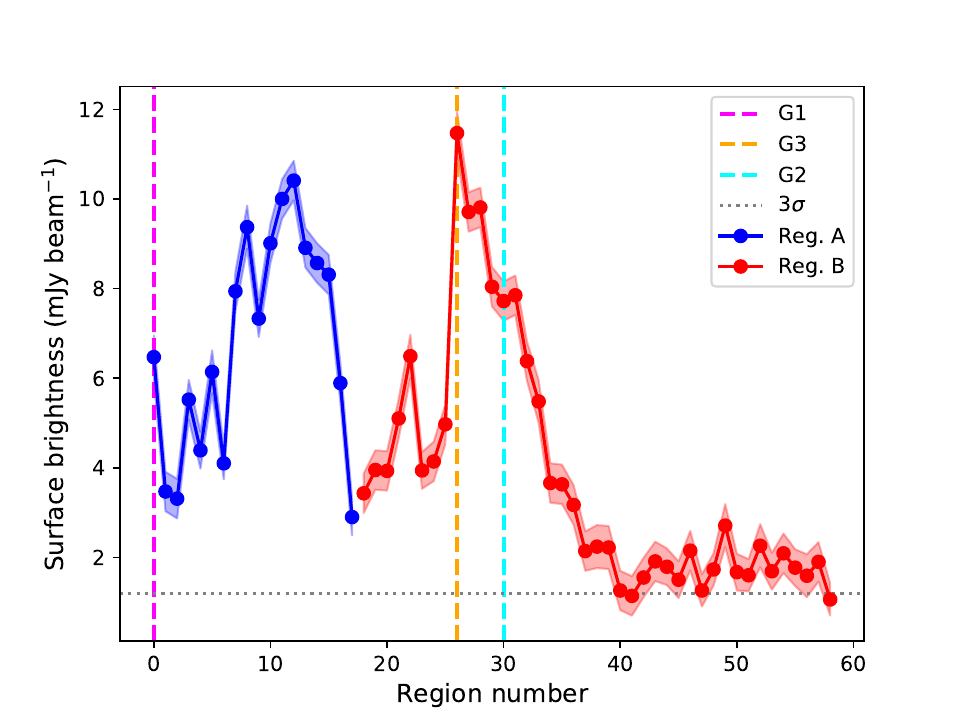} \\
\includegraphics[width=0.19\textwidth]{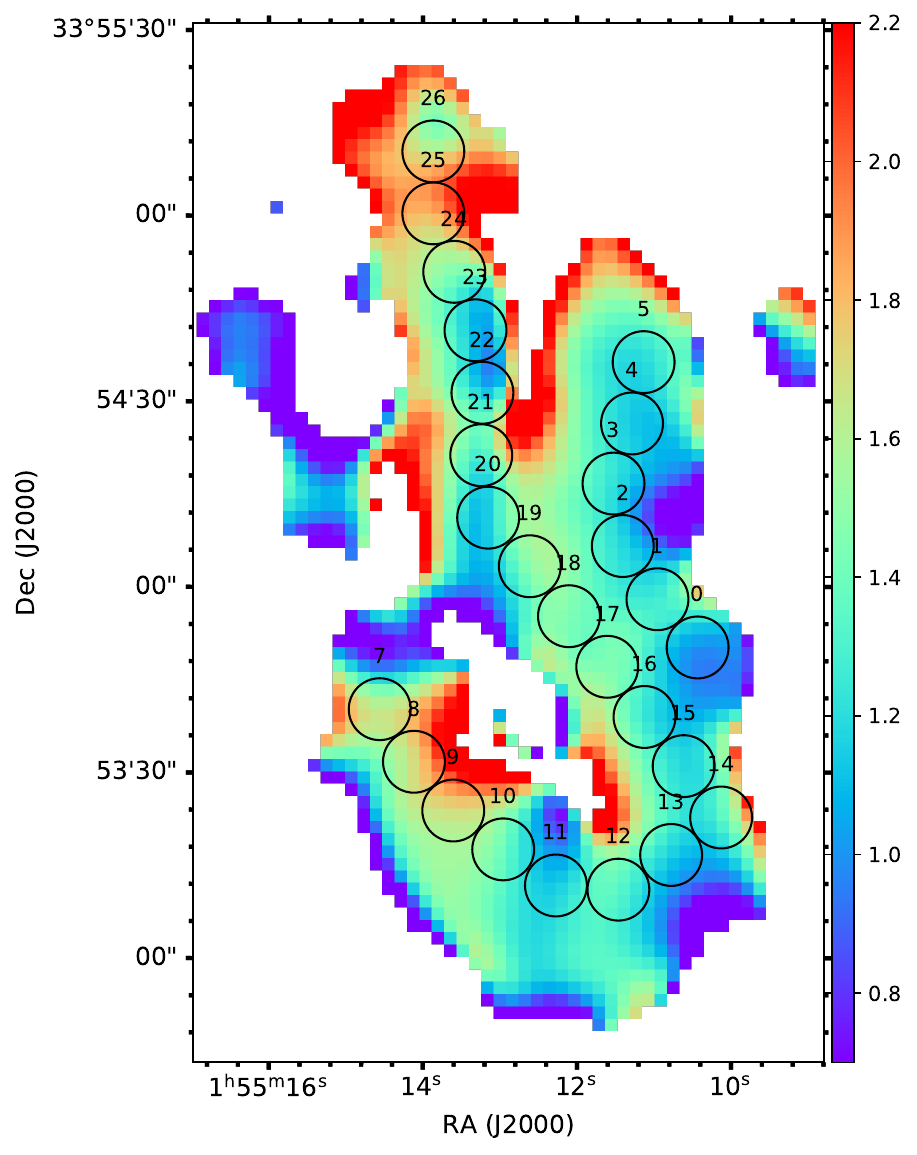}
\includegraphics[width=0.36\textwidth]{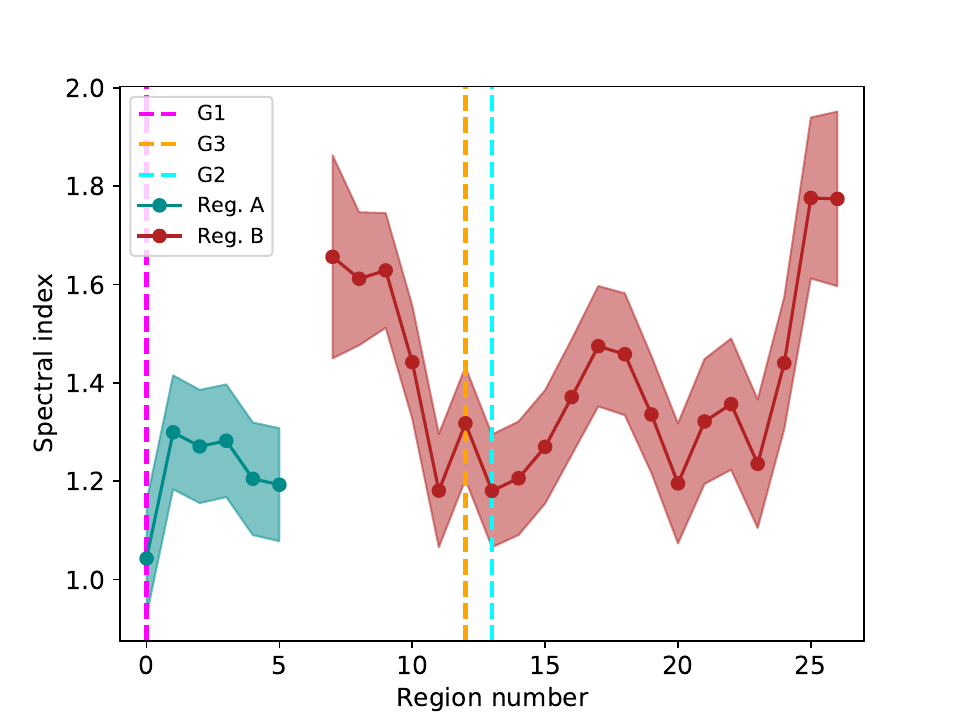}

        \caption{Profiles of the target in G137.  \textit{Top left}: 144 MHz image at resolution $\theta=4''\times3''$ sampled with beam-size circles ($4''=7$ kpc at $z=0.087$). \textit{Top right}: surface brightness profiles for source regions A and B. The horizontal dotted line indicates the $3\sigma$ level of the image. \textit{Bottom left}: spectral index map sampled with beam-size ($\theta=10''\times10''$) circles. \textit{Bottom right}: Spectral index profile for source regions A and B. In both right panels, the position of the galaxies G1, G2, and G3 is indicated by vertical dashed lines. }
        \label{fig: profiles G137}
\end{figure*}

\begin{figure*}[!h]
        \centering

\includegraphics[width=0.21\textwidth]{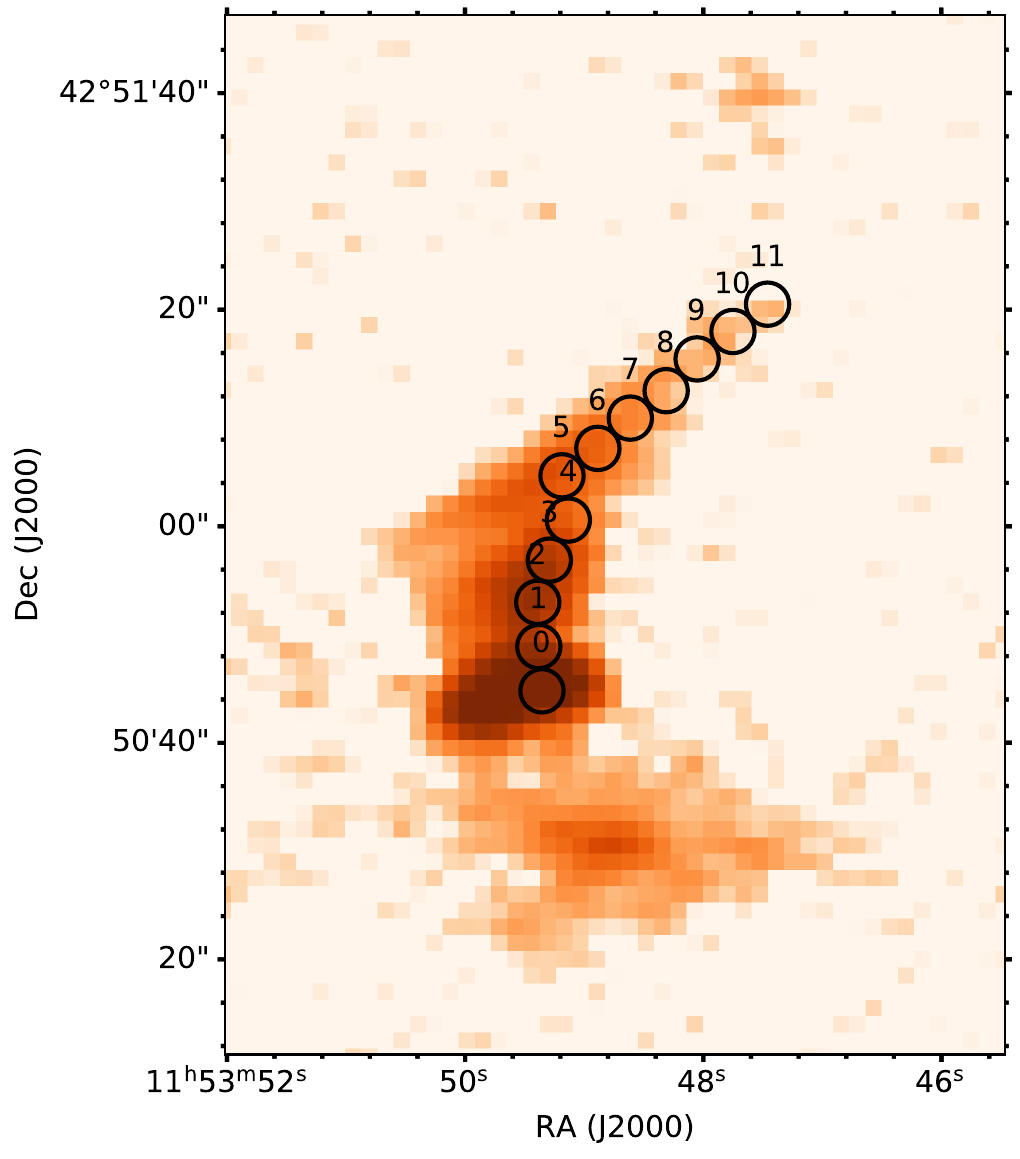}
\includegraphics[width=0.35\textwidth]{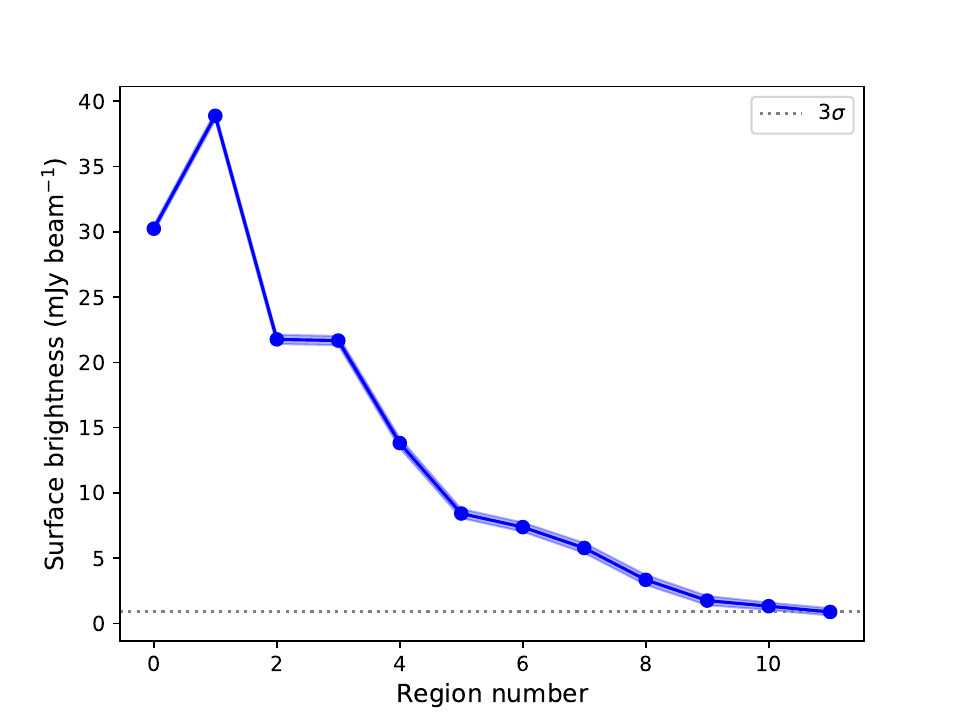} \\
\includegraphics[width=0.22\textwidth]{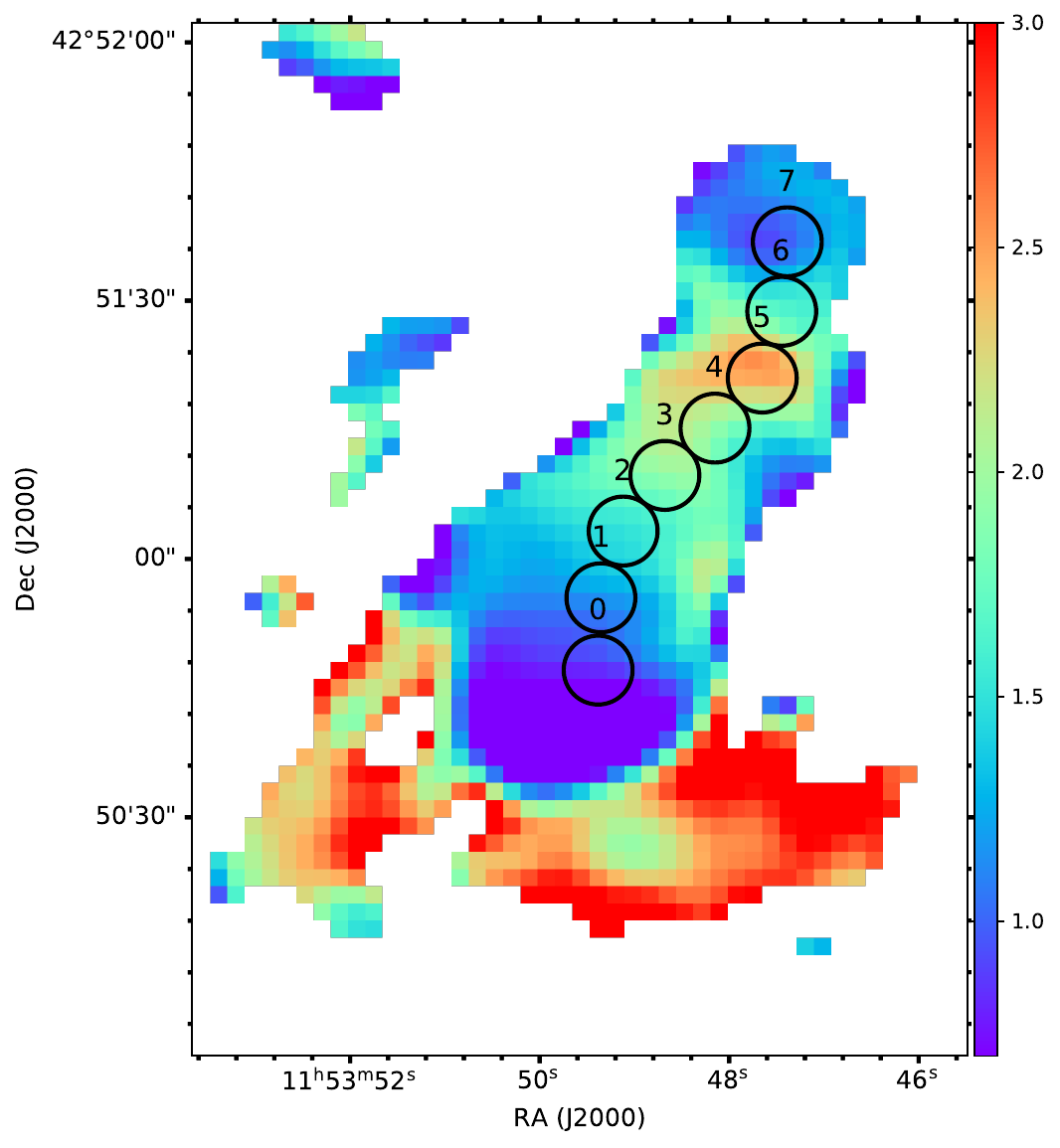}
\includegraphics[width=0.35\textwidth]{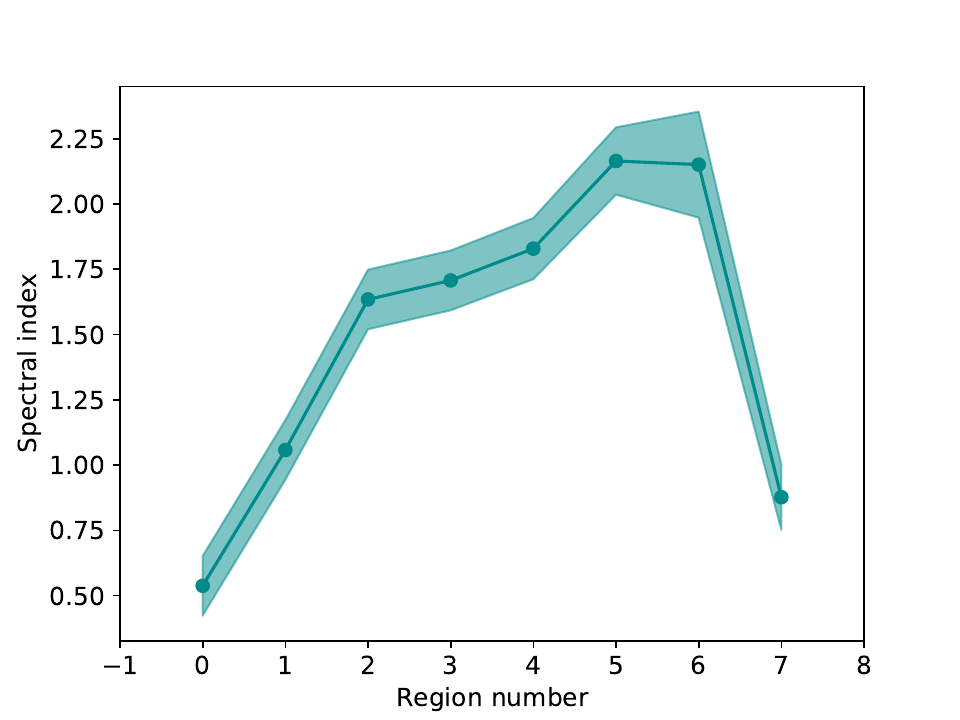}

\caption{Profiles of the target in G155.  \textit{Top left}: 144 MHz image at resolution $\theta=4''\times3''$ sampled with beam-size circles ($4''=19$ kpc at $z=0.333$). \textit{Top right}: surface brightness profile. The horizontal dotted line indicates the $3\sigma$ level of the image. \textit{Bottom left}: spectral index map ($\theta=8''\times8''$) sampled with beam-size circles. \textit{Bottom right}: Spectral index profile.}
        \label{fig: profiles G155}
\end{figure*}

In Figs. \ref{fig: profiles G071}-\ref{fig: profiles G155} we report surface brightness and spectral index profiles of the targets in G071, G088, G137, and G155. These profiles are discussed in Sect. \ref{sect: Morphological and spectral properties} to interpret the nature of the targets.

\FloatBarrier

\section{Possible hosts galaxies }
\label{sect: Properties of selected clusters}

In Table \ref{table: galaxies} we report redshift from the literature for galaxies identified as possible hosts of the radio sources (see Sect. \ref{sect: Morphological and spectral properties}). These have been identified by visual inspection of Pan-STARRS images, in regions close to the targets. In a few cases, the value of $z$ indicates that the galaxy is not a cluster member.

\begin{table}[!h]
 \fontsize{6.5}{6.5}\selectfont
 \centering
 \caption[]{Properties of the galaxies discussed in Sect. \ref{sect: Morphological and spectral properties}.}
 \label{table: galaxies}
 \begin{tabular}{cccccc}
 \hline
 \noalign{\smallskip}
 Cluster & Galaxy & ${\rm RA}_{\rm J2000}$ & ${\rm DEC}_{\rm J2000}$ & $z$ & Ref. \\ 
  &  & (deg) & (deg) &  &  \\ 
 \hline
 \noalign{\smallskip}
 G071 & G1 & 266.8148 & 45.1220 & 0.159 & 3 \\
 G071 & G2 & 266.8007 & 45.1252 & 0.157 & 3 \\
 G071 & G3 & 266.8111 & 45.1294 & $0.184 \pm 0.067$ & 1 \\
 G071 & G4 & 266.7858 & 45.1105 & $0.255 \pm 0.183$ & 1 \\
 G088 & G1 & 247.4121 & 58.5313 & 0.133 & 4 \\
 G088 & G2 & 247.4551 & 58.4898 & $0.141 \pm 0.012$ & 5 \\
 G088 & G3 & 247.3690 & 58.5637 & $0.172 \pm 0.027$ & 5 \\
 G113 & G1 & 2.9415 & 32.4832 & 0.125 & 3 \\
 G113 & G2 & 2.9264 & 32.4843 & 0.104 & 3 \\
 G113 & G3 & 2.9110 & 32.4744 & 0.099 & 1 \\
 G113 & G4 & 2.9115 & 32.4783 & 0.105 & 1 \\
 G137 & G1 & 28.7942 & 33.8967 & 0.085 & 2 \\
 G137 & G2 & 28.7951 & 33.8891 & $0.102 \pm 0.014$ & 1 \\
 G137 & G3 & 28.7979 & 33.8858 & $0.138 \pm 0.014$ & 1 \\
 G155 & G1 & 178.4482 & 42.8612 & 0.319 & 1 \\
 G155 & G2 & 178.4562 & 42.8453 & 0.327 & 1 \\
 G155 & G3 & 178.4536 & 42.8414 & $0.687 \pm 0.111$ & 1 \\
 G165 & G1 & 140.5134 & 51.9217 & $0.302 \pm 0.047$ & 5 \\
 G172 & G1 & 167.9317 & 40.8208 & 0.078 & 1 \\
 G172 & G2 & 167.9156 & 40.8401 & 0.074 & 1 \\
 G172 & G3 & 167.9347 & 40.8431 & $0.073 \pm 0.045$ & 5 \\
 \noalign{\smallskip}
 \hline
 \end{tabular}
 \begin{tablenotes}
 \item {\small \textbf{Notes}. Col. 1: cluster name. Cols. 2-4: galaxy ID as labelled in Figs. \ref{fig: mappefullres1}-\ref{fig: mappefullres7}, coordinates, redshift (values with and without errors are photometric and spectroscopic measurements, respectively). Col. 6: reference catalogue for $z$, being 1 for \cite{alam15}, 2 for \cite{wen&han15}, 3 for \cite{rines16}, 4 for \cite{dalya16,dalya18}, and 5 for \cite{duncan22}.}
 \end{tablenotes}
\end{table}

\FloatBarrier

\section{X-ray data and morphological parameters}
\label{sect: Xray data for G088}

\cite{zhang23} analysed X-ray data of PSZ2 clusters in LoTSS-DR2 with available \textit{Chandra} and/or XMM-Newton, produced the images shown in Figs. \ref{fig: mappefullres1}, \ref{fig: mappefullres3}, \ref{fig: mappefullres4}, and \ref{fig: mappefullres7}, and derived the centroid shift ($w$; \citealt{mohr93,poole06}) and concentration ($c$; \citealt{santos08}) morphological parameters. Such observations are not available for G088, G155, and G165 in our sample.

For G088, we retrieved pointed (11 ks long, ObsID: 800413, PI: A. Fabian) observations in the 0.1-2.4 keV band from the ROentgen SATellite carried out with the Position Sensitive Proportional Counter detector (\textit{ROSAT} PSPC). The \textit{ROSAT} PSPC exposure-corrected image (smoothed with a Gaussian function having ${\rm FWHM}\sim 100$ kpc) is shown in Fig. \ref{fig: mappefullres2}. Following similar procedures as those detailed in \cite{botteon22} and \cite{zhang23}, we measured $c$ and $w$ for G088 (Table \ref{tab: targets}). 

For G165, we retrieved and analysed \textit{Chandra} observations (10 ks long, ObsID: 7752, PI: C. Vignali). The 0.5-2 keV exposure-corrected image (smoothed with a Gaussian function having ${\rm FWHM}\sim 100$ kpc) is shown in Fig. \ref{fig: mappefullres6}. The sensitivity is insufficient to firmly claim the detection of the ICM, and thus determine the dynamical state of the cluster.

\begin{figure}[!h]
    \centering
    \begin{minipage}{0.45\textwidth} 
        \includegraphics[width=\linewidth]{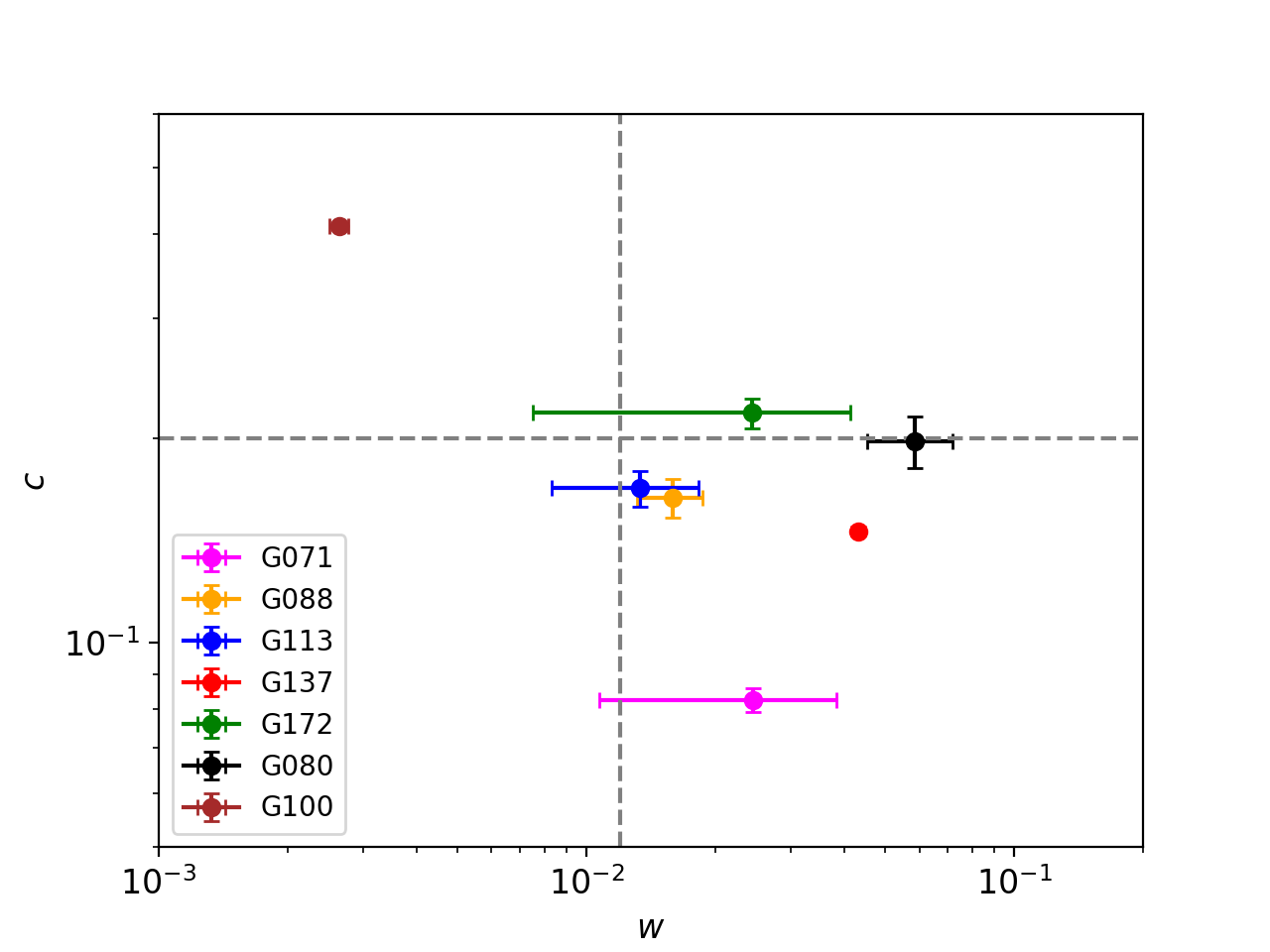}
    \end{minipage}%
    \hspace{1em}
    \begin{minipage}[t]{0.45\textwidth} 
        \captionof{figure}{Distribution of clusters with available X-ray data in the $c-w$ diagram.
        The dashed lines are the reference values $c=0.2$ and $w=0.012$ as in \citealt{cassano10A}. The clusters G080.41-33.24 and G100.45-38.42 are also reported (see details in Sect. \ref{sect: Sample selection}). }
        \label{fig: cw}
    \end{minipage}
\end{figure}

In Fig. \ref{fig: cw} we report the $c$ and $w$ distribution for clusters in our sample, considering in addition G080.41-33.24 and G100.45-38.42 (see details in Sect. \ref{sect: Sample selection}). The dashed lines ($c=0.2$, $w=0.012$; \citealt{cassano10A}) are references to distinguish between relaxed (high $c$, low $w$) and disturbed (low $c$, high $w$) systems. This plot shows that candidate revived fossils in our sample tend to inhabit disturbed systems. However, the low statistic does not allow us to provide firm conclusions. Moreover, we notice that G100.45-38.42, which is the host of the `Kite' revived source, is a relaxed system.

\FloatBarrier

\section{VLASS images}
\label{sect: VLASS images}

\begin{figure}[!h]
    \centering
    \begin{minipage}{0.33\textwidth} 
        \includegraphics[width=\linewidth]{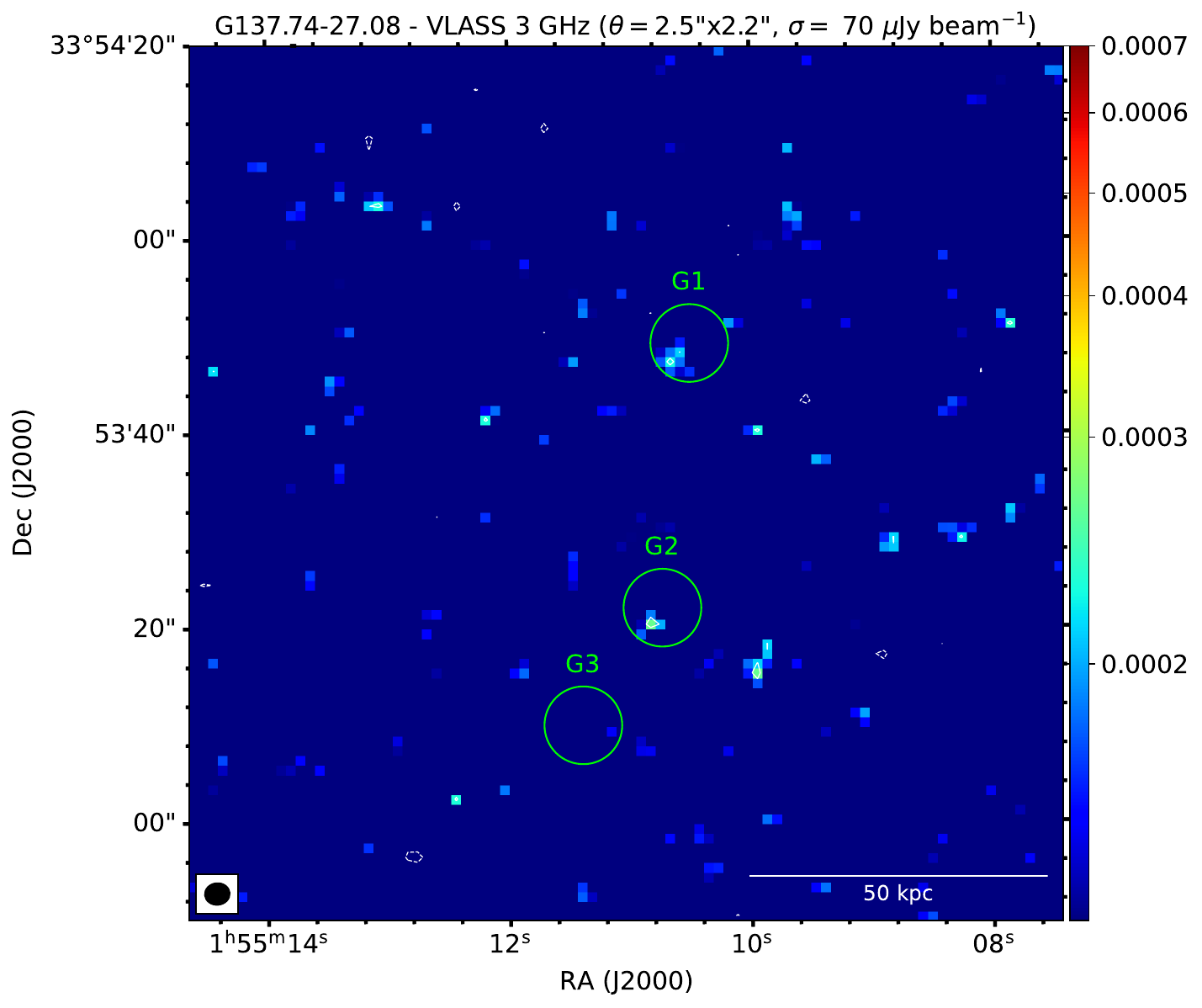}
    \end{minipage}%
    \hspace{1em}
    \begin{minipage}[t]{0.5\textwidth} 
        \captionof{figure}{VLASS radio image of G137.74-27.08 at 3 GHz (units are ${\rm Jy \; beam^{-1}}$). Resolution and noise are reported on top of the panel. The contour levels are $[\pm3, \;6, \;12,\; 24,\; ...]\times \sigma$. Green circles indicate optical sources discussed in Sect. \ref{sect: G137}. }
        \label{fig: vlass1}
    \end{minipage}
\end{figure}

\begin{figure}[!h]
    \centering
    \begin{minipage}{0.33\textwidth} 
        \includegraphics[width=\linewidth]{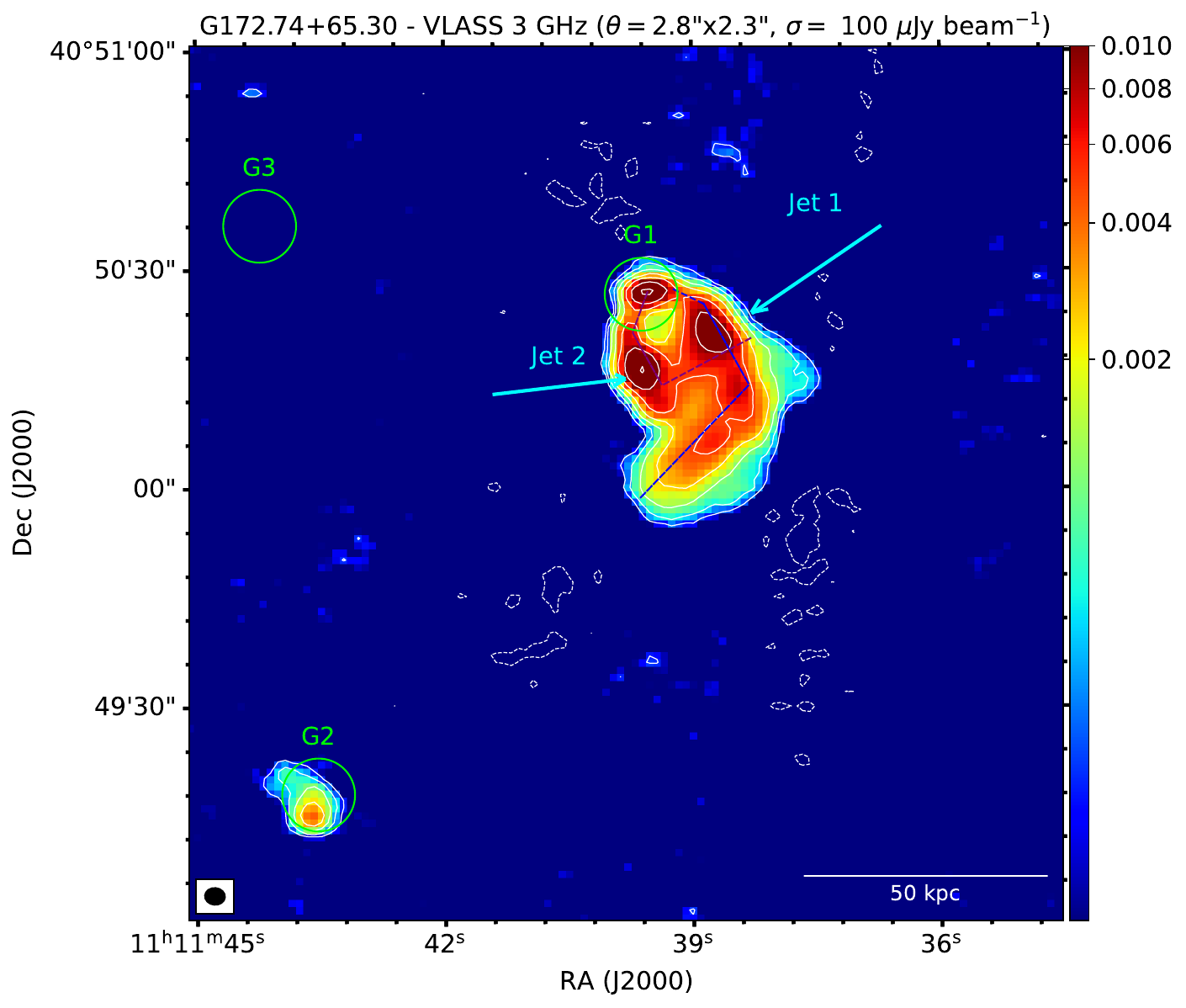}
    \end{minipage}%
    \hspace{1em}
    \begin{minipage}[t]{0.5\textwidth} 
        \captionof{figure}{VLASS radio image of G172.74+65.30 at 3 GHz (units are ${\rm Jy \; beam^{-1}}$). Resolution and noise are reported on top of the panel. The contour levels are $[\pm3, \;6, \;12,\; 24,\; ...]\times \sigma$. Green circles and cyan arrows indicate optical sources and regions discussed in Sect. \ref{sect: G172}. The blue and purple dashed lines roughly indicate the putative twisting path of jets 1 and 2, respectively.}
        \label{fig: vlass2}
    \end{minipage}
\end{figure}

In Figs. \ref{fig: vlass1} and \ref{fig: vlass2} we report VLASS images discussed in Sect. \ref{sect: G137} and Sect. \ref{sect: G172}.  

\end{appendix}

\end{document}